\documentclass[article]{aa}

\makeatletter
\let\AA@orig@enddocument\enddocument
\AtBeginDocument{\let\enddocument\AA@orig@enddocument}
\makeatother

\usepackage{graphicx}    
\usepackage[varg]{txfonts}             
\usepackage[dvipsnames]{xcolor}
\usepackage{hyperref}
\hypersetup{pdfborder = {0 0 0},
    colorlinks = true,
    linkbordercolor = {white},
    citecolor=NavyBlue,
    urlcolor=Green
}

\newcommand{\orcidlink}[1]{\protect\href{https://orcid.org/#1}{\protect\includegraphics[width=8pt]{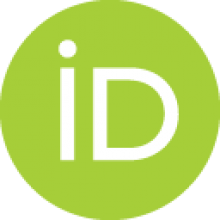}}}

\usepackage{amssymb}
\usepackage{amsmath}
\usepackage{xspace}
\usepackage{array,multirow}
\usepackage{enumitem}
\usepackage[english]{babel}
\usepackage{mathrsfs}
\usepackage{listings}
\usepackage{blindtext}
\usepackage{lipsum} 
\usepackage{booktabs}
\usepackage{url}
\usepackage{natbib,twoopt}
\usepackage[normalem]{ulem}  
\usepackage{placeins}

\newcommand{\ie}{i.e.\@\xspace} 
\newcommand{\eg}{e.g.\@\xspace} 

\renewcommand{\arraystretch}{1.}

\renewcommand{\eqref}[1]{Eq.~\ref{#1}}
\newcommand{\fref}[1]{Fig.~\ref{#1}}

\newcommand{\tref}[1]{Table~\ref{#1}}
\newcommand{\sref}[1]{Sect.~\ref{#1}}

\newcommand{\aref}[1]{Appendix~\ref{#1}}

\newcommand{\numax}{\ensuremath{\nu_{\rm max}}\xspace}
\newcommand{\dnu}{\ensuremath{\Delta\nu}\xspace}

\newcommand{\msol}{\ensuremath{\rm M_{\odot}}\xspace}

\newcommand{\kp}{\emph{Kepler}\xspace}

\newcommand{\teff}{\ensuremath{T_{\rm eff}}\xspace}

\newcommand{\logg}{\ensuremath{\log g}\xspace}
\newcommand{\feh}{\ensuremath{\rm [Fe/H]}\xspace}
\newcommand{\vsini}{\ensuremath{v\sin i}\xspace}

\numberwithin{equation}{section}

\makeatletter
\def\maketag@@@#1{\hbox{\m@th\normalfont\normalsize#1}}
\makeatother

\bibpunct{(}{)}{;}{a}{}{,} 

\makeatletter
\newcommand*\mysize{%
  \@setfontsize\mysize{5.7}{8.0}%
}
\makeatother

\makeatletter
\newcommand*\tabsize{%
  \@setfontsize\tabsize{7.}{8.0}%
}
\makeatother

\makeatletter
\newcommand\footnoteref[1]{\protected@xdef\@thefnmark{\ref{#1}}\@footnotemark}
\makeatother

\newcommand{\sn}{196\xspace}
\newcommand{\ssn}{128\xspace}

\newif\ifrefversion        
\refversionfalse         

\ifrefversion
  
  \DeclareRobustCommand{\revdel}[1]{\textcolor{red}{\sout{#1}}}

  \hypersetup{colorlinks,
              citecolor=blue,
              linkcolor=blue,
              urlcolor =blue}

\else
  
  \DeclareRobustCommand{\revdel}[1]{}
  
\fi

\begin{document}

   \title{Luminaries in the Sky: The TESS Legacy Sample of Bright Stars}
   \subtitle{I. Asteroseismic detections in naked-eye main-sequence \\and sub-giant solar-like oscillators}

   \titlerunning{The TESS Luminaries Sample}

   \author{Mikkel N. Lund\inst{\ref{I1}}\orcidlink{0000-0001-9214-5642} \and 
   Ashley Chontos\inst{\ref{IAsh}}\orcidlink{0000-0003-1125-2564} \and
   Frank Grundahl\inst{\ref{I1}}\orcidlink{0000-0002-8736-1639} \and
   Savita Mathur\inst{\ref{ISav}, \ref{ISav1}}\orcidlink{0000-0002-0129-0316}\and Rafael A. García\inst{\ref{IRaf}}\orcidlink{0000-0002-8854-3776} \and 
   {Daniel~Huber}\inst{\ref{haw},\ref{syd}}\orcidlink{0000-0001-8832-4488} \and
   Derek Buzasi\inst{\ref{DER}}\orcidlink{0000-0002-1988-143X} \and
   Timothy R. Bedding\inst{\ref{syd}}\orcidlink{0000-0001-5222-4661} \and
   Marc Hon\inst{\ref{MIT},\ref{haw}}\orcidlink{0000-0003-2400-6960} \and
    Yaguang Li\inst{\ref{haw}}\orcidlink{0000-0003-3020-4437}}
    
   \offprints{MNL, \email{mikkelnl@phys.au.dk}}          

    \institute{
    Stellar Astrophysics Centre, Department of Physics and Astronomy, Aarhus University, Ny Munkegade 120, DK-8000 Aarhus C, Denmark\label{I1} \and
    Department of Astrophysical Sciences, Princeton University, 4 Ivy Lane, Princeton, NJ 08540, USA\label{IAsh} \and 
    Instituto de Astrofísica de Canarias (IAC), E-38205 La Laguna, Tenerife, Spain\label{ISav} \and 
    Universidad de La Laguna (ULL), Departamento de Astrofísica, E-38206 La Laguna, Tenerife, Spain\label{ISav1} \and 
    Université Paris-Saclay, Université Paris Cité, CEA, CNRS, AIM, F-91191, Gif-sur-Yvette, France\label{IRaf} \and
    Institute for Astronomy, University of Hawai‘i, 2680 Woodlawn Drive, Honolulu, HI 96822, USA\label{haw} \and
    Sydney Institute for Astronomy (SIfA), School of Physics, University of Sydney, NSW 2006, Australia\label{syd} \and
    Department of Astronomy \& Astrophysics, University of Chicago, Chicago, IL 60637, USA\label{DER} \and
    Department of Physics and Kavli Institute for Astrophysics and Space Research, Massachusetts Institute of Technology, 77 Massachusetts Ave, Cambridge, MA 02139, USA\label{MIT}}

   \authorrunning{Lund et al.}
   \date{Received May 19, 2025; accepted July 15, 2025}
 
  \abstract
   {}
   {We aim to detect and characterise solar-like oscillations in bright naked‐eye ($V$<6) main-sequence and subgiant stars observed by the Transiting Exoplanet Survey Satellite (TESS). In doing so, we seek to expand the current benchmark sample of oscillators, provide accurate global asteroseismic parameters for these bright targets, and assess their potential for future detailed investigations -- including missions such as the Habitable Worlds Observatory (HWO) and PLAnetary Transits and Oscillations of stars (PLATO).}
   {Our sample of bright stars was selected from the Hipparcos/Tycho catalogues. We analysed TESS photometry from both 120‑s and 20‑s cadences using the standard TESS Science Processing Operations Center (SPOC) light curves and custom apertures extracted from target pixel files. After applying a filtering of the light curves, we extracted global asteroseismic parameters (\numax and \dnu) using the \texttt{pySYD} pipeline. Results were cross-validated with independent pipelines and compared to predictions from the Asteroseismic Target List (ATL), while noise properties were evaluated to quantify improvements from a 20-s observing cadence.}
   {We detect solar-like oscillations in a total of \sn stars—including \ssn new detections -- with extracted \numax and \dnu values showing strong conformity to expected scaling relations. This corresponds to an increase by more than an order of magnitude in the number of main-sequence stars with detection of solar-like oscillations from TESS. Importantly, our sample of newly detected solar-like oscillators includes nearly 40\% of the prime targets for HWO, paving the way for a systematic determination of asteroseismic ages that will be important for the possible interpretation of atmospheric biosignatures. Our analysis confirms that 20‑s cadence data yields lower high-frequency noise levels compared to 120‑s data. Moreover, the precise stellar parameters obtained through asteroseismology establish these bright stars as benchmarks for seismic investigations and provide useful constraints for refining stellar evolution models and for complementary analyses in interferometry, spectroscopy, and exoplanet characterisation.}
   {}

   \keywords{Asteroseismology -- Stars: oscillations -- Stellar properties -- Catalogues -- Binaries: general -- Planetary systems -- Methods: data
analysis}

   \maketitle
%

\section{Introduction}

Asteroseismology from space-based photometric missions has, in the last decade-and-a-half, heralded a revolution in our understanding of stars, exoplanet characterisation, Galactic archaeology, and many more fields within the broad landscape of astrophysics \citep{Chaplin2013,garcia2019}. This revolution was pioneered by the CoRoT \citep[Convection, Rotation, and planetary Transits,][]{Baglin2006,Michel2008} and \kp/K2 missions \citep{Borucki2010,Howell2014} and the field is now steadily maturing with the ongoing observations by the Transiting Exoplanet Survey Satellite \citep[TESS;][]{Ricker2014}, which we are hopeful will continue for many years to come.

Although TESS, with its larger pixel size, shorter continuous observation periods, and higher background noise compared to, \eg, \kp, has not (yet) yielded the same specific breakthroughs in asteroseismology of main-sequence solar-like oscillators, it offers several unique advantages. TESS provides nearly full-sky coverage, enabling the study of a vast array of bright, well-characterized stars, and for specific regions of the sky the total observing baseline has now surpassed the 4 years from \kp. A $10$-min observing cadence now enables asteroseismic studies of stars down to the subgiant regime based on full-frame images, and targeted $120$-s and $20$-s cadence observations have pushed the photometric asteroseismic studies into the K-dwarf regime \citep{Hon2024}. 

The bright stars that can be uniquely observed by TESS are crucial because they serve as important benchmarks for future missions, such as PLATO \citep[PLAnetary Transits and Oscillations of stars,][]{Rauer2024}, and many of them will be included in the target list for the future Habitable Worlds Observatory (HWO) mission \citep{Mamajek2024}. Moreover, their brightness makes them ideal for follow-up ground-based observations, which can significantly enhance our understanding of stellar physics.

With this work, we present a catalogue of \sn asteroseismic detections for the brightest solar-like main-sequence (MS) and subgiant (SG) oscillators observed by TESS. Hereafter, we refer to this cohort of stars as the ``TESS Luminaries Sample'' (TLS). Of these, \ssn are to our knowledge new detections, thereby more than doubling the number of known bright MS/SG oscillators, and include many of the stars predicted by \citet{Bedding1996} to show oscillation in the early days of asteroseismology. We have limited our sample to stars with $V<6$, hence stars that, for most people, will be visible to the naked eye under favourable observing conditions. For MS/SG stars in the brightness range of interest here, asteroseismic analyses based on TESS data have to date mainly focused on single or few stars \citep[see, \eg,][]{Ball2020,Ball2022,Metcalfe2020,Nielsen2020,Chontos2021,Huber2022} -- with the TLS catalogue we aim to provide a comprehensive asteroseismic catalogue of all the bright solar-like oscillators observed by TESS, and we will continue to update and extend the catalogue as more data from TESS become available. With its focus on MS/SG stars, the TLS catalogue can be seen as a complement to the HD-TESS catalogue by \citet{Hon2022} focusing on bright evolved stars observed by TESS.
The adopted brightness limit is admittedly somewhat arbitrary, but our analysis suggests that most fainter stars have already been covered by the recent impressive catalogues by \citet{Hatt2023} and \citet{Zhou2024}, both providing detections for thousands of solar-like oscillators from TESS, but mainly focusing on evolved solar-like stars (SG and RG). So, while we can generally confirm their detections for overlapping stars, we note that many of the brightest stars were not included in their analyses.

For the stars with detected oscillations, we provide measurements for the global asteroseismic parameters \dnu and \numax \citep{Chaplin2013,garcia2019}. With our results, we seek to highlight the opportunities for further detailed analysis from TESS for this cohort of stars that can prove valuable as benchmarks for stellar evolution theory and the asteroseismic method. Importantly, several of the stars that can be characterised asteroseismically by TESS, and in some cases by ground-based facilities such as SONG \citep{Grundahl2017}, will also be observed by the coming ESA PLATO mission and can here serve as key benchmarks for the quality of PLATO results (\sref{sec:plato}). 

The brightness of the stars in this catalogue enables a detailed characterisation of their properties. This is both in terms of asteroseismology, with the current and future observation from TESS, and crucially also from ground-based observations such as spectroscopy \citep[\eg,][]{Tautvai2022}, interferometry \citep{Vrard2024}, spectropolarimetry \citep{Metcalfe2023,Metcalfe2024}, measuring binary or multiple-star properties \citep{Malla2024}, and in some cases asteroseismic follow-up in radial velocity \citep{Kjeldsen2025}. 

The paper is structured as follows: in \sref{sec:sample} we outline the target sample, in \sref{sec:data} we describe our treatment of TESS data, while in \sref{sec:ast} we describe our asteroseismic analysis methodology. \sref{sec:res} is devoted to presenting the results of our analysis and highlighting the potential use cases and interesting aspects of the stars in our sample, including synergies with PLATO and HWO, exoplanets, interferometry, binarity, solar analogues, and individual cases.

This paper is the first in a series, and follow-up papers in development will focus on the detailed peak-bagging and stellar modelling for the highest quality targets in the sample, a detailed analysis of the targets that overlap with PLATO fields, an analysis of the detectability of solar-like oscillators near the red-edge of the classical instability strip, updates to exoplanet parameters and analysis of HWO targets, and a detailed analysis of several individual stars/systems.

\section{Targets}\label{sec:sample}
\begin{figure}
\centering
   \includegraphics[width=\columnwidth]{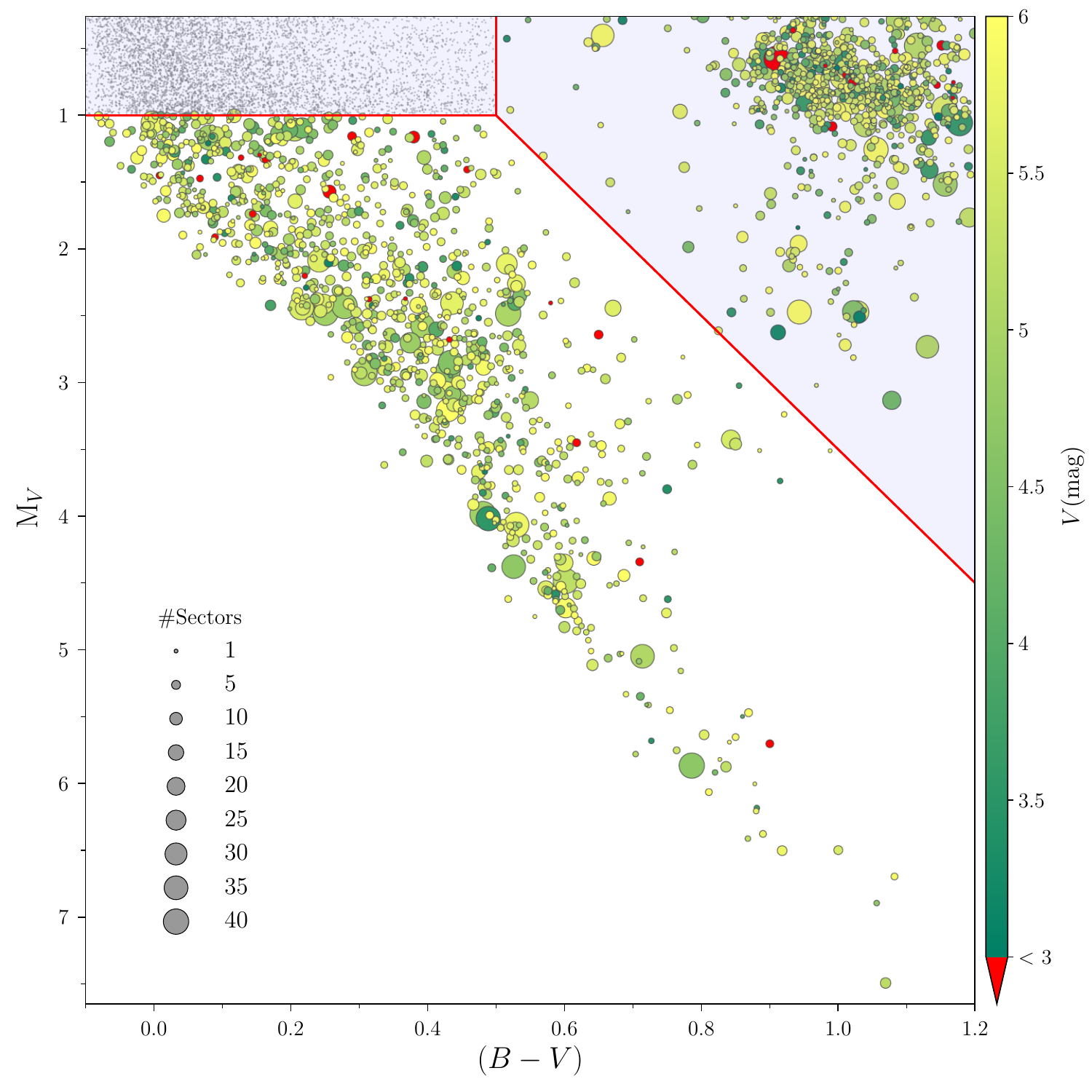}
   \caption{HR-diagram showing the criteria for our selection of targets. The red line shows our limits on ${M}_V$ and $(B-V)$, and targets in the lower non-shaded region were selected for analysis. The marker colour indicates the $V$-band magnitude for the stars, while the marker size indicates the number of sectors TESS will have observed a given star in Cycle 6 (up to and including Sector 83). The stars in the top left shaded box are not expected to show solar-like oscillations and here we have not indicated $V$ nor the number of sectors. The stars in the top-right shaded region contain evolved stars that may well show solar-like oscillations -- these will be the subject of a future study.}
   \label{fig:target_selection}
\end{figure}
The starting points for building our sample are the Hipparcos and Tycho catalogues \citep{Perryman1997,Hoeg1997,vanLeeuwen1997,hipparcos1997}. To limit our sample to stars visible to the naked eye, we first selected stars with a $V$-band magnitude below $6$. Based on Hipparcos parallaxes we computed the absolute $V$-band magnitude, $M_V$, and used this in combination with the $(B-V)$ colour as a proxy for the effective temperature to select stars that reside on the main-sequence (MS) or sub-giant (SG) branch -- our specific selection criteria are shown in \fref{fig:target_selection}. The adopted limits are conservative because many of the high-$M_V$ targets are expected to be classical pulsators. However, we wanted to ensure that no solar-like oscillators near the red edge of the classical instability strip were excluded.  

We also checked the Yale Bright Star Catalog \citep[5th ed.;][]{1995yCat.5050....0H} and the Gliese Catalogue of Nearby Stars \citep[3rd ed.;][]{1991adc..rept.....G} for targets that Hipparcos may have missed. However, none of the missing targets identified matched our selection criteria for $M_V$ and $B-V$. We note that for several targets that are members of binary systems, the Hipparcos (HIP) ID generally refers to the system rather than the individual components, as opposed to the TESS input catalogue \citep[TIC;][]{Stassun2019}, which builds on \textit{Gaia} \citep{GaiaDR2_2018} and the Two Micron All Sky Survey \citep[][]{Cutri2003} -- for these targets we manually assigned the TIC ID for the main component of the binary to the HIP ID.

Finally, we excluded known oscillators $\alpha$ Cen A and~B \citep{Bouchy2001,Bouchy2002,Carrier2003, Kjeldsen2005}, and $\alpha$ CMi \citep[Procyon;][]{Brown_1991, Martic1999,Arentoft2008} from our analysis as these extremely bright targets require special treatment in the form of a recalibration of the target pixel files and the smear data\footnote{the \texttt{Python} module \texttt{mundey} (\url{https://github.com/hvidy/mundey}) by White \& Pope could possibly achieve this.}. 

From data availability, our sample is limited to stars with $120$-s or $20$-s cadence observations in at least one TESS sector up to and including Sector $77$. These criteria leave us with $1060$ stars to be analysed (see \fref{fig:target_selection}), and of these $311$ have at least one sector with $20$-s cadence observations. 

By the end of TESS Cycle 7 (up to and including Sector $92$) an additional $69$ targets that match our $M_V$ and $B-V$ selection criteria will have been observed (the observing cadence will depend on their inclusion in Guest Investigator proposals), among which are well-known asteroseismic targets such as $18$ Sco \citep{Bazot2011} and 70 Oph \citep{Carrier2006_70Oph}. These targets are primarily located near the ecliptic plane in the constellations of Libra, Scorpius, Ophiuchus, and Sagittarius, and none of them overlap with the PLATO fields (\sref{sec:plato}). These targets will generally only be observed during a single sector and, in a few cases, 2--3 sectors, so for the most promising of these targets having access to $20$-s cadence observations will be particularly important. Finally, $19$ Hipparcos targets that match our $M_V$ and $B-V$ criteria have not been, and will not be, observed by TESS up to and including TESS Cycle 7. \tref{tab:all_miss} in \aref{app:missing} provides an overview of these targets.

\section{Data}\label{sec:data}
\begin{figure*}
\centering
   \includegraphics[width=\textwidth]{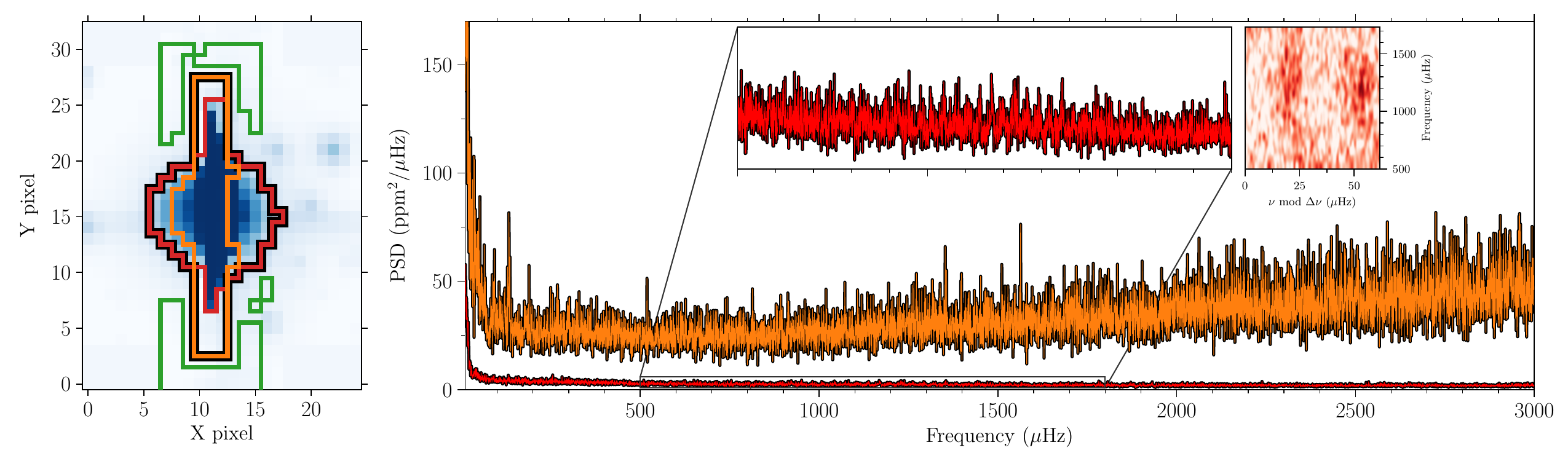}
   \caption{Example of the effect of adopting custom apertures, here for \object{$\psi^1$ Dra A} (\object{TIC 441804568}) as observed during Sector 15 in $120$-s cadence, where the SPOC aperture is missing several high-flux pixels. Similar apertures are seen for $\psi^1$ Dra A in other sectors. Left: The adopted aperture, shown with a black outline, combines the SPOC aperture in orange and the K2P$^2$ aperture in red. The green outline shows the aperture used to estimate the background. In blue the median pixel flux levels are shown on a log-scale. Right: Segments of the power-density spectra of the filtered SAP light curves from the apertures on the left, where the PSD obtained from the SPOC aperture is shown in orange and the one from the adopted custom aperture is shown in red. The inset shows a zoom of the region with identified oscillations from the custom aperture data. The small inset to the right shows the échelle diagram of the zoomed region after correcting for the background using a robust Siegel slope estimator.}
   \label{fig:custom_aperture}
\end{figure*}
Our main source of data comes in the form of \texttt{PDCSAP} light curves generated by the TESS Science Processing Operations Center \citep[SPOC;][]{Jenkins2016}. We used light curves with both a 120-s and, when available, 20-s cadence (introduced from Sector 27 onwards). In sectors where both 20- and 120-s light curves were available, we also used the 20-s data binned to a 120-s cadence. As demonstrated by \citet{Huber2022}, the 20-s data from TESS generally has lower noise than 120-s data, especially for bright stars, and with a reduction of scatter of the order ${\sim}25\%$ at TESS magnitude $6$ (see \sref{sec:detect}).
All data were downloaded from the Mikulski Archive for Space Telescopes (MAST) using functionalities provided by \texttt{Astroquery} \citep{astroquery2019} and \texttt{lightkurve} \citep{Lightkurve2018}.

For the targets where a detection of oscillations was made from an initial inspection of the data, and/or where a detection was expected\footnote{using the conservative criterion of a detection probability of $P_{\rm det}\geq50\%$.} from the asteroseismic target list \citep[ATL;][]{Schofield2019} in the updated version provided by \citet{Hey2024}\footnote{available via the \texttt{Python} module \texttt{tess-atl} (v0.1) at \url{https://github.com/danhey/tess-atl}}, we also analysed the data contained in the star's target pixel files (TPFs; see \sref{sec:ast}). In the TPF analysis, we specifically focused on the light curve improvement from constructing custom apertures. We used the K2P$^2$ pipeline \citep{Lund2015} to create apertures via the density-based clustering algorithm \texttt{DBSCAN}, as implemented in \texttt{scikit-image} \citep{scikit-image}, combined with a watershed algorithm to separate close targets. The clustering algorithm was applied to pixels with flux values greater than $3$ times the standardised median absolute deviation (MAD) from the median flux level. This routine is, by construction, excellent at defining apertures that include pixels that are grouped, but the occasional long bleed trails from the blooming of brighter saturated stars are more difficult to handle. However, while the apertures defined by SPOC are often small near the centroid of the star, the bleed trails are generally well-covered. Therefore, in the end, we opted to construct apertures by combining those from K2P$^2$ and SPOC. 

In certain instances, we could only make a positive seismic detection after adopting the custom joint aperture and, for several cases, the custom apertures significantly improved the data quality. \fref{fig:custom_aperture} provides an example of a problematic SPOC aperture for the star \object{$\psi^1$ Dra A} (\object{TIC 441804568}). The scatter in the light curve introduced by the SPOC aperture, which is missing several bright pixels, inhibits the detection of a seismic signal. As seen, when adopting the larger custom aperture, the noise level in the power spectral density (PSD) is significantly reduced, and oscillations can be detected.

We noticed that the pixel stamp available for 20-s cadence data was sometimes significantly smaller than the corresponding 120-s cadence pixel stamp -- examples of this are \object{$\delta$ Eri} (\object{TIC 38511251}), \object{$\gamma$ Lep} (\object{TIC 93280676}), and \object{$\epsilon$ Eri} (\object{TIC 118572803}). For these stars, the saturated bleed trails extend beyond the 20-s cadence stamp, causing the resulting light curves to have high noise levels. This inhibits the detection of oscillations from the 20-s data, while detections from the 120-s data are possible. 

In some cases, mainly for the brightest stars in the sample (\eg \object{ $\theta$ UMa}; \object{TIC 150226696}), SPOC light curves are not available for all sectors. In these cases, TPF data are needed to make full use of the TESS observations. We refer to \aref{app:comp} for additional details on the comparison of SPOC and custom apertures.

Before searching for oscillations, we processed the light curves using the KASOC filter \citep{Handberg2014} to correct for any spurious signals or long-term trends \citep[see, \eg,][]{Campante2019,Jiang2020,Ong2021}. In this filtering and the construction of new apertures, we followed the suggestion by \citet{Huber2022} to adopt the \texttt{default} quality bitmask defined by \texttt{lightkurve}\footnote{\url{https://github.com/nasa/Lightkurve/blob/master/lightkurve/utils.py\#L19}} for 120-s cadence data and the \texttt{hard} bitmask for 20-s cadence data.

\section{Asteroseismic analysis}\label{sec:ast}

To identify targets of interest for asteroseismic analysis, an initial inspection of SPOC and TPF data (\sref{sec:data}) was conducted using a variety of methods, including visual inspection of all data, autocorrelation function (ACF) analysis of the PSD, échelle diagrams, comparison to expectations from the ATL, etc. To obtain the first estimates of the global asteroseismic parameter \numax (the frequency of maximum oscillation power) we used a 2D-ACF method \citep[see][]{Verner2011}, as implemented in \texttt{lightkurve}.

For stars with detected oscillations (or an expected detection of oscillations from the ATL) we used the \texttt{pySYD} pipeline \citep{PySyd} to extract the global asteroseismic parameters \dnu and \numax. The \texttt{pySYD} pipeline is an open-source adaptation of the closed-sourced \texttt{SYD} pipeline \citep{SYD}, which was extensively tested and benchmarked to \kp LEGACY results \citep{Lund2017}. To summarise, \texttt{pySYD} first performs an automated optimisation to identify the best-fit background model due to stellar granulation on different timescales and with varying amplitudes, which can ultimately bias parameter estimates if not properly accounted for. The best-fit background model is then subtracted from the PSD to calculate a background-corrected power spectrum (BCPS), from which \numax and \dnu can be measured. The frequency corresponding to maximum power (\numax) is adopted as the frequency with peak power in the heavily-smoothed BCPS. An autocorrelation function is then used to identify the characteristic frequency spacing (\dnu), which corresponds to the average frequency separating modes of the same spherical degree ($l$) and consecutive radial order ($n$). Due to highly correlated data and the stochastic nature of solar-like oscillations, parameter uncertainties are estimated through a bootstrapping technique discussed in more detail in \citet{SYD}.

As mentioned in \sref{sec:data}, a detection of oscillations is not always possible from all data products, either because of problems with the aperture and systematics in the light curve or simply from too high noise levels from 120-s cadence data compared to 20-s cadence data. To this end, \texttt{pySYD} was applied to PSD prepared from each of the SPOC and custom aperture light curves. We tested on both 120-s cadence and, if available, 20-s cadence data. We calculated a variance-weighted PSD from the full light curves, as well as a PSD from the weighted PSD of individual sectors (weighted by the average variance in each sector). From the results obtained from the different data sets, we identified the ones that agreed within 3 standardised MAD of the median of all results in both \numax and \dnu, checked that these agreed within errors with the initial assessment of \numax, and that \dnu and \numax were in correspondence with the known relation between these parameters \citep{Hekker2009, Stello2009, Huber2011}. From the measurements meeting these requirements, we computed the variance-weighted averages of \dnu and \numax as our final reported parameters; see \sref{sec:astero_params} for details. 

As an additional validation of the identified detections, we also applied the \texttt{A2Z} pipeline \citep{Mathur2010} to our data. Briefly, \texttt{A2Z} first blindly searches the modes by computing the power spectrum of the power spectrum in sliding boxes along the PSD. This allows us to measure \dnu along with the frequency range where the modes are detected. The background is then fitted with one Harvey model \citep{Harvey1985} for the granulation and a component for the photon noise. After subtracting the background, a Gaussian function is fitted around the frequency range found in the first step to obtain \numax. For cases with low SNR, the search for the modes was forced to the expected \numax.
The 196 stars in the sample represent the consolidated cohort of stars analysed.  

\section{Results}\label{sec:res}
\begin{figure}
\centering
   \includegraphics[width=\columnwidth]{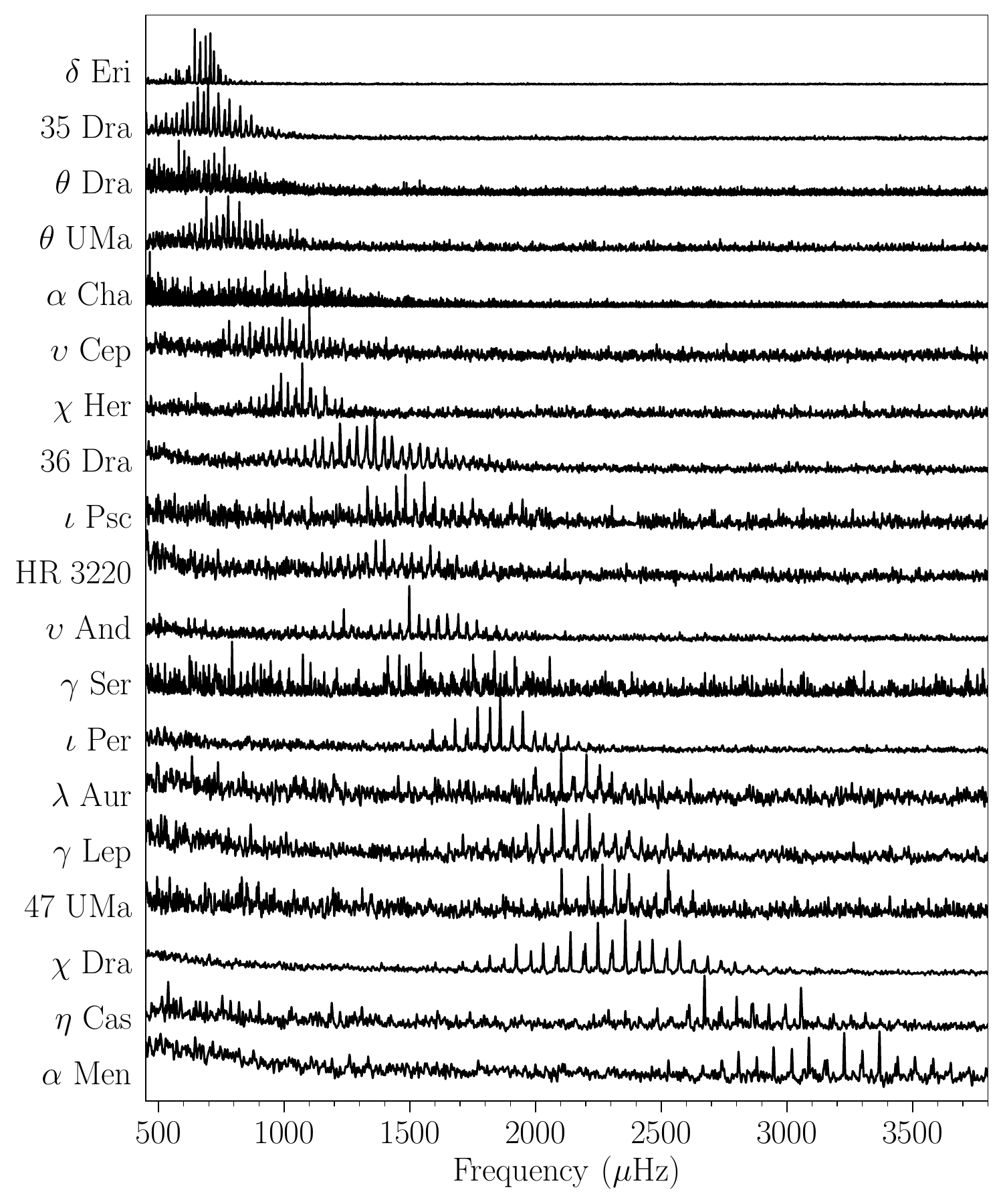}
   \caption{Examples of PSD for a small subset of stars with detected oscillations, arranged according to increasing \numax (see \tref{tab:all_seis} for details). The spectra have been smoothed by an Epanechnikov kernel \citep{Epanechnikov} with a width of $\dnu/20$.}
   \label{fig:examples}
\end{figure}
\begin{figure*}
\centering
   \includegraphics[width=\textwidth]{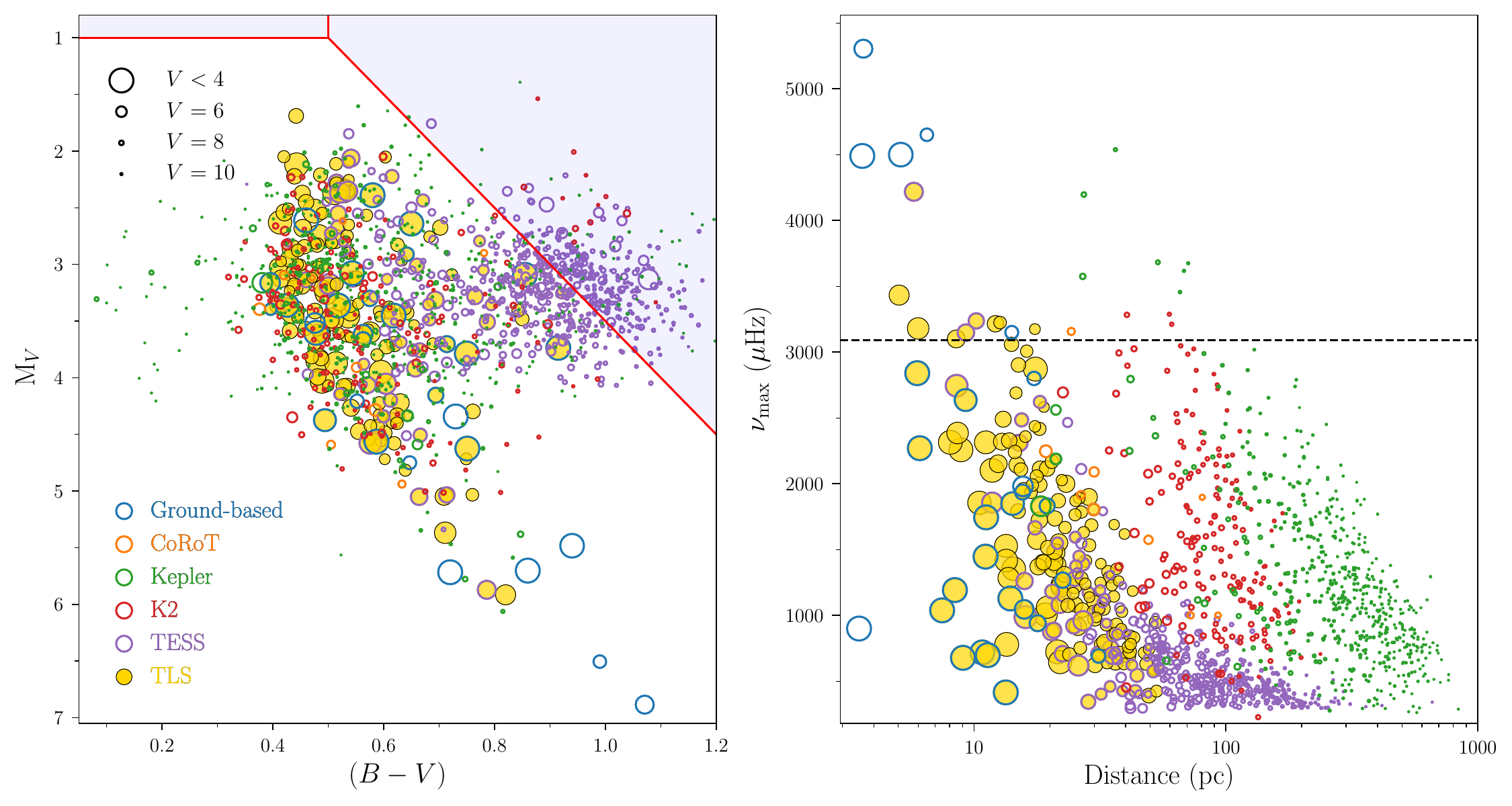}
   \caption{Comparison of the TLS with solar-like detections in MS/SG stars from other missions. Left: position of solar-like oscillators in the HR-diagram, with an indication of the selection criteria in $M_{V}$ and $(B-V)$ used to define our sample (\sref{sec:sample}). The marker size indicates the $V$-band magnitude of the stars, while the marker edge colour indicates how or by which mission oscillations were first detected. Any stars with a detection of oscillations from this work are shown with a filled yellow marker. Stars with ground-based detections were identified from individual cases in the literature (see \tref{tab:all_seis} and \sref{sec:comp}); the \kp comparison sample was constructed from the compilations of \citet{Lund2017}, \citet{Serenelli2017}, \citet{Mathur2022}, in addition to Kepler-444 \citep{Campante2015} and $\theta$ Cyg \citep{Guzik2016}; the 9 stars from CoRoT were identified from individual cases in the literature \citep{Barban2009,Barban2013,Appourchaux2008,Mosser2009,Mathur2010c,Mathur2013,Ballot2011,Boumier2014,Castro2021}; the stars forming the K2 sample are obtained from \citet{Keystone2016,Keystone2024}; while the TESS sample was obtained from the catalogues of \citet{Hatt2023}, \citet{Zhou2024}, and \citet[][considering only their confident detections; their Table~1]{Corsaro2024}, in addition to individual cases from the literature (see \tref{tab:all_seis}).
   For the TESS and K2 comparison samples, we have limited these to stars with $\numax<284\,\mu\rm Hz$.   
   Right: distribution of the stars in terms of distance and \numax, using only stars that in the left plot fall within the $M_{V}$ and $(B-V)$ boundaries defined in our target selection. We note that $\alpha$ Cen A+B, at a distance of ${\sim}1.35$ pc, have been omitted from the plot. Distances and magnitudes used in this plot were adopted from the TESS Input Catalog \citep[TICv8.2;][]{TIC82_2021}. The horizontal dashed line indicates the solar \numax for comparison.}
   \label{fig:sample}
\end{figure*}
From our asteroseismic analysis, we detected oscillations in a total of \sn stars and provide the extracted global asteroseismic parameters \dnu and \numax in \tref{tab:all_seis}. Examples of PSD for some targets with detected oscillations, many of which are new detections, are shown in \fref{fig:examples}. In \fref{fig:sample}, the TLS is compared to the cohorts obtained from ground-based efforts, other studies based on TESS data, and the previous missions of CoRoT, \kp, and K2. 

As seen from \fref{fig:sample}, the main overlap with the TLS is from oscillators identified in ground-based efforts, and partly from the \citet{Hatt2023} and \citet{Zhou2024} samples, where our new detections extend these works to brighter stars. In terms of proximity, and thereby brightness, we see that the oscillators found from TESS are complementary to those found from \kp, with K2 and CoRoT nearly closing the gap, at least for sub-giants and more evolved stars. Consequently, the overlap with these missions is very limited, and only includes the CoRoT target \object{HD 49933} \citep{Mosser2005,Appourchaux2008,Garcia2010} and the two \kp targets \object{$\theta$ Cyg} \citep{Guzik2016} and \object{16 Cyg A} \citep{Metcalfe2012}. With continued observations from TESS, the fainter magnitude limit for seismic detections on the MS will further close the gaps between the samples \citep{Campante2016}. Although we did not investigate the seismic detectability in stars fainter than $V=6$ in this study, a comparison with the extensive \citet{Hatt2023} and \citet{Zhou2024} catalogues suggests that very few additional MS targets would be detected from the current amount of data. All TLS stars with an earlier seismic detection are identified in \tref{tab:all_seis}, with reference given to the discovery paper.

\subsection{Asteroseismic parameters}\label{sec:astero_params}

\begin{figure*}
\centering
   \includegraphics[width=\textwidth]{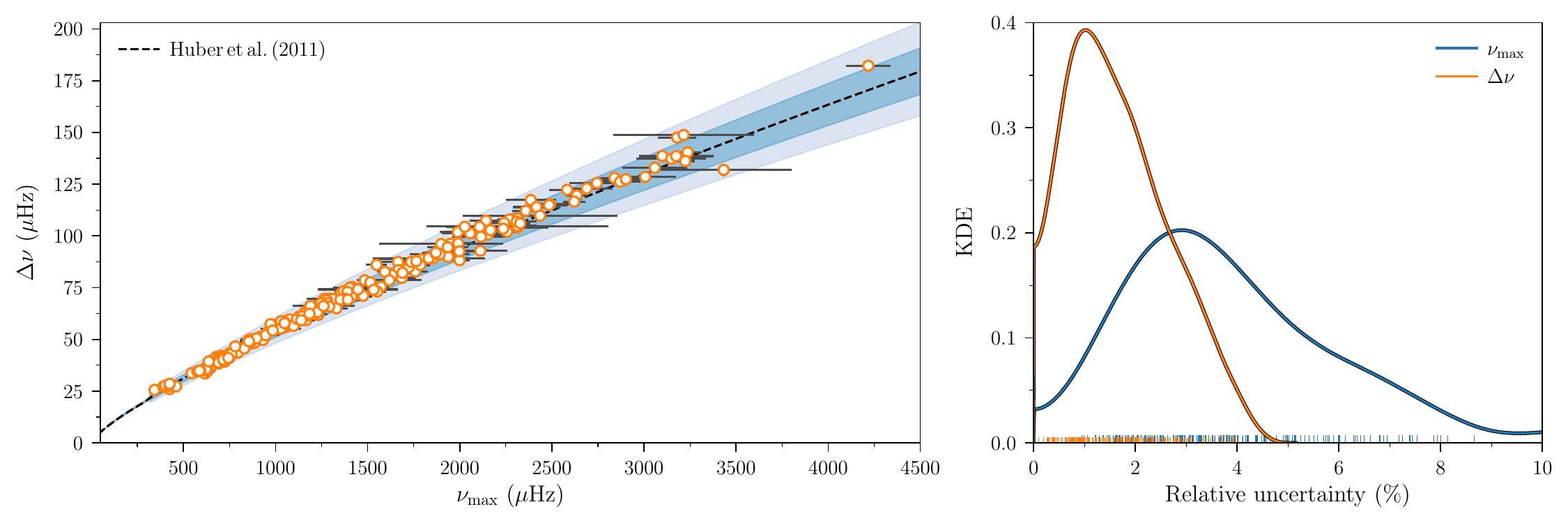}
   \caption{Left: correlation between the measured global asteroseismic parameters \dnu and \numax. The dashed line indicate the empirical relation from \citet{Huber2011} together with the $1$- and $2$-$\sigma$ confidence bands on their relation. Right: KDE of the relative uncertainties on \dnu and \numax for the sample, with median values of ${\sim}1.6\%$ in \dnu and ${\sim}3.7\%$ in \numax. The ticks at the bottom of the panel indicate the individual values, coloured according to the legend.} 
   \label{fig:dnu_numax}
\end{figure*}
The correlation between the extracted \dnu and \numax values is shown in \fref{fig:dnu_numax}, together with the expected empirical relation obtained from \kp by \citet{Huber2011}. The measured values are generally well within expectations, indicated by the 1- and 2-$\sigma$ bands to the empirical relation. The right panel of \fref{fig:dnu_numax} shows the kernel density estimates of the relative \dnu and \numax uncertainties, with median values of ${\sim}1.6\%$ in \dnu and ${\sim}3.7\%$ in \numax. These typical uncertainties are comparable to similar analyses in the literature of MS/SG stars observed with Kepler and K2 \citep{Verner2011B,Viani2019,Keystone2024,Sayeed2025}. From our analysis, we find that the typical RMS deviation amongst the different values used in the reported weighted average is well below the typical uncertainties, at ${\sim}0.5\%$ in \dnu and ${\sim}1.4\%$ in \numax. Notably, we also find no indications of systematic biases between the different analyses.

We note that the uncertainties obtained for \dnu and \numax from the individual data products are generally lower (${\sim}0.8\%$ in \dnu and ${\sim}2\%$ in \numax) for $20$-s cadence data (when available) or the combination of $120$-s and binned $20$-s data (depending on the amount of $20$-s cadence data available). While these data naturally contribute proportionally more to the final value in the weighted average, the uncertainties are increased slightly by the inclusion of $120$-s data and the averaged PSD. However, considering that our main focus is the detection of oscillations and given that not all stars have $20$-s cadence data available, we opted to include these data products in the reported averages. Indeed, in some cases the detection and measurement of \dnu and \numax were only possible when using the averaged PSD (like for  \object{19 Dra} (\object{h Dra}/\object{TIC 289622310}) and \object{$o^2$ Eri} (\object{TIC 67772871}; \sref{sec:indv})).    

In addition to the global asteroseismic parameters, we estimate\footnote{This fraction is a heuristic estimate based on previous experience with analysis of comparable \kp/K2 data.} that ${\sim}63\%$ (or ${\sim}125$ stars) will be amenable to peak-bagging and the extraction of individual mode parameters \citep[\eg,][]{Handberg2011,Davies2016,Lund2017,Corsaro2020,Nielsen2021}. Such a peak-bagging effort and modelling of the asteroseismic parameters will be the subject of future works. 

\subsubsection*{Comparison to the literature}\label{sec:comp}

As a check of our global asteroseismic parameters, we compared our values to those of \citet{Hatt2023}, \citet{Zhou2024}, and \citet{Corsaro2024} for the stars we have in common. In addition, we also compared to the predicted values from the ATL (which was included in the initial pruning of our sample), both versions 1 \citep{Schofield2019} and 3 \citep{Hey2024}.

\begin{figure*}
\centering
   \includegraphics[width=\textwidth]{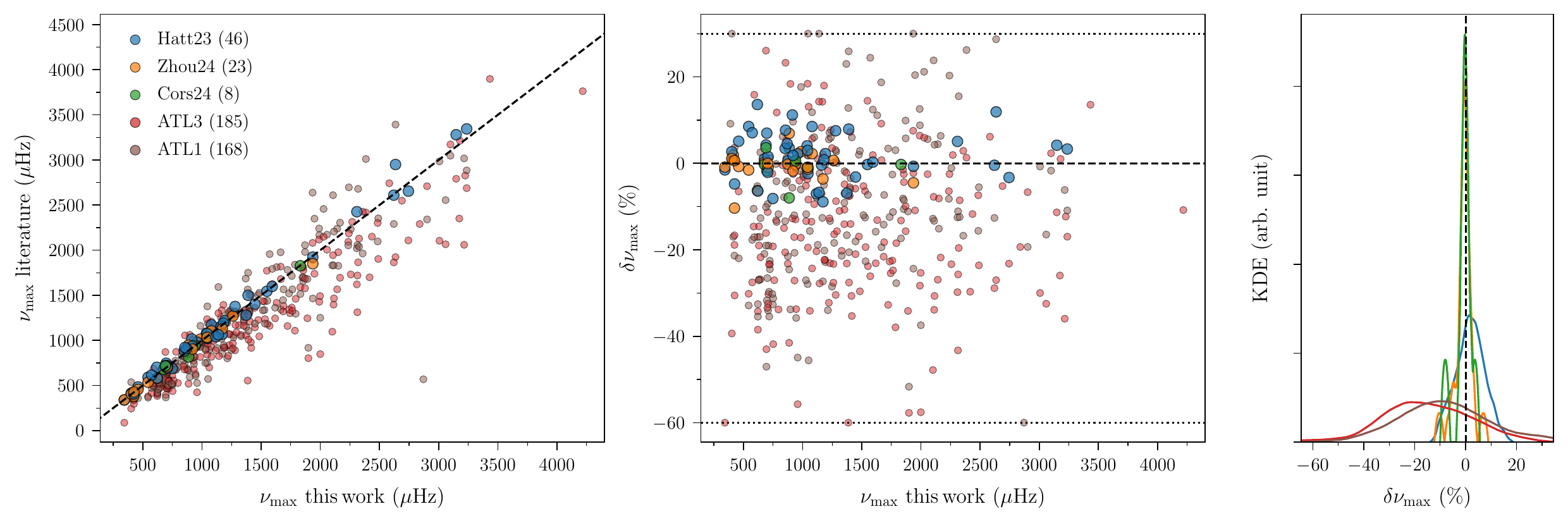}
   \caption{Comparison of \numax values for the stars in our sample that overlap with those of \citet{Hatt2023}, \citet{Zhou2024} and \citet[][considering only their confident detections; their Table~1]{Corsaro2024}, and with the predictions from the Asteroseismic Target List (ATL) versions 1 \citep{Schofield2019} and 3 \citep{Hey2024}. Left: a direct comparison between our values and those published in the literature or predicted in the ATLs. The colour indicates the comparison source (see legend), while the numbers in the legend indicate how many stars are in common with the different comparison sources. Middle: the relative differences between the values. Values beyond either $+30\%$ or $-60\%$ have been adjusted to these values (dotted lines) for a better visual rendition. Right: the KDE of the relative differences.} 
   \label{fig:numax_comp}
\end{figure*}

Our comparison for \numax is given in \fref{fig:numax_comp}, and shows good overall agreement with the other studies using TESS data. The differences are generally within uncertainties, and we see no apparent bias in differences against \numax. We find the ATL typically underestimates \numax compared to our values, at a level of ${\sim}10\%$ for ATL1 and ${\sim}16\%$ for ATL3. For ATL3, we can trace this to a general offset in the \logg and \teff values adopted from Gaia DR3 \citep{GaiaDr3_2023}. We refer to \sref{sec:detect} and \aref{app:atl} for further discussion on the ATL comparison, which should be considered in future target selection efforts.

In terms of our ability to detect oscillations, we note that for all stars in the \cite{Hatt2023} and \citet{Zhou2024} catalogues that overlap with our sample, we also obtain a detection. Indeed, we confirm all previous detections from single-star analyses of TESS data that we could find in the literature, except for two stars. These are \object{HD 4628} (\object{HIP 3765}) and \object{111 Tau} (\object{HIP 25278}), listed as confident detections in \citet{Corsaro2024} from TESS observations, for which we could not confirm a detection in our analysis. 

Among the ten stars flagged as having moderate, weak, or inconclusive seismic detections by \citet[][their Table~2]{Corsaro2024} that overlap our sample, we obtain clear detections for $\eta$ Cas, $47$ UMa, HD $30562$, and HD $9562$, thereby corroborating the seismic nature suggested by those authors. Interestingly, these four cases all have negative ln-Bayes factors from the Bayesian model comparison in \citet{Corsaro2024}, suggesting a preference for the null hypothesis with no seismic power excess.

Among stars with ground-based detections, we have identified only seven matching our selection criteria where a positive detection could not be made, namely \object{18 Sco} (\object{HIP 79672}), \object{70 Oph} (\object{HIP 88601}), \object{$\tau$ Boo} (\object{HIP 67275}), \object{$\epsilon$ Indi} (\object{HIP 108870}), \object{$\tau$ Cet} (\object{HIP 8102}), \object{$\iota$ Hor} (\object{HIP 12653}), and \object{HD 219134} (\object{HIP 114622}). No TESS observations up to Sector 77, as considered in our analysis, are available for \object{18 Sco} \citep{Bazot2011} and \object{70 Oph} \citep{Carrier2006_70Oph}, but both will be observed for one sector during TESS Cycle 7 (\tref{tab:all_miss}). For \object{$\tau$ Boo} \citep{Borsa2015} data are available from two sectors, although only with a 120-s cadence; \object{$\epsilon$ Ind} \citep[][]{Campante2024,Lundkvist2024} has data from two sectors, including one of 20-s cadence, and with more observations scheduled for Sector 95; \object{$\tau$ Cet} \citep{Teixeira2009} also has data from two sectors, one with 20-s cadence, but unfortunately the 20-s cadence stamp is too small to contain the star's saturation trails, which adds significant noise to the photometry; \object{$\iota$ Hor} \citep{Vauclair2008} has five sectors of data, of which three are of 20-s cadence; \object{HD~219134} \citep{Li2025} has four sectors of data, of which two are of 20-s cadence, and with additional observations scheduled for Sector 84. Of these five non-detections, \object{$\epsilon$ Ind}, \object{$\tau$ Cet}, and \object{HD 219134} are cool dwarfs (spectral types G8, K5, and K3, respectively) and therefore have very low oscillation amplitudes \citep{Kjeldsen1995,Corsaro2013}. 

\subsection{Detectability and noise properties}\label{sec:detect}

\begin{figure*}
\centering
   \includegraphics[width=\textwidth]{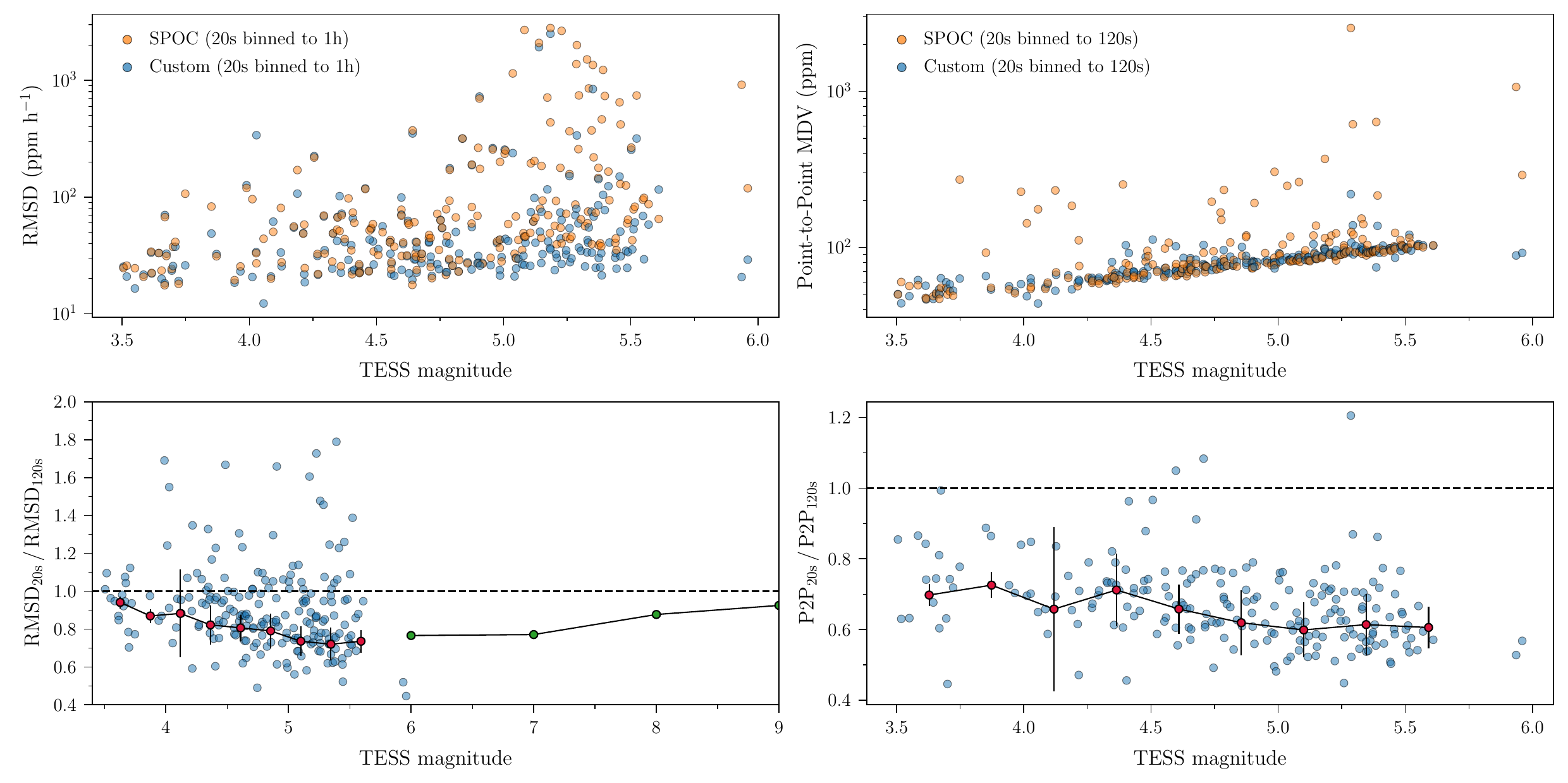}
   \caption{Comparison of noise statistics between different light curve sources and observing cadences, including all stars with 20-s cadence observations where a custom aperture was constructed (cf. \sref{sec:data}). Top left: Root-mean-square deviation (RMSD) of 20-s light curve flux binned to 1-hour against TESS magnitude for both SPOC and custom aperture data. Top right: Point-to-point (P2P) median difference variability (MDV) of 20-s light curve flux binned to 120-s against TESS magnitude. Bottom left: Ratio between the 1-hour RMSD from 20-s and 120-s cadence custom aperture data. The red markers indicate median-binned values, with uncertainties given by the standardised MAD, while the green markers give the ratios provided by \citet{Huber2022} (their Table~1). Bottom right: Ratio between P2P-MDV from 20-s (binned to 120-s) and 120-s cadence custom aperture data. Red markers again indicate median-binned values, with uncertainties given by the standardised MAD. } 
   \label{fig:noise_compare}
\end{figure*}

From our asteroseismic analysis, we find that our two sources of light curves, those from SPOC and those extracted from TPFs using custom apertures (\sref{sec:data}), greatly complement each other. Generally, the photometric quality is comparable, but with some outliers and with ranges in visual magnitude where the custom data performs better, typically from an improved aperture that better conserves the stellar flux. As an example of this, \fref{fig:noise_compare} provides a comparison between 20-s cadence light curves (including all stars where custom apertures were computed) from SPOC and those from custom apertures. The two metrics shown are the 1-hour root-mean-square deviation (RMSD; calculated as the standardised MAD of the data binned to 1-hour) and the point-to-point (P2P) median difference variability (MDV; calculated as the median of the absolute point-to-point flux differences, binned to 120-s cadence), both plotted as a function of TESS magnitude. 
The 1-hour RMSD is similar to the metric estimated in \citet{Huber2022} (their Fig. 2b), and we find that the levels from our sample agree well with those from \citet{Huber2022} in the magnitude range around $T_{\rm mag}{\sim}5.5$ where the samples overlap. The P2P-MDV metric is included to measure the high-frequency noise, which is particularly important for the detection of oscillations in MS/SG stars. 

We confirm the lower noise of 20-s vs. 120-s cadence data, as demonstrated by \citet{Huber2022}, and we can extend their reported improvement of $20$--$30\%$ to stars down to $T_{\rm mag}{\sim}4.3$. Brighter than this $T_{\rm mag}$, we observe less improvement, but note that the sample size in this regime is limited to only $39$ stars with data from both cadences. For the high-frequency noise, captured by the PTP-MDV, the improvement seems to be even greater, at a level of $30$--$40\%$ in the median and remaining nearly constant across the magnitude range covered by our sample.

\begin{figure*}
\centering
   \includegraphics[width=\textwidth]{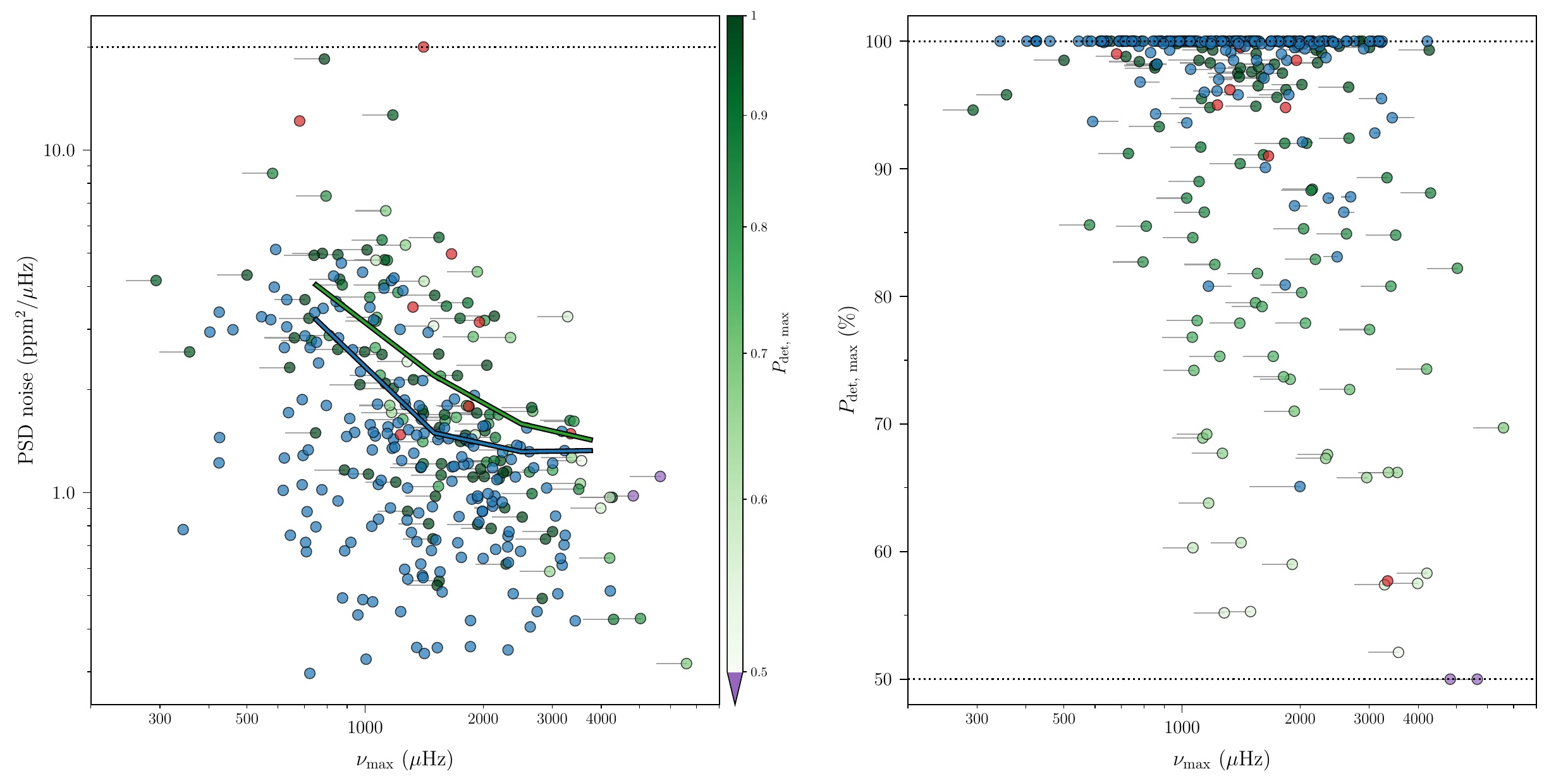}
   \caption{Comparison of noise levels and predicted detectability in ATL3 for stars with and without seismic detections. Left: correspondence between the global asteroseismic \numax parameters and the high-frequency PSD noise level for stars with positive seismic detections in blue, for stars identified as classical pulsators (\eg $\delta$ Sct/$\gamma$ dor) in red, and for stars without a seismic detection in green. The \numax and colouring of the non-detection cases are given by the \numax and detection probability ($P_{\rm det}$) returned by the ATL3 \citep[][]{Hey2024}, and except for a few cases we have only included stars with $P_{\rm det}>50\%$. The non-detection \numax values have been offset (horizontal line indicates the offset from the original position) by increasing \numax by $16\%$, corresponding to the apparent overall bias of the ATL3 values, as seen from \fref{fig:numax_comp}. Stars with a noise level above $20\,\rm ppm^2/\mu Hz$ have been offset to this value (dotted line). The coloured lines give the median binned noise levels of stars with (blue) and without (green) detections against \numax. Right: correspondence between the maximum ATL3 $P_{\rm det}$ (for either 20- or 120-s cadence) against \numax, with the same colouring and \numax as in the left panel. Stars with a $P_{\rm det}$ below $50\%$ have been offset to this value (dotted line). For the stars with seismic detection (blue) we have indicated with the small horizontal lines the ATL3 predicted \numax }
   \label{fig:detect_comp}
\end{figure*}

In \fref{fig:detect_comp} we show the minimum high-frequency noise levels (median between $\rm 3000\, \mu Hz<\nu<4000\, \mu Hz$) from the power density spectra considered in our analysis against the measured (blue) or predicted (green) \numax for the stars with a seismic detection or a detection probability in ATL3 of $P_{\rm det}\geq50\%$ (based on whichever is highest if both 20-s and 120-s cadence observations are available). As expected, we see that stars with a positive seismic detection at a given \numax generally have lower noise levels than stars without a detection. 
If we look at the ATL3 detection probabilities, we see that only a single star (HD 49933) with a firm detection has a detection probability below $P_{\rm det}\sim 80\%$, and the vast majority has a rounded probability of $100\%$. As noted in \sref{sec:comp} the ATL3 \numax values are generally underestimated, and hence $P_{\rm det}$ is overestimated, suggesting that most of the stars without a detection, but with high predicted detection probabilities, may indeed be overestimated. We also see that the stars with confirmed detections, but detection probabilities below $100\%$ generally have slightly overestimated \numax values from ATL3 (hence underestimated detection probabilities). We refer again to \aref{app:atl} for further details on the ATL comparison. 
It is also worth noting that the ATL detectability calculation does not take into account that some targets could have an increased activity level, which is known to suppress the amplitudes of oscillation modes \citep{Garcia2010,Chaplin2011b,Bonanno2014,Mathur2019,Sayeed2025}. This could also explain why oscillations are not detected in several stars where the noise level should be low enough for detection. 


\subsection{TLS and PLATO}\label{sec:plato}

Because they are---or have the potential to become---extremely well-characterised, the stars of the TLS are of particular interest as calibrators and benchmarks for future asteroseismology missions, such as the ESA PLATO mission scheduled for launch in late 2026 \citep{Rauer2024}.

PLATO's planned observing strategy is currently focused on two so-called ``Long-duration Observation Phase'' (LOP) fields, each of which will cover a $49^{\circ}\times 49^{\circ}$ region of the sky \citep{Rauer2024}. While subject to potential changes following the evaluation of initial observations, the nominal plan is to conduct observations for each of the LOP fields for two years. The 24 ``normal'' cameras will observe at a cadence of 25 sec, while a cadence of 2.5 sec will be used for the two ``fast'' cameras, capable of observing towards the centre of the FOV.
At the time of writing, only the southern LOP field (LOPS2; centred on $l=255.9375^{\circ}$, $b=-24.62432^{\circ}$ in Galactic coordinates) is fully defined\footnote{\url{https://www.cosmos.esa.int/web/plato/first-sky-field}} and this is where observations are scheduled to start \citep{Nascimbeni2025}. We provisionally adopt the northern LOP candidate field LOPN1 ($l=81.56250^{\circ}$, $b=24.62432^{\circ}$) as defined by \citet{Nascimbeni2022}. Until the PLATO Science Working Team (PSWT) makes its final field selection, however, there is no assurance that LOPN1—or any other northern variant—will ultimately be observed.

\begin{figure*}
\sidecaption
   \includegraphics[width=12cm]{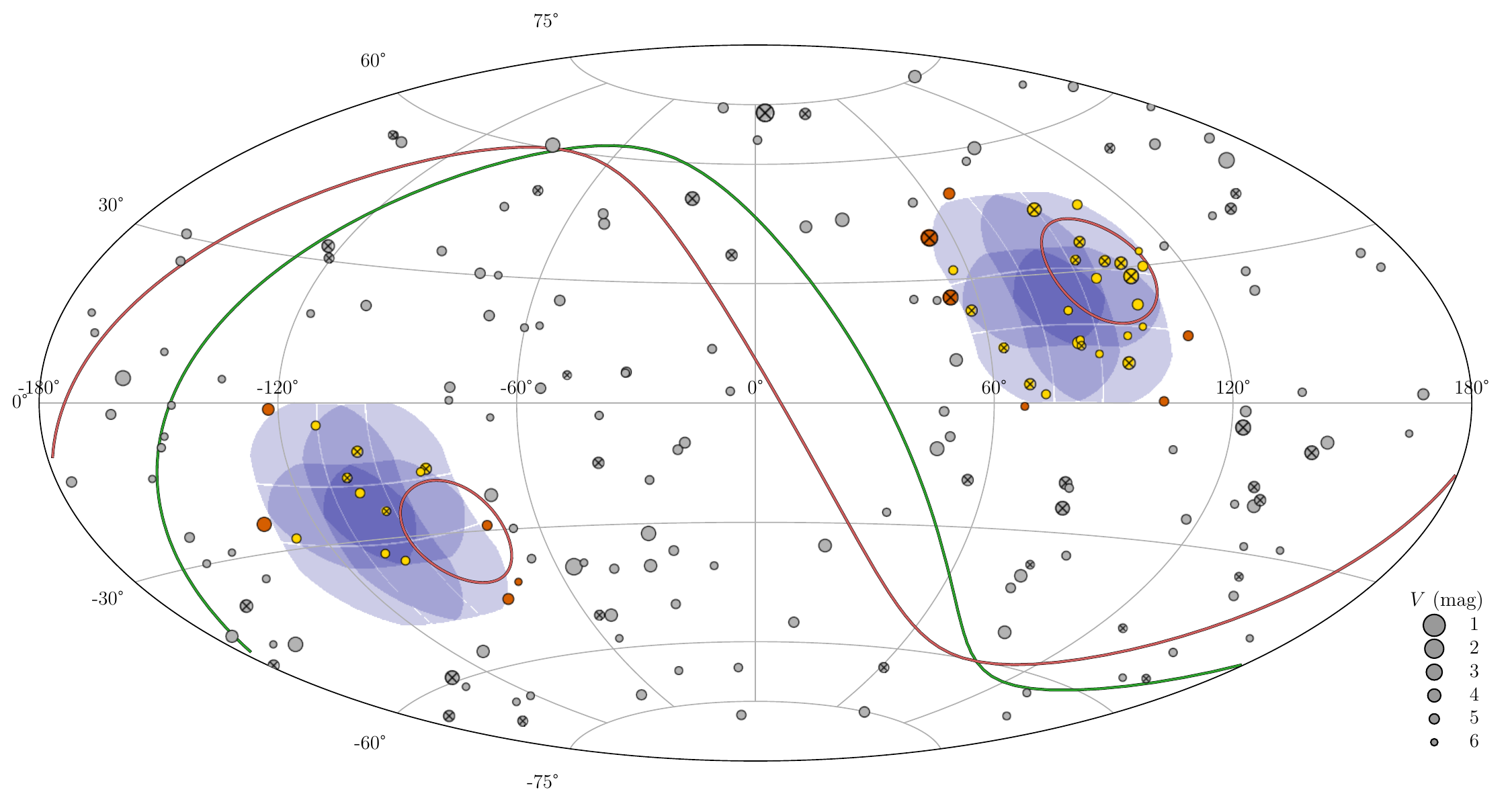}
   \caption{Aitoff sky projection in Galactic coordinates showing (circular markers) the TLS, with the marker size corresponding to the star's visual magnitude (see legend key). The PLATO long-stare (LOP) fields are shown in blue (with shade corresponding to the number of either 24, 18, 12, or 6 overlapping cameras). Yellow markers indicate targets identified as being within current PLATO LOP field definitions (see text), while orange markers indicate targets within $5^{\circ}$ of the LOP field boundaries. Stars that are in binary systems and listed in \tref{tab:binary} are further marked with a cross ($\times$). The TESS CVZs are given by the red circles that partly overlap the PLATO fields. The red line indicates the sky's equatorial plane, while the green line gives the ecliptic plane. }
   \label{fig:platooverlap}
\end{figure*}

In \fref{fig:platooverlap} we show the sky distribution of the TLS, including the overlap with the PLATO LOP fields\footnote{The PLATO field plots are made using functionalities of the \texttt{platopoint} code by Hugh P. Orborn, in the modified version used by \citet{Boettner2024} (\url{https://github.com/ChrisBoettner/plato/tree/main/plato/instrument/platopoint.py})}. \tref{tab:plato} provides an overview of the targets within the current LOP field definitions, and also those near the fields (defined as being within $5^{\circ}$ of the field edges). To check if a given target will be within the LOP fields, we tested against the field edges obtained from the LOP field versions \texttt{pLOPN1PIC2.0.0.1-t} and \texttt{pLOPSsPIC2.0.0.1-t} of the PLATO Input Catalog \citep[PIC;][]{Montalto2021}.
We found that $10$ targets are within the LOPS2 field, while $24$ targets are within the current LOPN1 field. Interestingly, we identified several targets as spectroscopic and/or visual/astrometric binaries, where independent constraints can be placed on the stellar masses (see \sref{sec:binary}, \tref{tab:binary}) and, in some cases, also on radii from interferometry (see \sref{sec:interfer}, \tref{tab:interfer}).

Based on the brightness of our sample and our target selection strategy (\sref{sec:sample}), most (if not all) of the stars overlapping the PLATO fields should meet the requirements for the P2 ``bright-star'' sample \citep{Montalto2021,Rauer2024,Goupil2024}. We note, however, that the stars will generally be best suited for observations with the two fast cameras, whose dynamic range is $4<V<8$. For the normal cameras, optimised for stars fainter than $V=8$, it remains to be seen how well photometric observations can be extracted for heavily saturated stars, \eg from extended imagettes, which will experience significant blooming \citep[\eg,][]{Jannsen2024}. Interestingly, observing TLS targets with the fast cameras will deliver simultaneous blue (505–700 nm) and red (665–1000 nm) photometry. Access to these two passbands could provide new insights on mode physics and on the links between asteroseismic observables and fundamental or dynamical stellar properties, \eg, rotation, convection, and magnetic activity \citep[\eg,][]{Houdek2015,Santos2019,Sreenivas2025}.

It would be instructive for the stars that, in addition to now being known asteroseismically, are well-characterised binaries, to be considered for the science calibration and validation PIC \citep[scvPIC;][]{patrick_gaulme_2023_8107108} and possibly for the so-called ``prime'' sample of stars that will receive the highest priority throughout the mission \citep{Rauer2024}. 
The TLS should be taken into account when/if the PSWT finalizes a Northern LOP field. In the provisional northern layout, TLS could provide a crucial addition to the sample of stars where stellar magnetic activity can be studied using asteroseismology \citep[\eg,][]{Garcia2010,Chaplin2014,Salabert2016b,Kiefer2017,Santos2018}--particularly valuable because most ground-based activity surveys \citep[Mount Wilson, Lowell SSS, TIGRE, STELLA, etc.; see, \eg,][]{Jeffers2023} also operate in the northern hemisphere. Furthermore, \object{$\mu^1$~Her} (\object{HIP 86974}), monitored for more than 10 years 
in radial velocity by SONG \citep{Grundahl2017} and on track to become one of the best characterised benchmark asteroseismic sub-giants, lies near (\ie, $\leq5^{\circ}$ away from) the current LOPN1 field edge.
A dedicated in-depth analysis of the TLS stars in or near the PLATO fields will be the subject of subsequent analysis (Panetier et al., in prep.). We also refer the reader to \citet{2024Eschen} and \citet{Nascimbeni2025} for additional analysis of the target content of the PLATO LOP2S field.

\subsection{TLS and the Habitable Worlds Observatory (HWO)}\label{sec:hwo}

HWO is a planned NASA 6-meter-class UV/optical/IR space observatory capable of high-contrast imaging and spectroscopic characterisation of potentially habitable exoplanets in reflected light. The HWO concept was the top priority of the 2020 Decadal Survey on Astronomy and Astrophysics and is currently envisioned to launch in the early 2040s. 

The primary targets for HWO are nearby, bright Sun-like stars for which the inner working angle allows the detection of planets on angular separations consistent with the habitable zone. Possible targets for HWO are described in a catalogue of HWO Precursor Science Stars by the NASA Exoplanet Exploration Program (ExEP)  \citep{Mamajek2024} and in the HWO Input Catalog \citep{tuchow24}. The former consists of 164 stars divided into tiers (A, B, or C), based on their expected suitability for detecting Earth-like exoplanets. 

\begin{figure}
\centering
   \includegraphics[width=\columnwidth]{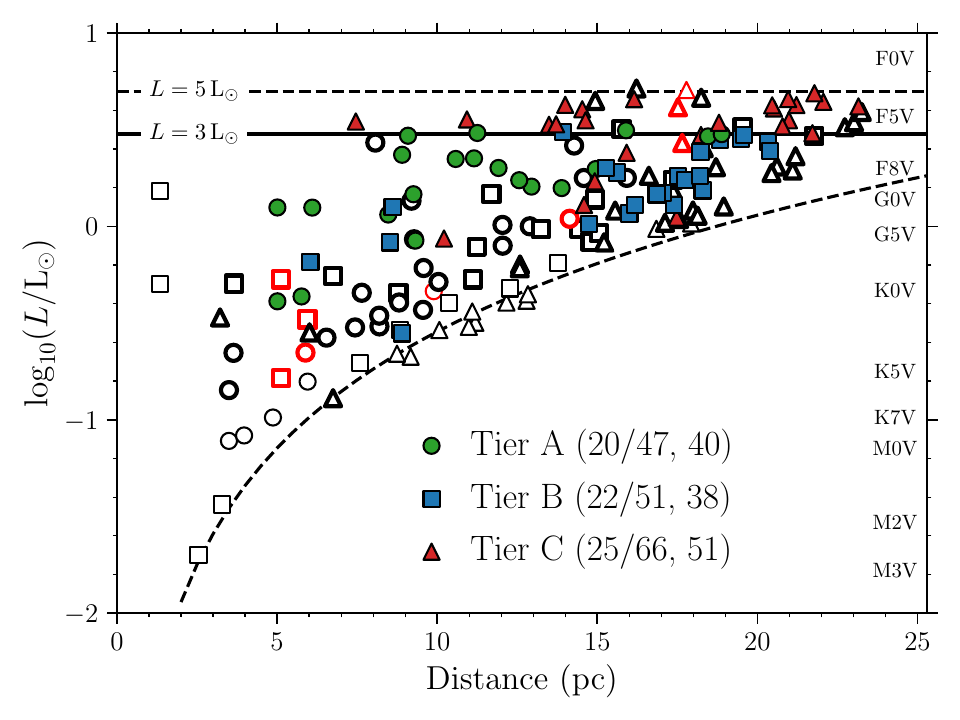}
   \caption{HWO target stars in terms of their luminosity and distance, with values adopted from the tables of \citet{Mamajek2024}. Filled markers indicate stars with detected oscillations, with the colour and shape indicating the HWO tier (see legend). Stars with unfilled black thick-edged markers were considered for analysis, while black thin-edged ones did not conform to the selection criteria for our sample. Stars with red thick-edged markers were considered but did not have data available before S77, as considered in this analysis (but will later in Cycle 7), while red thin-edged ones are not scheduled for observations with TESS. The legend showing the different tiers gives in parentheses the number of detections out of the total number of stars in the tier, and lastly the number of stars from each tier that we analysed. }
   \label{fig:hwo}
\end{figure}

Of the $164$ ExEP HWO targets, we have identified $139$ matching the selection criteria adopted in this analysis (\sref{sec:sample}). The remaining stars include $\alpha$ Cen A+B, which were not included in our analysis, as well as stars fainter than our $V=6$ selection cut that mostly have spectral types later than K0V (hence, with low probability of detecting oscillations). Of the $139$ stars matching our selection criteria, two stars ($\tau^6$ Eri and $12$ Oph) have not been, and are not scheduled to be, observed by TESS, while eight stars\footnote{HD 131977, 18 Sco, 36 Oph A+B, $\xi$ Oph, 58 Oph, and 70 Oph A+B} did not have any data up until Sector 77, as considered in this analysis. From the remaining $129$ stars with data, we detected oscillations in $67$ stars, including $20$ from tier A, $22$ from tier B, and $25$ from tier C. The HWO sample is shown in \fref{fig:hwo} in terms of luminosity against distance (following the illustration of the sample in \citealt{Mamajek2024}), with indications of which stars have detections. In \tref{tab:all_seis}, we have for all stars with seismic detections indicated their tier if they overlap with the HWO sample. 

The detections presented here, combined with asteroseismology of cooler HWO targets with extreme precision radial velocities \citep[\eg,][]{Campante2024,Hon2024,Li2025}, will allow the systematic determination of precise ages of HWO targets, which are critical for interpreting possible biosignatures from directly imaged planets \citep{bixel20}. A future paper in this series will focus on the asteroseismic age distributions for the HWO sample (Chontos et al., in prep.). Continued 20-s data of bright nearby stars in future TESS extended missions will be important to expand the sample of detections presented here.

\subsection{Exoplanets, disks, and substellar objects}\label{sec:exo}
Detailed stellar characterisation from asteroseismology of known exoplanet hosts is of great interest, including mass estimates of RV exoplanets, information on system ages, and obliquities. \citep{VanEylen2014,Huber2018,Lundkvist2018,Lund2019}.

Among the oscillating stars in the TLS (\tref{tab:all_seis}), we have identified 24 as known exoplanet hosts, of which 13 are also in the HWO target list (\sref{sec:hwo}). In 12 of these 24 systems, ours are the first asteroseismic measurements, including one brown dwarf host (\object{HD 46588}/\object{HIP 32439}), one star that is positioned within the current definition of the PLATO LOPN1 field  (\object{HD 184960}/\object{HIP 96258}; \sref{sec:plato}), and several notable multiplanet systems, such as \object{$\upsilon$ And} (\object{HIP 7513}/\object{Titawin}), \object{47 UMa} (\object{HIP 53721}/\object{Chalawan}), \object{61 Vir} (\object{HIP 64924}), and \object{82 Eri} (\object{e Eri}/\object{HIP 15510}), which \citet{Nari2025} recently found to host a super-Earth orbiting in the star's habitable zone. 

In \tref{tab:exo} we provide an overview of the known exoplanet systems in the TLS, whose derived planet properties can be tightened thanks to the precise stellar radii, masses and ages delivered by an asteroseismic analysis. In addition to exoplanets, the table also lists any sub-stellar companions (such as brown dwarfs), and information on binarity (see also \tref{tab:binary}). The planetary data provided in \tref{tab:exo} was primarily obtained from the ``Planetary Systems'' of the NASA Exoplanet Archive\footnote{\url{https://exoplanetarchive.ipac.caltech.edu}}, and supplemented by information from the Extrasolar Planets Encyclopedia\footnote{\url{https://exoplanet.eu/}}.

Also worth mentioning is $\psi^1$ Dra B (\object{HIP 86620}), which is known to host a long-period giant exoplanet ($\psi^1$ Dra Bb) with a minimum mass of $1.5$ $M_{\rm Jup}$ and an orbital semi-major axis of $4.4$ AU \citep{Endl2016}. Though we did not identify oscillations in $\psi^1$ Dra B, we did in its companion $\psi^1$ Dra A (see \tref{tab:binary} and \fref{fig:custom_aperture}) and asteroseismology can therefore anchor the age for the whole system. We note also that $\psi^1$ Dra A/B are within the current definition of the PLATO LOPN1 field. 

Finally, several of our identified asteroseismic stars are known to host debris disks \citep{Hughes2018,Pearce2024}. Notable examples include binary systems (\tref{tab:binary}) such as \object{99 Her} (\object{b Her}/\object{HIP 88745}), which hosts a nearly polar-aligned circumbinary debris disk \citep{Kennedy2012,Smallwood2020}, and \object{HD 121384} (\object{HIP 68101}) \citep{Rhee2007,Rodriguez2012}; \object{HD 132254} (\object{HIP 73100}) and \object{110 Her} (\object{HIP 92043}) have been identified as hosts of cold debris disks \citep{Krivov2013,Marshall2013}; and known exoplanet hosts, such as, \eg, \object{82 Eri} (\object{e Eri}/\object{HIP 15510}) \citep{Pepe2011,Montesinos2016} and \object{61 Vir} (\object{HIP 64924}) \citep{Wyatt2012} also exhibit debris disks.

Access to well-characterised stellar parameters from asteroseismology, particularly stellar ages, is crucial for understanding the evolution of debris disks. These parameters provide a critical context for interpreting the current state of the disks, including their composition, structure, and dynamical processes, while also constraining the history of planet formation and interactions within these systems \citep{Trilling2008,Montesinos2016}. Moreover, accurate stellar characterisation enables meaningful comparisons across different systems, thereby enhancing our ability to discern patterns and trends in the evolution of planetary systems.

\subsection{Interferometry for angular stellar diameters}\label{sec:interfer}

Long-baseline interferometric observations for the measurement of stellar angular diameters provide an essential ingredient for obtaining stellar fundamental parameters of the highest precision that are nearly model-independent (apart from a small dependence on the adopted limb-darkening). Most importantly, an independent estimate is provided for the stellar linear radius by incorporating the distance (\eg from Gaia). Similarly, combining an estimate of the stellar bolometric flux with the measured angular diameter provides an independent \teff that, when combined with the independent radius and \numax, provides an estimate of the stellar mass \citep[][]{Pijpers2003,Cunha2007,Creevey2007,Bruntt2010,Bazot2011,Huber2012,White2013,White2018}.

Identifying stars with both interferometric and asteroseismic measurements is therefore essential for calibrating asteroseismology, both in terms of the application of scaling relations and the fine details of model physics.
In our search for existing interferometric stellar diameter measurements, we cross-referenced our sample against the compilation of \citet{Baines2023} (their Table~9), the Jean-Marie Mariotti Center \citep[JMMC;][]{2014Bourges} Measured Stellar Diameters Catalog \citep[JMDC, Cat II/345/jmdc; ][introduced as part of \citet{Chelli2016}]{Duvert2016} (updated last 13 Sep. 2021), the list of CHARA published interferometric diameters\footnote{\url{https://www.chara.gsu.edu/tables/interferometric-diameters}}, and the samples of \citet{Rains2020}, \citet{Karovicova2022}, and \citet{North2007}.

\begin{figure}
\centering
   \includegraphics[width=\columnwidth]{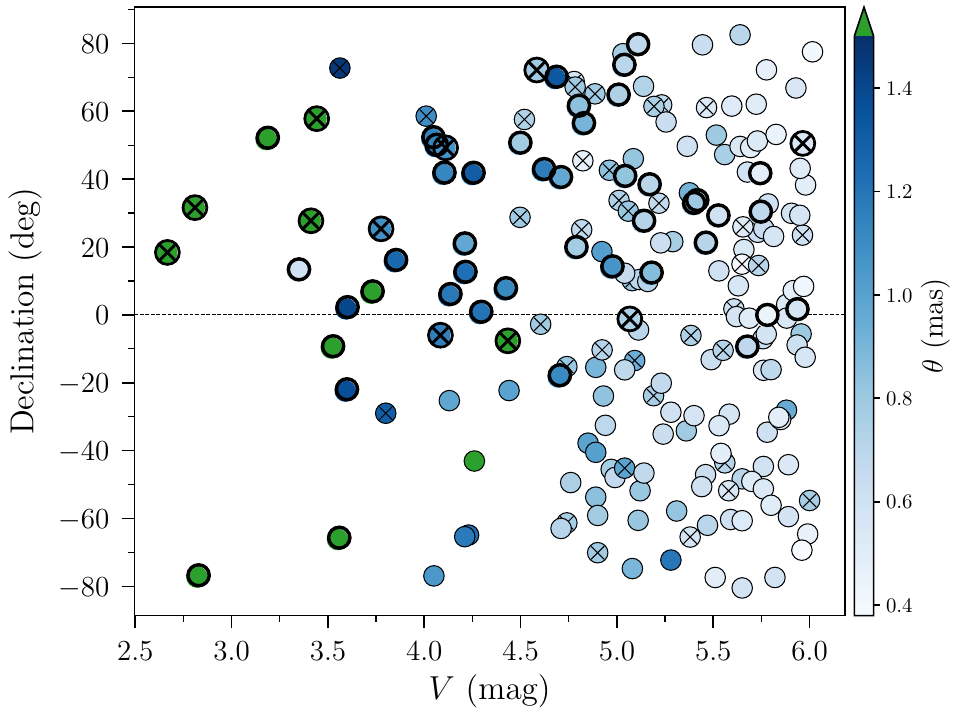}
   \caption{TLS stars plotted in terms of visual magnitude against declination, with the colour indicating the expected angular diameter (see colourbar). Stars with a thick outline have a published interferometric angular diameter (\tref{tab:interfer}), and stars with a cross ($\times$) are binaries listed in \tref{tab:binary}.}
   \label{fig:interfer}
\end{figure}

In \tref{tab:interfer} we list the 54 stars for which a published measurement could be found from the adopted compilations. We identify 31 of these as being new oscillators,  significantly expanding the cohort of MS/SG stars suitable for testing asteroseismology with interferometry. In \fref{fig:interfer} we provide an overview of the TLS in terms of visual magnitude and declination, and indicate the stars with existing interferometric measurements for stellar diameter determination. Brighter than $V=4.5$, we identify only 6 targets (\object{82 Eri}, \object{$\zeta$ Tuc}, \object{$\alpha$ Cha}, \object{$\gamma$ Pav}, \object{$\psi$ Cap}, and \object{HD~60532}) where we could not identify an interferometric measurement in the literature. 
For each star, \fref{fig:interfer} also provides a simple estimate of the expected angular diameter from combining distance (from the TIC) with a seismic radius obtained from scaling relations using our measured \numax and \dnu in combination with a \teff from \citet{Casagrande2011} (or the TIC if not available). 

We note that all stars have a predicted angular diameter above ${\sim}0.38$ mas. Combined with the brightness of the sample, all stars in the TLS (except some binaries) should be accessible for interferometric observations with either CHARA \citep[Centre for High Angular Resolution Array;][]{Brummelaar2005} or NPOI \citep[Navy Precision Optical Interferometer;][]{Armstrong1998} for the more northern targets, and VLTI \citep[Very Large Telescope Interferometer;][]{Glindemann2001} for the more southern targets. 

\subsection{Binarity}\label{sec:binary}

Stars exhibiting solar-like oscillations whilst also being members of binary systems are particularly important if they can be characterised spectroscopically (preferably with solutions for both components), in addition to either being eclipsing or having their orbit traced on the sky (either from resolved or interferometric observations for visual binaries or from astrometric observations). The constraint offered by the binarity in providing nearly model-independent estimates for the stellar mass (individual masses in the best cases and minimum masses in the worst) is of paramount importance for testing the masses provided by asteroseismology \citep{Serenelli2021}.

To date, when focusing on constraining the asteroseismology of solar-like oscillators, the effort has mainly been on eclipsing spectroscopic binary systems typically containing one or two evolved red giants
\citep[\eg,][]{Gaulme2016,Brogaard2018,Brogaard2022,Benbakoura2021,Thomsen2022}. In cases where oscillations are detectable in both components of a binary system containing MS and/or SG stars \citep{Miglio2014}, the binary period is often too long or the orbital configuration cannot be constrained to a degree that allows independent constraints to be placed on the masses \citep{Metcalfe2015,White2017,Li2018,Joyce2018}, whereby only the asteroseismic ages can be tested against the assumed coevality of the stars. This lack of added constraint on the stellar mass is also, in general, the case for the known binary systems with one well-characterised MS/SG oscillating component  \citep{Kjeldsen1995_etaboo,Deal2017,Grundahl2017,Metcalfe2021,Ball2022}. Given these challenges, only very few asteroseismic analyses of MS/SG stars have currently benefited from the added constraints of binarity in the stellar modelling \citep[][]{Appourchaux2015,Metcalfe2020}, or have been able to serve as benchmarks of asteroseismology.

Given the brightness of the stars in our sample, many have been subjected to extensive studies and are generally well-characterised. To identify known binary stars in our sample that could become important benchmarks of asteroseismology, we cross-referenced against the Washington Double Star catalog \citep[WDS;][]{2014WDS}, the Ninth Catalogue of Spectroscopic Binary Orbits \citep[SB9;][]{SB9_2004}, the Sixth Catalog of Orbits of Visual Binary Stars \citep[ORB6;][]{ORB6}, and the Observatorio Astronómico Ramón María Aller Catalog of Orbits and Ephemerides of Visual Double Stars \citep[OARMAC;][]{Docobo2001,OARMAC}. We note that many stars in the sample have a WDS designation. However, we are interested in systems where constraints can be placed on the orbits rather than simply having information on the existence of companions, and we require as a minimum an estimate of the orbital period. In addition to consulting the above catalogues, we conducted an extensive literature search of all the asteroseismic targets. The result of our search is given in \tref{tab:binary}. As an additional check, we matched our sample against the \textit{Gaia} DR3 non-single star (NNS) catalogue \texttt{gaiadr3.nss\_two\_body\_orbit} \citep{Gaia_NNS2023,Holl2023}. We found seven stars\footnote{HIPs 39903, 69226, 80686, 37606, 86036, 5081, and 50319} in the NNS catalogue, but with no new additions beyond the ones already identified from the other catalogues. Of the seven stars, all but two (HIPs 86036 and 5081) have orbital periods from the \textit{Gaia} NNS catalogue in agreement with the values listed in \tref{tab:binary}, and both poorly matching stars have \texttt{goodness\_of\_fit} and/or \texttt{significance} values indicating a poor NNS solution. Only in the case of 35 Leo (HIP 50319) can the \textit{Gaia} solution bring to bear information on the orbit (in the form of Campbell elements\footnote{from the conversion of the provided Thiele-Innes elements \citep[see][Appendix A]{Gaia_NNS2023}.}) not available from existing observations.

In addition to catalogue identifiers, period measurements and semi-amplitudes (for spectroscopic systems), we also indicate which of the stars are located within or near the PLATO LOP fields (\sref{sec:plato}). We also provide literature estimates of \vsini, since they can influence the quality of ground-based spectroscopic follow-up efforts. 
Unless otherwise stated in \tref{tab:binary}, notes via a letter reference to the spectroscopic periods (``$P$ (Spec)''), and semi-amplitudes (``$K_1(/K_2)$'') were obtained from SB9, adopting the latest entry if multiple exist (a numerical reference to this entry is provided in the table). The visual/astrometric period (``$P$ (Orb)'') was generally obtained from ORB6, if not otherwise indicated with a letter reference, and uncertainties in parentheses provide any non-zero root-mean-square-deviation between multiple entries in ORB6 and/or OARMAC.
The table is meant to give an overview of the feasibility of follow-up observations of the stars and their potential use as benchmarks. Therefore, we have generally omitted uncertainty estimates on the parameters (see table notes).  

Of the 48 asteroseismic stars with available binary orbital information listed in \tref{tab:binary}, a significant fraction could become valuable benchmark systems for asteroseismology, and several are likely to be observed by PLATO (\sref{sec:plato}). Seven stars are listed as SB2 systems with additional orbital constraints from being visual/astrometric binaries; for these individual components, masses can be directly determined and compared to the asteroseismic values. With additional follow-up observations, more systems could potentially be identified as SB2, and later \textit{Gaia} releases should provide orbital constraints for more of the spectroscopically characterised systems.   

One facility that is well-suited to providing follow-up spectroscopic observations of the binary orbits for bright stars, and in many cases also for asteroseismology, is the Stellar Observations Network Group \citep[SONG;][]{Grundahl2017}. In \tref{tab:binary} we provide an overview of the current observations conducted for these stars using SONG\footnote{We note that upon joining the SONG community (\url{https://soda.phys.au.dk/}), the spectra in the SONG database are freely available for members to use.} and note that most of the stars with periods below ${\sim}7$ yr have already been scheduled for long-term monitoring. Some of the stars with high numbers of existing spectra from SONG will be the subject of future dedicated analysis of both the binary and asteroseismic data. This includes such stars as \object{$\omega$ Dra} (\object{HIP 86201}), \object{$\iota$ Peg} (\object{HIP 109176}), and \object{$\chi$ Dra} (Rudrasingam et al., in prep.).
In \aref{app:bin_indv} we provide notes concerning binarity for several individual stars, in some cases to elaborate on the information in \tref{tab:binary} and in some to clarify why stars identified as binaries, \eg on \texttt{SIMBAD}, have been excluded.  

\subsection{Solar analogues with seismic detections}\label{sec:analog}

Identifying and characterising stars that resemble the Sun is important because they provide an essential context for understanding the Sun in terms of evolution, activity, and chemistry, and they are naturally of interest in the search for exoplanets. Sun-like stars come in different categories, depending on their resemblance to the Sun, where ``solar twins'' are restricted to having near-solar parameters on all fronts, while ``solar analogues'' are more loosely defined as having parameters ``similar'' to the Sun \citep{Hardorp1982,Cayrel1996}. 
The definition of a solar analogue is not very stringent, and the criteria used in the literature for their identification vary. However, to give an overview of the subsample of (potential) interest for solar-analogue studies, we applied the criteria of \teff within $\pm500$ K of the Sun, \feh within $\pm 0.3$ dex (corresponding to a metallicity within a factor of two of solar), and $M_V$ within $\pm1$ magnitude of the solar at $M_{V, \odot}=4.83$ \citep{Soderblom1998}\footnote{\url{http://www2.lowell.edu/users/jch/workshop/drs/drs-p1.html}}. We have not restricted the subsample to be without close companions, but refer to \sref{sec:binary} (\tref{tab:binary}) for information on binarity.

\begin{figure*}
\centering
   \includegraphics[width=\textwidth]{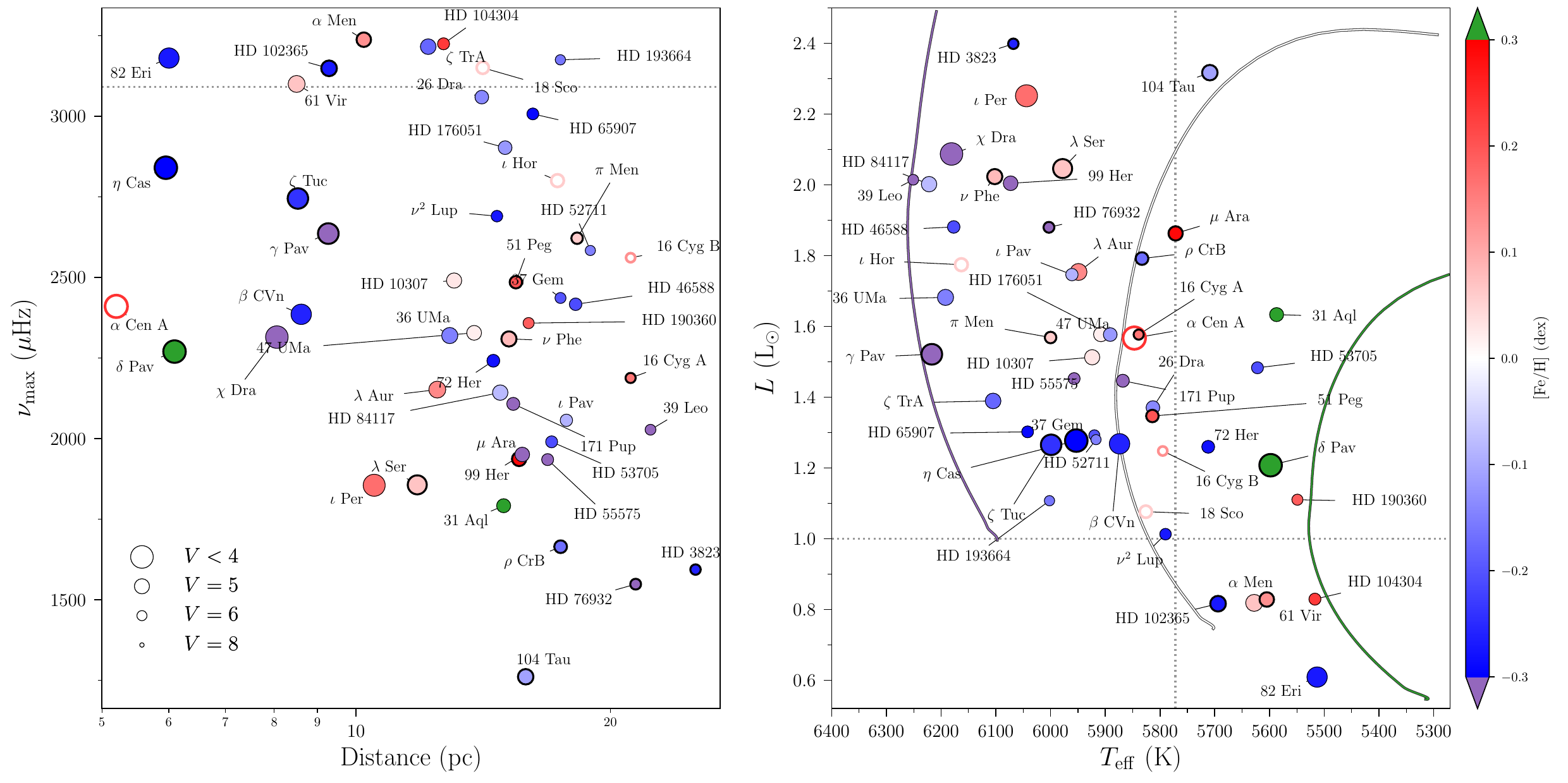}
   \caption{Stars identified to match our solar-analogue criteria (see \sref{sec:analog}). The marker size indicates the visual magnitude (see legend), while the colour gives the $\feh$-value from \citet{Casagrande2011}. Empty markers indicate known bright oscillators from ground-based surveys (and 16 Cyg B from \kp) that match the solar-analogue criteria, but where no detection was obtained in this study. Markers with a thick black edge indicate the stars that are already known oscillators (see \tref{tab:all_seis}). Left: identified (potential) solar analogues in terms of distance and \numax ($\alpha$ Cen A at a distance of ${\sim}1.35$ pc is moved to a higher distance for a better display of the sample). The horizontal dotted line gives the solar $\numax$-value for reference. Right: identified (potential) solar analogues in terms of \teff and luminosity, using magnitudes and distances from the TESS Input Catalog \citep[TICv8.2;][]{TIC82_2021}. The dotted lines provide the solar values for reference. For reference, we also show $1\,\rm M_{\odot}$ MIST\protect\footnotemark evolutionary tracks \citep{MIST1_2016,MIST2_2016} with $\feh$-values of $-0.3$, $0$, and $+0.3$ dex (increasing with tracks from left to right; also see colorbar).} 
   \label{fig:analogs}
\end{figure*}

To have a homogeneous source for the stellar \teff and \feh, we adopted values from the Geneva-Copenhagen Survey \citep[][]{Nordstrom2004} in the revised version by \citet{Casagrande2011}. Distances and magnitudes were adopted from the TESS Input Catalog \citep[TICv8.2;][]{TIC82_2021}. In \fref{fig:analogs} we show the subsample that meets the above criteria (though we have still included stars with \feh beyond the limits). As seen, our sample contains dozens of potential solar analogues that can now be characterised asteroseismically, many of which are well-known from several spectroscopic compilations of solar analogues \citep[\eg,][]{Cayrel1996,ramirez2009,Porto2014,Datson2015}.

We can tighten the required resemblance to the Sun, in line with the definition sometimes used for a solar-twin \citep[\eg,][]{Adibekyan2017}, to $\teff=5772\pm100$ K and $\logg = 4.44\pm 0.1$ dex \citep[calculated from $\teff$ and $\numax$, see][]{Keystone2024}. This identifies four stars, namely:
\object{$\nu^2$ Lup} \citep[\object{HIP 75181}; a known multi-planet system, see][]{Udry2019,Kane2020},  
\object{51 Peg} \citep[\object{HIP 113357}; a known exoplanet host, ][]{Mayor1995}, \object{26 Dra} (\object{HIP 86036}; a known long-period visual and spectroscopic binary, see \tref{tab:binary}), and \object{HD 102365} \citep[\object{HIP 57443}; a known exoplanet host, see][]{Tinney2011}. Two of these have already been studied with asteroseismology: \object{51 Peg} \citep{Metcalfe2024} and \object{$\nu^2$ Lup} \citep[][]{Delrez2021,Weeks2025}. As suggested by the \citet{Casagrande2011}, $\feh$ values for these stars (\fref{fig:analogs}) are all outside the limit of $\feh\pm0.1$ dex corresponding to the tightened limits on $\teff$ and $\logg$ (but all within $\feh\pm 0.3$ dex). In general, we find a broad consensus between the $\feh$-values from \citet{Casagrande2011} with those found from the literature, except for \object{26 Dra} which in many other studies are found to have a near-solar metallicity \citep[see, \eg,][]{ramirez2013,Tautvai2020,Fuhrmann2008,Soubiran2022}. Of interest for future studies, \object{26 Dra} is also inside the current northern PLATO long-stare field (see \sref{sec:plato}).  

\footnotetext{using the \texttt{isochrones} Python module (\url{https://github.com/timothydmorton/isochrones}).}

\subsection{Notes on individual targets}\label{sec:indv}
In the following, we list a few targets of potentially high interest that have not been discussed in detail in the previous sections. These targets are not necessarily mentioned because of the quality of their seismology, but for the potentially improved understanding we may gain from these stars/systems through the information offered by an asteroseismic analysis. 

\paragraph{\object{$\theta$ UMa} (\object{HIP 46853}):}
A very bright ($V=3.17$) sub-giant solar-like oscillator (see \fref{fig:examples}), comparable in brightness to $\eta$ Boo, $\beta$ Hyi, $\zeta$ Her, and $\mu$ Her. While predicted by \citet{Bedding1996} to show oscillations, it has until now escaped a detailed observational investment for an asteroseismic characterisation. \object{$\theta$ UMa} has independent interferometric measurements of its diameter (\tref{tab:interfer}), and is currently being observed in RV with SONG for asteroseismic analysis.

\paragraph{\object{$\chi$ Dra} (\object{HIP 89937}):}
A bright ($V=3.55$) newly detected solar-like oscillator (F7V; see \fref{fig:examples}) that is a member of a double-lined spectroscopic (SB2) and visual binary system (see \tref{tab:binary}), with a K1V companion. The binary orbit ($P=280.5$ d) is extremely well-determined from RV observations, including many observations from SONG, from which individual component spectra can be disentangled for abundance analysis. In addition to the SB2 characterisation the visual orbit is well constrained from astrometric observations \citep[\eg,][]{Hartkopf2001}, allowing individual masses to be measured, and interferometric observations have been obtained using CHARA for independent constraint on the stellar radius. Importantly, $\chi$ Dra is included in PLATO's current northern LOP field and promises to become a key benchmark star for asteroseismology. A detailed system analysis will be presented by Rudrasingam et al. (in prep.). 

\paragraph{\object{HR 3220} (\object{HIP 39903}/\object{B Car}):} A known single-lined spectroscopic binary \citep{Murdoch1991,1993Obs...113..126M} and visual binary \citep{Goldin2006,Goldin2007} with an orbital period of $P\sim 900$ d (see \tref{tab:binary} and \fref{fig:examples}).
\citet{Fuhrmann2011b} identified \object{HR 3220} as a field blue straggler with a white dwarf companion. This was based on a match between the estimated secondary mass and the \citet{Rappaport1995} white dwarf mass-period relationship, and the measured $\rm [Fe/Mg] = -0.27$ abundance (and metallicity of $\feh=-0.27$), which suggest an old ($\tau\sim 8$--$10$ Gyr) star \citep[\eg,][]{Nissen2020}.
We note that \citet{Brown2000} indicated the potential existence of a substellar companion, based on IR-excess observed with the Hubble Near-Infrared Camera and Multi-Object Spectrometer (NICMOS) Camera 2 coronagraph (see also \citet{Schultz2014}, who suggested that the companion is a rare brown T-type dwarf, which is very uncommon to find around an F-star like HR 3220). 
Finally, we note that HR 3220 is within the PLATO LOPS2 field (\tref{tab:plato}).

\paragraph{\object{$\iota$ Peg} (\object{HIP 109176}) and \object{$\omega$ Dra} (\object{HIP 86201}):}
Both of these stars are members of well-characterised visual and SB2 systems (\tref{tab:binary}) with short orbital periods \citep[\eg,][]{Boden1999,Morel2000,Fekel2009,Konacki2010,2011AJ....142....6B}. Similar to \object{$\chi$ Dra}, it should therefore be readily possible to provide dynamical mass estimates to test the results from asteroseismology (with the caution that the assumption of isolated stellar evolution in an asteroseismic modelling effort could be questionable). \object{$\omega$ Dra} is furthermore within the current PLATO LOPN1 field (\tref{tab:plato}).

\paragraph{\object{171 Pup} (\object{HIP 37853}):} This star is the primary of a wide binary system containing the common proper motion companion star \object{VB3} \citep{VB1961}, identified by \citet{Kunkel1984} to be a low-luminosity white dwarf of spectral type DC9-11 \citep{Wesemael1993}. \object{VB3} is often identified as \object{WD 0743-336} (and sometimes \object{WD 0743-340}, \object{GJ 288B} or \object{NLTT 18414B}) and is one of the coolest WD stars and widest binaries known amongst Sirius-like-systems, which comprise a WD and a star of spectral type K or earlier \citep{Holberg2013} \citep[see also][]{Bergeron2001,Holberg2008,McCook2016}. \citet{Holberg2013} provides a period for the system of $1.38$ Myr, a separation of $14682.4$ AU, and masses of $1.08$ and $0.59\msol$ for the components. The MS A-component is found to consist of a close pair (WDS discoverer designation TOK 193 Aa,Ab; resolved in speckle interferometry by \citet{Hartkopf2012} and \citet{Tokovinin2012}) for which \citet{Tokovinin2014a} gives a period of 8.258 yr (the ORB6 database provides a period of 23.10 yr based on later observations by Tokovinin).
We find that this star/system would be very interesting to study in relation to comparing the asteroseismic age of the MS star 171 Pup A with the estimated cooling age of VB3, which could help to empirically constrain the age determination for the oldest WDs. From the age difference, the pre-WD lifetime could be estimated, which could provide the WD progenitor mass and thereby help to constrain the initial-final mass relation. 
To date, only very few binary systems containing a WD and an oscillating companion have been identified, and these often tight systems contain an early-type MS primary star that does not readily allow an age to be determined using asteroseismology. Examples are the EL CVn binaries consisting of an A- or F-type primary and a low-mass helium white dwarf (WD) secondary, where the primary occasionally is found to oscillate as a $\delta$ Scuti pulsator \citep[\eg,][]{Maxted2014,Guo2017}. 

\paragraph{\object{$o^{2}$ Eri} (\object{HIP 19849}/\object{40 Eridani A}\protect\footnote{to Trekkies, potentially better known as the planet \textit{Vulcan's} host star in the \textit{Star Trek} universe.})} This star is part of a tripe-star system, orbited ($P{\sim}8000$ yr, \tref{tab:binary}) by a binary (B and C components) consisting of a WD ($o^{2}$ Eri B/WD 0413-077; type DA2.9 and the first ever recognized WD) and an M-dwarf ($o^{2}$ Eri C) in a $230$-year-long orbit \citep{Bond2017,Mason2017}. As for 171 Pup, an asteroseismic age determination could be compared with the WD age, as obtained from cooling tracks and initial--final mass relations.

\paragraph{\object{31 Aql} (\object{HIP 95447}/\object{b Aql}):} With a metallicity of $\feh{\sim} 0.35$ dex, this star was analysed by \citet{Mishenina1996} and \citet{Feltzing2001} (among others) in the context of being a "super-metal-rich" (SMR) star.

\paragraph{\object{HD 76932} (\object{HIP 44075}):} This star was listed by \citet{Nissen2011} as a thick-disk star, and \citet{Fuhrmann2017} found it to be discrepant in $\rm [Ba/Fe]$ vs. $\rm [Fe/H]$, suggesting it to have a WD companion.

\paragraph{\object{HD 65907} (\object{HIP 38908}):} \citet{Fuhrmann2012} analyzed this star in the context of being an old Pop II star based on its abundance but found this to conflict with the age derived from evolutionary tracks, which they explained as being caused by a former mass transfer. An asteroseismic age could potentially help resolve the issue.

\paragraph{\object{HD 81809} (\object{HIP 46404}):} This star is well-studied in the context of stellar activity, and long-term X-ray monitoring has shown a well-defined chromospheric activity cycle with a period of $8.05\pm0.07$ yr \citep[][see also \citealt{Orlando2017,Egeland2018}]{Bonanno2022}, and found to have a suggested dynamo action similar to that of the Sun. Moreover, the star is a member of a well-defined visual and spectroscopic (SB2) binary system (see \tref{tab:binary}). The star was studied asteroseismically (\tref{tab:all_seis}) by \citet{Corsaro2024} in the context of oscillation amplitude suppression from magnetic activity.

\paragraph{HD 156098 (HIP 84551):} Analysis by \citet{Feng2022} listed this star as the host of two potential exoplanets with periods of $21.85\pm0.01$ and $7841^{+247}_{-517}$ days. The star is also commonly used as a comparison star in the analysis of the GRO J1655-40 system \citep{Foellmi2009}, where the companion was found to be a black hole \citep{Orosz1997,Mirabel2002}. 

\paragraph{HD 186155 (HIP 96825; HR 7495; KIC 9163520):} This star was included in our initial selection and was identified as a so-called ``Hump-and-Spike'' (H+S) star by \citet{Pope2019} based on long-cadence (30-min) smear data from \kp. It has been analysed recently in this context by \citet{Henriksen2023} and \citet{Antoci2025}. Based on the star's H+S classification and its location in the HR diagram (positioned in a region where solar-like oscillations are not expected), this star is not included in the final sample. However, we note that clear excess power akin to solar-like oscillations was identified at $\rm \nu \sim 250 \, \mu Hz$, in addition to the low-frequency peaks associated with the H+S characteristics. The origin of this excess power remains to be understood.

\section{Conclusions}

With the TLS, we have provided detections of asteroseismic signals for a total of \sn MS/SG stars, visible to the naked eye ($V\leq6$). Of these, to the best of our knowledge, \ssn are new detections. Given the brightness of this sample compared to most asteroseismic stars from \kp, it is possible to obtain a high-quality characterisation of the stars from ground-based efforts, and many of the stars already have extensive literature from decades of scrutiny. 
Our goal with this analysis has been to report on the asteroseismic detections, to encourage continued observations from TESS, and to highlight the many potential uses of the sample in future in-depth analyses.

In processing the TESS data, we used the products produced by SPOC \citep{Jenkins2016} and the light curves extracted from the TPF using custom apertures (\sref{sec:data} and \aref{app:comp}). While we found neither data product to be superior in general, we consider them highly complementary; the custom apertures allowed us in several cases (see, \eg, \fref{fig:custom_aperture}) to make detections of oscillations (or improve upon these) where the SPOC apertures were ill-defined.
From our processed light curves, we confirmed the apparent superior quality of 20-s over 120-s cadence data identified by \citet{Huber2022} and found this to extend to the brighter stars in the TLS (\sref{sec:detect}). 

We have provided values for the global asteroseismic parameters \dnu and \numax from the \texttt{PySYD} pipeline for all stars with identified oscillations (\sref{sec:astero_params}, \tref{tab:all_seis}), and all reported detections have been confirmed by three independent pipelines. We found excellent overall agreement with values from the literature for the stars with previous detections (\sref{sec:comp}) and, with only two exceptions, we were able to confirm all previous detections based on TESS data (\sref{sec:detect}).
In our comparison of measured \numax values with those expected from the ATL3, we have identified an apparent underestimation of \numax from the latter, which leads to overly optimistic detection probabilities (\sref{sec:comp} and \fref{fig:numax_comp}). The bias in \numax estimates from ATL3 is probably caused by biases in the Gaia DR3 estimates of \teff and \logg (which is especially clear for the brightness range covered by our sample) used in the \numax calculation (\aref{app:atl}).

The TLS also contains several groups that are of potential interest, as follows. We identified a total of 34 stars overlapping the current LOPN1 or LOPS2 field definitions of PLATO, and thereby of interest for the calibration or validation of the asteroseismic parameters returned from this upcoming mission (\sref{sec:plato}).  Several TLS stars are of potential interest to studies of solar-analogues (\sref{sec:analog}). We identified 54 stars that have long-baseline interferometric observations (\sref{sec:interfer}, \tref{tab:interfer}), providing independent measurements of their radii. We identified 48 stars that are members of stellar binaries where an orbital characterisation has been possible. Of these, 23 (with 9 being SB2 systems) have both spectroscopic and visual/astrometric constraints providing independent constraints on the stellar masses (\sref{sec:binary}, \tref{tab:binary}). We identified 24 exoplanet-host stars (\tref{tab:exo}), 12 of which are without previous detections of oscillation (\sref{sec:exo}), and all except 11 systems are also included in HWO target list. Finally, we identified several individual stars/systems where the detection of oscillations could be of particular interest for dedicated in-depth studies (\sref{sec:indv}), including the bright ($V=3.17$) sub-giant \object{$\theta$ UMa} and the well-characterised SB2 system $\chi$ Dra (Rudrasingam et al., in prep.).

In addition to our measurements of the global asteroseismic parameters, we estimate that ${\sim}63\%$ of the stars are amenable to peak-bagging for the analysis of individual mode parameters -- this analysis and stellar modelling of the sample will be the subject of a future analysis.
In future work, we will also extend our analysis to the correspondingly bright evolved stars observed by TESS (\fref{fig:sample}), and we will provide updates and extensions to the TLS following the continued collection of data from TESS.


\section{Data Availability}
Tables~\ref{tab:all_miss}, \ref{tab:all_seis}, \ref{tab:plato}, \ref{tab:exo}, \ref{tab:interfer}, and \ref{tab:binary} are also available in electronic form at the CDS via anonymous ftp to \url{cdsarc.u-strasbg.fr} ($130.79.128.5$) or via \url{http://cdsweb.u-strasbg.fr/cgi-bin/qcat?J/A+A/}

\begin{acknowledgements}

We thank the anonymous referee for useful comments that helped improve the initial version of the paper. 
The authors acknowledge the dedicated team behind the TESS mission, without whom this work would not have been possible. We recognise the PIs of the TESS Guest Investigator proposals that, over the years, have ensured the 120- and 20-s cadence observations of the stars analysed here (see \url{https://heasarc.gsfc.nasa.gov/docs/tess/approved-programs.html}). 

We are grateful to Daniel Hey for discussions on the ATL, to Pierre Maxted, Hugh Osborn, Valerio Nascimbeni, and Juan Cabrera for discussions on the PLATO LOP fields, and to Vichi L. Antoci for discussions on individual classical pulsators in the sample.
M.N.L. acknowledges support from the ESA PRODEX programme (PEA 4000142995).
S.M. acknowledges support from the Spanish Ministry of Science and Innovation with the grants number PID2019-107061GB-C66 and PID2023-149439NB-C41, and through AEI under the Severo Ochoa Centres of Excellence Programme 2020--2023 (CEX2019-000920-S). 
R.A.G. acknowledges the support from the GOLF and PLATO Centre National D'{\'{E}}tudes Spatiales grants. 
T.R.B. acknowledges support from the Australian Research Council (FL220100117).
D.H. acknowledges support from the Alfred P. Sloan Foundation, the National Aeronautics and Space Administration (80NSSC22K0303, 80NSSC23K0434, 80NSSC23K0435, 80NSSC21K0652) and the Australian Research Council (FT200100871).

This research has made use of the Washington Double Star Catalog maintained at the U.S. Naval Observatory.

This research has made use of the SIMBAD database \citep{wegner2000}, operated at CDS, Strasbourg, France.

This research has made use of the Jean-Marie Mariotti Center (JMMC) Measured Stellar Diameters Catalogue (available at \url{http://www.jmmc.fr/jsdc}) and the \texttt{OiDB} service (available at \url{http://oidb.jmmc.fr}).

This research has made use of the NASA Exoplanet Archive, which is operated by the California Institute of Technology, under contract with the National Aeronautics and Space Administration under the Exoplanet Exploration Program.

This research has made use of data obtained from or tools provided by the portal exoplanet.eu of The Extrasolar Planets Encyclopaedia.

This work presents results from the European Space Agency (ESA) space mission Gaia. Gaia data are being processed by the Gaia Data Processing and Analysis Consortium (DPAC). Funding for the DPAC is provided by national institutions, in particular the institutions participating in the Gaia MultiLateral Agreement (MLA). The Gaia mission website is \url{https://www.cosmos.esa.int/gaia}. The Gaia archive website is \url{https://archives.esac.esa.int/gaia}.

This research made use of the SONG database SODA (\url{https://soda.phys.au.dk/}), operated and maintained at Aarhus University, DK.      

We acknowledge the use of the following Python-based software modules: \texttt{Astropy} \citep{Astropy}, \texttt{PyAstronomy} \citep{pya},  \texttt{Lightkurve} \citep{lightkurve}, \texttt{KDEpy} \citep{KDEpy}, \texttt{scikit-image} \citep{scikit-image}, \texttt{platopoint}, \texttt{tess-atl} \citep{Hey2024}, \texttt{tpfi} \citep{tpfi2024}, and \texttt{pySYD} \citep{PySyd}.

\end{acknowledgements}

\bibliographystyle{aa} 
\bibliography{biblio} 

\newpage

\begin{appendix}
\onecolumn

\section{Stars without observations from TESS}\label{app:missing}
\tref{tab:all_miss} provides an overview of the stars that met our selection criteria in terms of $M_V$ and $B-V$, but did not have TESS observations up to and including Sector 77 as considered in this analysis. \tref{tab:all_miss} lists both the $69$ stars that will be observed during Cycle 7 and the $19$ stars that are not scheduled for observations.  

\setlength\tabcolsep{0pt} 
\begin{table*} 
\centering 
\caption{Stars fulfilling TLS selection criteria, but with no data before Sector 78} 
\label{tab:all_miss}
\scriptsize 
\renewcommand{\arraystretch}{0.8}  
\begin{tabular*}{\linewidth}{@{\extracolsep{\fill}}lllccccccc@{}} 
\toprule 
Name & TIC & HIP & $V$ & Constellation & RA & DEC &  HWO & Seis. ref. & Sectors\\ 
  &      &  & (mag) & & (deg) & (deg) & (tier) & & \\ 
\midrule 
$\alpha$ Oph & 289643770 & 86032 & 2.08 & Ophiucus & $263.7$ & $12.6$ &  - & \tablefootmark{(a)} & 79\\   
$\delta$ Cap & 155842257 & 107556 & 2.85 & Capricornus & $326.8$ & $-16.1$ &  - & - & 92\\   
$72$ Oph & 24592444 & 88771 & 3.71 & Ophiucus & $271.8$ & $9.6$ &  - & - & 80\\   
$\gamma$ Oph & 324016035 & 87108 & 3.75 & Ophiucus & $267.0$ & $2.7$ &  - & - & 80\\   
$\rho$$^1$ Sgr & 334177803 & 95168 & 3.92 & Sagittarius & $290.4$ & $-17.8$ &  - & \tablefootmark{(b)} & 92\\   
$70$ Oph & 398120047 & 88601 & 4.03 & Ophiucus & $271.4$ & $2.5$ &  B & 1 & 80\\   
$44$ Oph & 87238691 & 85340 & 4.16 & Ophiucus & $261.6$ & $-24.2$ &  - & - & 91, 92\\   
$36$ Oph\tablefootmark{(c)} & 79454735 & 84405 & 4.33 & Ophiucus & $258.8$ & $-26.6$ &  B & - & 91\\   
$\xi$ Oph & 75899957 & 84893 & 4.39 & Ophiucus & $260.3$ & $-21.1$ &  C & - & 91\\   
$\zeta$ Ser & 104572330 & 88175 & 4.62 & Serpens & $270.1$ & $-3.7$ &  - & - & 80\\   
$\theta$$^1$ Ser & 227271980 & 92946 & 4.62 & Serpens & $284.1$ & $4.2$ &  - & - & 80\\   
$20$ Oph & 181290095 & 82369 & 4.64 & Ophiucus & $252.5$ & $-10.8$ &  - & - & 91\\   
$\omega$ Sgr & 209188615 & 98066 & 4.70 & Sagittarius & $299.0$ & $-26.3$ &  - & - & 92\\   
$\mu$ Aqr & 23913506 & 103045 & 4.73 & Aquarius & $313.2$ & $-9.0$ &  - & - & 81, 92\\   
$\rho$ Cap & 429135124 & 101027 & 4.77 & Capricornus & $307.2$ & $-17.8$ &  - & - & 92\\   
$\tau$ Oph & 204010915 & 88404 & 4.77 & Ophiucus & $270.8$ & $-8.2$ &  - & - & 80\\   
$\eta$ Cap & 418824772 & 104019 & 4.82 & Capricornus & $316.1$ & $-19.9$ &  - & - & 92\\   
$58$ Oph & 238115675 & 86736 & 4.86 & Ophiucus & $265.9$ & $-21.7$ &  C & - & 91, 92\\   
$\psi$ Sco & 420895269 & 79375 & 4.93 & Scorpius & $243.0$ & $-10.1$ &  - & - & 91\\   
$\theta$$^2$ Ser & 227271997 & 92951 & 4.98 & Serpens & $284.1$ & $4.2$ &  - & - & 80\\   
$55$ Sgr & 422736099 & 96950 & 5.06 & Sagittarius & $295.6$ & $-16.1$ &  - & - & 92\\   
$\mu$ Cap & 206238826 & 108036 & 5.08 & Capricornus & $328.3$ & $-13.6$ &  - & - & 92\\   
HD $170680$ & 186642657 & 90806 & 5.12 & Sagittarius & $277.9$ & $-18.4$ &  - & - & 92\\   
$42$ Cap & 155706950 & 107095 & 5.16 & Capricornus & $325.4$ & $-14.0$ &  - & - & 92\\   
$15$ Sgr & 243895751 & 89439 & 5.29 & Sagittarius & $273.8$ & $-20.7$ &  - & - & 91\\   
$\chi$ Cap & 99389357 & 104365 & 5.30 & Capricornus & $317.1$ & $-21.2$ &  - & - & 92\\   
$\mu$ Lib & 386858986 & 72489 & 5.32 & Libra & $222.3$ & $-14.1$ &  - & - & 91\\   
HD $171802$ & 371026327 & 91217 & 5.38 & Ophiucus & $279.1$ & $9.1$ &  - & - & 80\\   
$\eta$ Lib & 71859994 & 77060 & 5.41 & Libra & $236.0$ & $-15.7$ &  - & - & 91\\   
HD $155078$ & 145740988 & 83962 & 5.43 & Ophiucus & $257.4$ & $-10.5$ &  - & - & 91\\   
HD $171834$ & 371127111 & 91237 & 5.43 & Ophiucus & $279.2$ & $6.7$ &  - & \tablefootmark{(d)} & 80\\   
$16$ Sco & 420896293 & 79387 & 5.43 & Scorpius & $243.0$ & $-8.5$ &  - & - & 91\\   
$95$ Vir & 19924794 & 68940 & 5.46 & Virgo & $211.7$ & $-9.3$ &  - & \tablefootmark{(e)} & 91\\   
$18$ Aqr & 288404080 & 105668 & 5.48 & Aquarius & $321.0$ & $-12.9$ &  - & - & 92\\   
$18$ Sco & 135656809 & 79672 & 5.49 & Scorpius & $243.9$ & $-8.4$ &  A & 2 & 91\\   
HD $186185$ & 422856892 & 97063 & 5.49 & Sagittarius & $295.9$ & $-15.5$ &  - & - & 92\\   
HD $138413$ & 73852746 & 76106 & 5.50 & Libra & $233.2$ & $-19.7$ &  - & - & 91\\   
HD $124683$ & 428790951 & 69658 & 5.53 & Virgo & $213.8$ & $-18.2$ &  - & - & 91\\   
HD $150453$ & 287619474 & 81754 & 5.55 & Ophiucus & $250.5$ & $-19.9$ &  - & - & 91\\   
HD $175317$ & 352487147 & 92882 & 5.56 & Sagittarius & $283.9$ & $-16.4$ &  - & - & 92\\   
HD $181240$ & 12056898 & 95077 & 5.59 & Sagittarius & $290.2$ & $-22.4$ &  - & - & 92\\   
HD $166285$ & 338412799 & 89000 & 5.67 & Ophiucus & $272.5$ & $3.1$ &  - & - & 80\\   
HD $173638$ & 145429390 & 92136 & 5.69 & Scutum & $281.7$ & $-10.1$ &  - & - & 80\\   
$37$ Cap & 441018038 & 106559 & 5.70 & Capricornus & $323.7$ & $-20.1$ &  - & - & 92\\   
$19$ Aqr & 187064019 & 105761 & 5.71 & Aquarius & $321.3$ & $-9.7$ &  - & - & 92\\   
$73$ Oph & 18517778 & 88964 & 5.71 & Ophiucus & $272.4$ & $4.0$ &  - & - & 80\\   
HD $131977$\tablefootmark{(f)} & 287157634 & 73184 & 5.72 & Libra & $224.4$ & $-21.4$ &  A & - & 91\\   
$26$ Oph & 18809756 & 83196 & 5.74 & Ophiucus & $255.0$ & $-25.0$ &  - & - & 91\\   
HD $171130$ & 433218066 & 90991 & 5.74 & Scutum & $278.4$ & $-14.9$ &  - & - & 80, 92\\   
$34$ Peg & 265917815 & 110785 & 5.76 & Pegasus & $336.7$ & $4.4$ &  - & - & 82\\   
HD $163318$ & 135485021 & 87836 & 5.76 & Sagittarius & $269.2$ & $-28.1$ &  - & - & 91, 92\\   
HD $162917$ & 277303601 & 87558 & 5.77 & Ophiucus & $268.3$ & $6.1$ &  - & - & 80\\   
HD $172051$ & 1544160 & 91438 & 5.85 & Sagittarius & $279.7$ & $-21.1$ &  - & - & 92\\   
HD $152311$\tablefootmark{(g)} & 220765386 & 82621 & 5.86 & Ophiucus & $253.4$ & $-20.4$ &  - & - & 91\\   
$236$ Vir & 46095850 & 69929 & 5.86 & Virgo & $214.7$ & $-18.7$ &  - & - & 91\\   
HD $124425$ & 147755337 & 69493 & 5.89 & Virgo & $213.4$ & $-0.8$ &  - & - & 91\\   
HD $199443$ & 442481598 & 103460 & 5.89 & Capricornus & $314.4$ & $-16.0$ &  - & - & 92\\   
$5$ Aql A & 225849715 & 92117 & 5.89 & Aquila & $281.6$ & $-1.0$ &  - & - & 80\\   
$17$ Cap & 422364664 & 102487 & 5.91 & Capricornus & $311.5$ & $-21.5$ &  - & - & 92\\   
$33$ Oph & 142696134 & 83478 & 5.91 & Hercules & $255.9$ & $13.6$ &  - & - & 79\\   
HD $151862$ & 147752950 & 82350 & 5.91 & Hercules & $252.4$ & $13.3$ &  - & - & 79\\   
HD $171856$ & 842202 & 91347 & 5.93 & Sagittarius & $279.5$ & $-21.4$ &  - & - & 92\\   
HD $163336$ & 210131774 & 87813 & 5.93 & Serpens & $269.1$ & $-15.8$ &  - & - & 80, 91, 92\\   
DV Aqr & 442508593 & 103545 & 5.95 & Aquarius & $314.7$ & $-14.5$ &  - & - & 92\\   
HD $163624$ & 62916034 & 87875 & 5.95 & Ophiucus & $269.3$ & $0.1$ &  - & - & 80\\   
$45$ Cap & 155772218 & 107302 & 5.96 & Capricornus & $326.0$ & $-14.7$ &  - & - & 92\\   
$16$ Sgr & 243891646 & 89440 & 5.96 & Sagittarius & $273.8$ & $-20.4$ &  - & - & 91\\   
\midrule 
$\xi$ Sco & 49725171 & 78727 & 4.16 & Scorpius & $241.1$ & $-11.4$ &  - & - & -\\   
$\tau$$^6$ Eri & 121078878 & 17651 & 4.22 & Eridanus & $56.7$ & $-23.2$ &  C & - & -\\   
$47$ Oph & 125165048 & 85365 & 4.53 & Ophiucus & $261.7$ & $-5.1$ &  - & - & -\\   
$\upsilon$ Oph & 163938779 & 80628 & 4.62 & Ophiucus & $247.0$ & $-8.4$ &  - & - & -\\   
$\sigma$ Ser & 270639162 & 80179 & 4.82 & Serpens & $245.5$ & $1.0$ &  - & - & -\\   
$60$ Her & 263798437 & 83613 & 4.89 & Hercules & $256.3$ & $12.7$ &  - & - & -\\   
$\alpha$$^1$ Lib & 386886976 & 72603 & 5.15 & Libra & $222.7$ & $-16.0$ &  - & - & -\\   
HD $158614$ & 164139204 & 85667 & 5.31 & Ophiucus & $262.6$ & $-1.1$ &  - & - & -\\   
HD $158352$ & 348108914 & 85537 & 5.41 & Ophiucus & $262.2$ & $0.3$ &  - & - & -\\   
$49$ Lib & 410696277 & 78400 & 5.47 & Libra & $240.1$ & $-16.5$ &  - & - & -\\   
$47$ Her & 276880793 & 82402 & 5.48 & Hercules & $252.6$ & $7.2$ &  - & - & -\\   
HD $159170$ & 164542749 & 85922 & 5.61 & Ophiucus & $263.4$ & $-5.7$ &  - & - & -\\   
$4$ Ser & 461273024 & 74689 & 5.62 & Serpens & $229.0$ & $0.4$ &  - & - & -\\   
HD $164402$ & 110325484 & 88298 & 5.72 & Sagittarius & $270.5$ & $-22.8$ &  - & - & -\\   
$14$ Oph & 282056074 & 81734 & 5.74 & Ophiucus & $250.4$ & $1.2$ &  - & - & -\\   
$12$ Oph & 58092025 & 81300 & 5.77 & Ophiucus & $249.1$ & $-2.3$ &  A & - & -\\   
HD $145148$ & 277548101 & 79137 & 5.93 & Serpens & $242.3$ & $6.4$ &  - & - & -\\   
o Cap A & 72429235 & 101123 & 5.94 & Capricornus & $307.5$ & $-18.6$ &  - & - & -\\   
V2213 Oph & 297236135 & 83601 & 6.00 & Ophiucus & $256.3$ & $0.7$ &  - & - & -\\   
\bottomrule
\end{tabular*} 
\tablefoot{\tiny The table provides an overview of the stars matching the selection criteria for the TLS, but without observations up until and including Sector 77, including at the bottom stars without any scheduled observations in Cycle 7. Stars are sorted according to their visual magnitude (``$V$"). The first three columns provide identifiers for the stars in the form of their Bayer/Flamsteed designation (or primary name according to \texttt{SIMBAD}) in addition to their TESS (``TIC") and Hipparcos (``HIP") IDs. The ``HWO" column indicates if the star is found in the HWO target list and provides the tier (A, B, or C). The last two columns provide references to the primary sources of previous seismic detections if they exist, and list the Sectors with scheduled TESS observations.}
\tablefoottext{a}{Fast-rotating A-star with oscillations measured by \citet{Monnier2010} and \citet{Mirouh2014};}\tablefoottext{b}{Listed as a $\delta$ Sct variable in the GCVS \citep{Samus2017} based on observations by \citet{Baglin1973};}\tablefoottext{c}{Triple-star system of K-dwarfs \citep{Strobel1989};}\tablefoottext{d}{Identified as a potential $\gamma$ Dor pulsator by \citet{Uytterhoeven2011};}\tablefoottext{e}{Identified as a $\delta$ Sct variable by \citet{Paunzen2017};}\tablefoottext{f}{A quaternary star system with the B and C components forming their own binary system, and with component D being a T-type brown-dwarf (Gl 570 D) in a 1500 au wide orbit \citep{Burgasser2000,Geballe2001};}\tablefoottext{g}{Long period ($P{\sim}2713$ d) visual and spectroscopic binary \citep{Tokovinin2012,Fekel2018}.} 
\tablebib{\tiny (1) \citet{Carrier2006_70Oph}; (2) \citet{Bazot2011}.} 
\end{table*}


\section{Comparing SPOC and custom apertures}\label{app:comp}

\begin{figure*}[h!]
\centering
   \includegraphics[width=\textwidth]{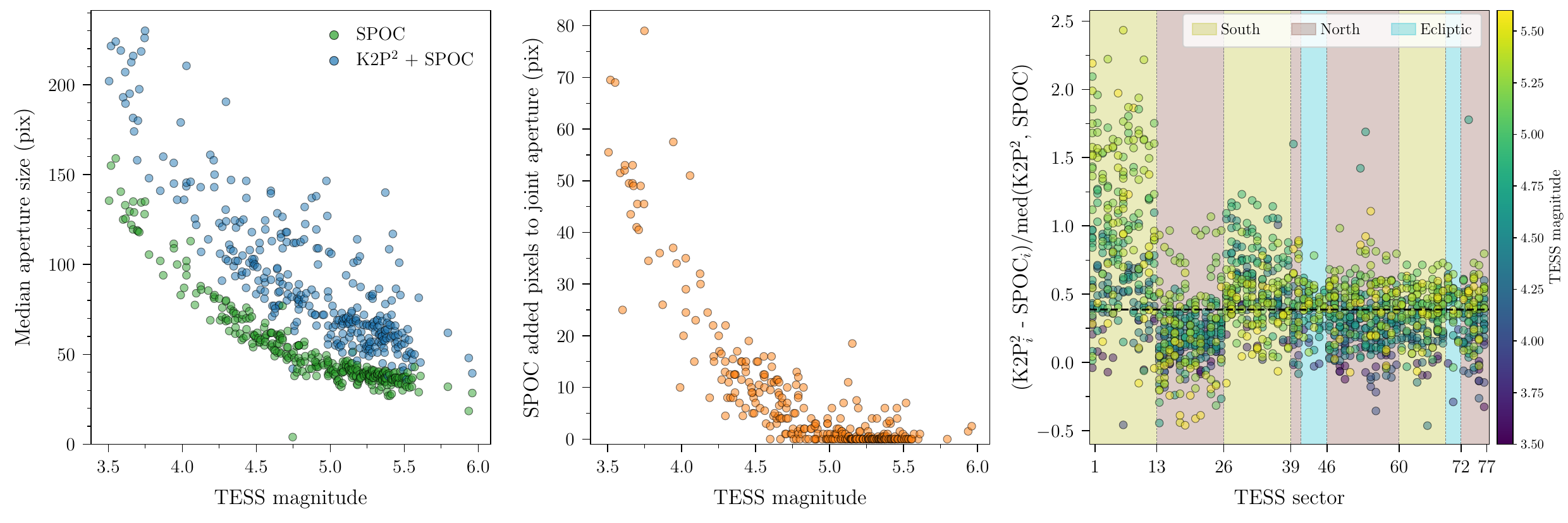}
   \caption{Comparison between SPOC and K2P$^2$ apertures. Left: Median (across sectors) aperture size in pixels against TESS magnitude for stars observed in 120-s cadence for apertures defined by SPOC and the adopted joint custom K2P$^2$ + SPOC apertures (see legend). Middle: Number of pixels contributed to the custom apertures by SPOC (\ie, number of pixels not contained in the K2P$^2$ aperture) as a function of TESS magnitude. Right: Difference in K2P$^2$ and SPOC aperture size relative to the median, with points colour-coded according to TESS magnitude. The dashed horizontal line provides the median level across sectors, while vertical lines indicate shifts in pointing (see legend), typically coinciding with different sectors.} 
   \label{fig:apertures1}
\end{figure*}

As mentioned in \sref{sec:data}, we use, in addition to data from SPOC light curves extracted from the TESS TPFs using custom apertures made from the combination of SPOC and K2P$^2$ apertures \citep{Lund2015}. In \fref{fig:apertures1} we compare the joint apertures and those from SPOC based on 120-s cadence data, and as seen the aperture sizes follow the expected trend against magnitude\footnote{The SPOC outlier near ${T_{\rm mag}}{\sim}4.75$ is \object{TIC 20601206} (\object{HIP 93017}) which for several sectors in 120-s cadence data have an aperture defined by a few disjoint pixels. For the corresponding 20-s cadence data, the apertures look as expected.}. Per definition, the joint apertures will always be equal to or larger than the one from SPOC alone, but we can see that from a TESS magnitude (${T_{\rm mag}}$) of ${\sim}4.7$ and below the aperture from SPOC contributes an increasing fraction of pixels to the joint aperture -- this is the magnitude region where the blooming trails from saturation become significant. Above ${T_{\rm mag}}{\sim}4.7$, the aperture from K2P$^2$ will typically fully contain the SPOC aperture.  
As expected, we see (not shown) little ($\pm{1-2}$ pixels) to no difference between the apertures defined for 120-s vs. 20-s cadence data.

\begin{figure*}
\centering
   \includegraphics[width=\textwidth]{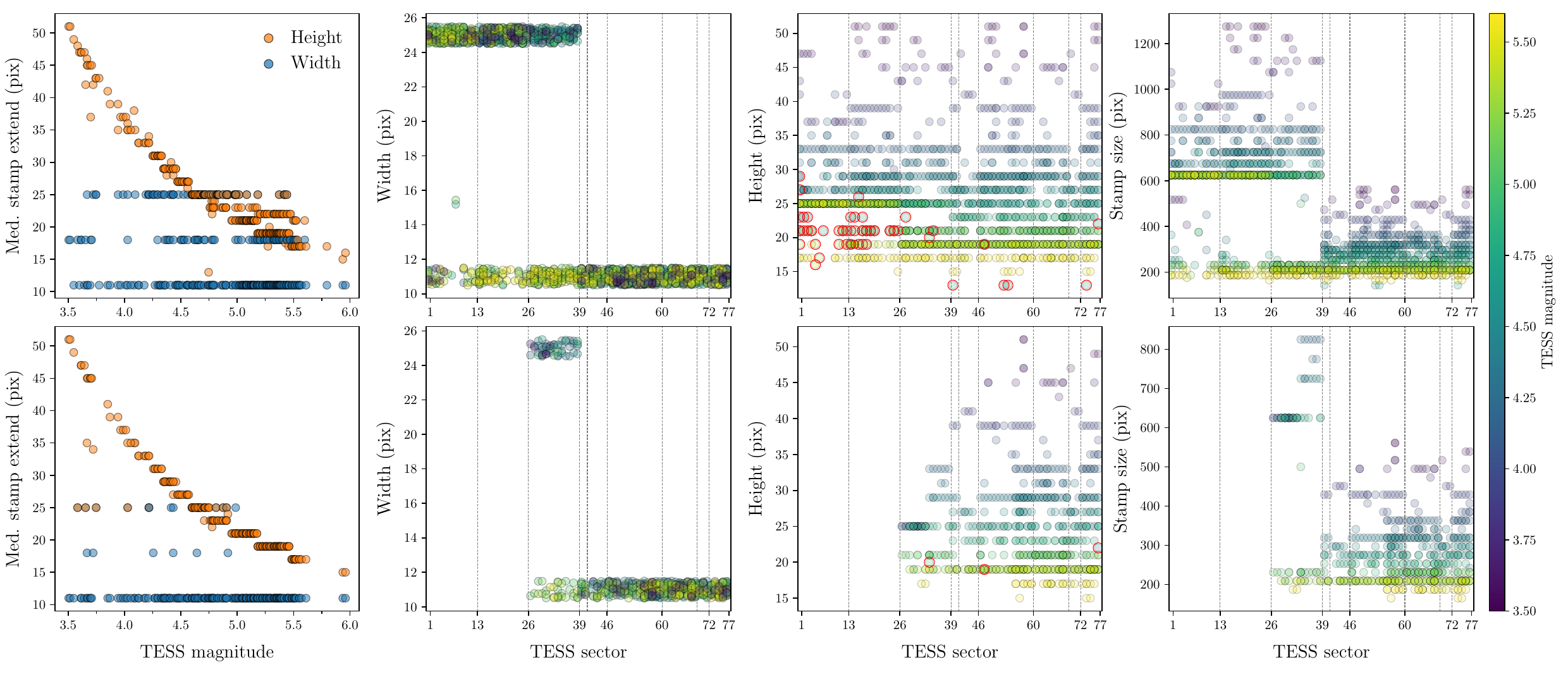}
   \caption{Properties of TESS pixel stamps for 120-s (top row) and 20-s (bottom row) cadence observations. First column: Median pixel stamp width and height across sectors for individual stars as a function of their TESS magnitude. Second (third) column: Stamp width (height) as a function of sector, coloured by TESS magnitude. Width values have been dithered by up to $\pm0.5$ pixels to better show any dependence on magnitude. Vertical lines indicate changes in pointings (see \fref{fig:apertures1}, right panel). For the heights, red markers indicate stars that are found to have a pixel stamp height that is smaller than 2 or more fainter stars, \ie, breaking with the general monotonic increase in stamp height with decreasing TESS magnitude seen on the first column. Fourth column: Total stamp size as a function of sector, colour-coded according to TESS magnitude. } 
   \label{fig:apertures2}
\end{figure*}

The aperture sizes in the left and middle panels of \fref{fig:apertures1} are median values if stars are observed in multiple sectors. For a given star, there is some scatter in the aperture sizes defined for different sectors. There can be many causes for such scatter, \eg, from the variation in TESS PSFs across the focal plane which could add to the scatter for a given star from varying boresight distances \citep{Vanderspek2018}, and this effect would also add to the width of the aperture size distribution of the given ${T_{\rm mag}}$. 
From the right panel of \fref{fig:apertures1} we see that at least part of the scatter in aperture sizes for a given star is related to the observing sector. The plot shows for a given star the difference in aperture sizes between SPOC and K2P$^2$ for a given sector ($i$) compared to the median of the SPOC and K2P$^2$ apertures across all observing sectors for the star. We see that on this scale (and with the adopted settings for the threshold for selecting pixels of interest), the K2P$^2$ apertures are in median ${\sim}40\%$ larger than the SPOC ones across all magnitudes and sectors. There are, however, clear variations between different sectors, or rather between different pointings/cycles -- in Cycle 1 (southern ecliptic hemisphere; sectors 1-13) the K2P$^2$ apertures are relatively larger than SPOC as compared to Cycle 2 (northern ecliptic hemisphere; sectors 14-26), and the largest effect is seen for the fainter stars in the sample. One might speculate that this relates to different observing conditions in the different sectors/pointings, and indeed variations in scattered light or pointing stability could and might have an effect, but an important aspect also seems to be the TPF stamp sizes (\fref{fig:apertures2}). 

In \fref{fig:apertures2} we show the variation in TPF stamp widths and heights as a function of magnitude and observing sector. As expected, a strong dependence on the stamp height (direction of potential blooming trails) on magnitude is seen, while the width has a value of $11$ or $25$ pixels. In the early cycles (1-3, sectors 1-39) there is a general tendency for brighter targets observed in 120-s cadence to have widths of $25$ pixels while fainter targets adopt widths of $11$ pixels. For 20-s cadence observations (introduced from Cycle 3, starting with Sector 27), this tendency is mainly observed for Cycle 3 (sectors 27-39) -- later sectors almost exclusively adopt widths of $11$ pixels. Concerning the stamp heights, we see that a value of $25$ pixels is representative in median across cycles and cadences.    

\section{Comparison to the ATL}\label{app:atl}

The apparent underestimation of \numax in the ATL (\fref{fig:numax_comp}) means that the calculated detection probability will generally be overestimated, as oscillation amplitudes correlate inversely with \numax. The ATL was built to guide the target selection process for the TESS Asteroseismic Science Consortium (TASC). Though certainly not intentional, the slightly optimistic detection probabilities ensure that a minimal number of targets showing oscillations, if observed, are overlooked in the observing proposals. Conversely, this could explain the generally low fraction of positive seismic detections obtained from TESS, at least early in the mission. 

The ATL in its different versions is built on the methodology laid out by \citet{Chaplin2011}, and the ATL3 version \citep{Hey2024} relies primarily on the Gaia DR3 \citep{GaiaDr3_2023} estimates of \teff and \logg to calculate \numax following:
\begin{equation}\label{eq:numax}
    \numax = \frac{g}{g_{\odot}} \sqrt{\frac{T_{\rm eff, \odot}}{\teff} } \, \nu_{\rm max, \odot}\, .
\end{equation}
The \teff (\logg) in ATL3 is first and foremost taken as Gaia's \texttt{teff$\_$gspphot} (\texttt{logg$\_$gspphot}), followed by the \teff (\logg) in the TIC \citep[which can originate from a variety of sources; see][]{Stassun2019} and finally Gaia's \texttt{teff$\_$gspspec} (\texttt{logg$\_$gspspec}), depending on availability. We note that the detection probability calculation adopted by \citet{Keystone2016,Keystone2024} for K2 targets follows the same overall methodology, but here adopting principally 2MASS $(J-K_S)$ colours and the relation of \citet{Casagrande2010} to estimate \teff, combined with a luminosity (from Hipparcos) and the stellar mass from a simple mass-luminosity relation to estimate \numax following $\numax = M/L \left(\teff/T_{\rm eff, \odot}\right)^{3.5}  \times \nu_{\rm max, \odot} $. However, in this case, no systematic offset is seen between predicted and measured \numax values (see their Figure~11 and F.1, respectively), suggesting that the Gaia inputs used to calculate \numax are somehow biased.

To assess the \numax underestimation from the current ATL3 we made a comparison of ATL3 \numax against values from the catalogues of \citet{Hatt2023}, \citet{Zhou2024}, \citet{Keystone2024} (K2; Keystone), and the \kp targets from \citet{Lund2017}, \citet{Mathur2022}, and \citet{Serenelli2017}. 
\begin{figure*}
\centering
   \includegraphics[width=\textwidth]{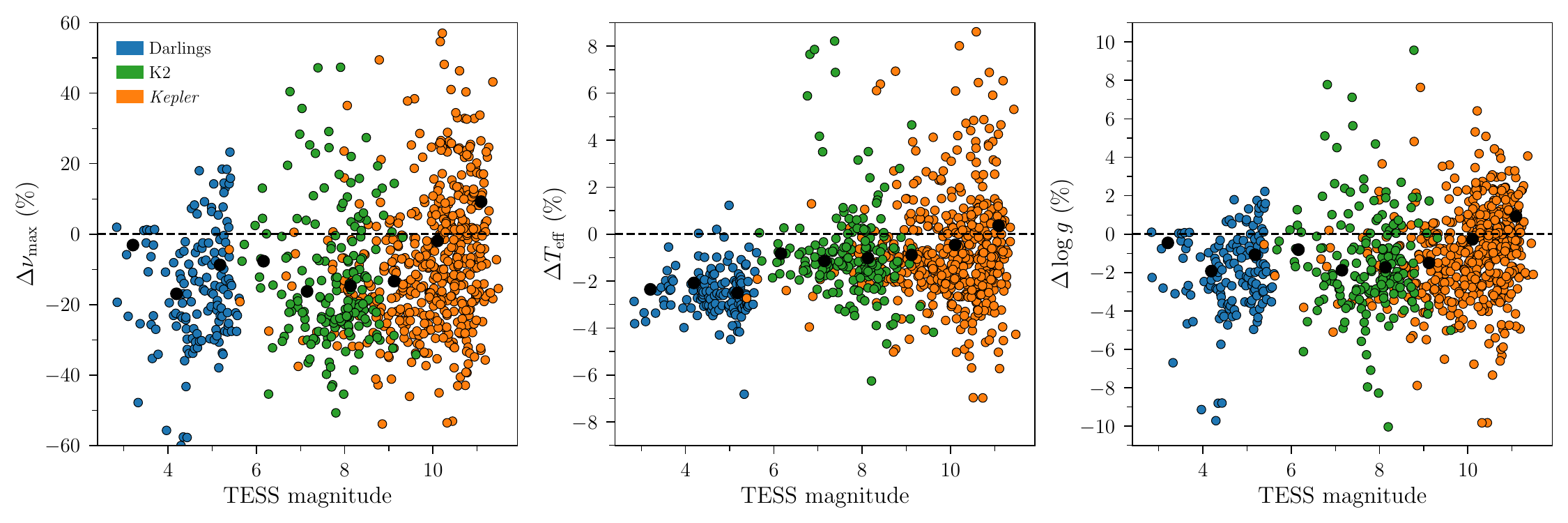}
   \caption{Comparison between ATL3 input/output and measured values for the TLS and literature values for the K2 \citep{Keystone2016,Keystone2024} and \kp \citep{Lund2017,Mathur2022,Serenelli2017} samples. Left: relative difference (measured-predicted) between measured and predicted values of \numax from ATL3 with its standard priority of inputs on \teff and \logg (primarily from Gaia DR3's \texttt{gspphot}) against TESS magnitude. Middle (Right): relative difference on \teff (\logg) for the three samples. For the K2 and \kp samples, the measured values are obtained from the literature and adopted from \citet{Casagrande2011} for the TLS. The black markers indicate median binned values of the combined samples.}
   \label{fig:atl3_offset}
\end{figure*}
For the K2 and \kp samples, we have estimates of both \teff and \logg (from \teff and \numax) from the above publications, while we for the current TLS adopt the GCS values of \citet{Casagrande2011}.
 
For all samples, we see a general underestimation of \numax from ATL3, ranging between $12-20\%$, and being the most significant for the TLS and K2 samples. For the samples where we have \teff and \logg (\kp, K2, and TLS) in addition to \numax, \fref{fig:atl3_offset} illustrates that a general median underestimation of ${\sim}1.2\%$ and ${\sim}1.6\%$ is obtained for the \teff and \logg values adopted in ATL3 (primarily from Gaia DR3's \texttt{gspphot}) -- this was also noted in the analysis of $\sigma$ Dra by \citet{Hon2024}. For both parameters, especially \teff, we see a tendency for the underestimation to increase with the stellar brightness.   

We also compared the offset in \numax for the different possible sources of \teff and \logg used in ATL3. Our analysis suggests that for the brighter stars given by the TLS and K2 samples with TESS magnitudes in the approximate range $2.7-8.5$, a significantly better agreement is obtained by preferentially adopting inputs from the TIC (followed by \texttt{teff$\_$gspphot}/\texttt{logg$\_$gspphot} and lastly \texttt{teff$\_$gspspec}/\texttt{logg$\_$gspspec}) -- this order of preferred inputs reduced the \numax underestimation from ${\sim}16-17\%$ to ${\sim}6-8\%$ in median. However, for the fainter \kp sample, typically with TESS magnitudes ${>}8$, the current order adopted by the ATL3 does provide the smallest \numax underestimation at ${\sim}9\%$ in median. 


\section{Tables}
This appendix provides the tables referred to in the main text of the paper.

\longtab[1]{                     
\begingroup
\setlength\tabcolsep{0pt}
\scriptsize 
\renewcommand{\arraystretch}{0.8}  
\begin{longtable}{@{\extracolsep{\fill}}lllccllccccc@{}}
\caption{\label{tab:all_seis}TLS global asteroseismic parameters}\\
\toprule
Name & TIC & HIP & $V$ & Constellation & RA & DEC &
$\nu_{\rm max}$ & $\Delta\nu$ & HWO & Seis.\ ref. & Also in Table\\
      &     &     & (mag) &               & (deg) & (deg) &
($\mu$Hz) & ($\mu$Hz) & (tier) & &\\
\midrule
\endfirsthead               
\caption{continued.}\\
\toprule
Name & TIC & HIP & $V$ & Constellation & RA & DEC &
$\nu_{\rm max}$ & $\Delta\nu$ & HWO & Seis.\ ref. & Also in Table\\
\midrule
\endhead

\midrule\endfoot           
\bottomrule\endlastfoot    
$\eta$ Boo & 367758676 & 67927 & 2.68 & Bo{\"o}tes & $208.7$ & $18.4$ &  $697.9 \pm 17.8$ & $39.2 \pm 0.9$ & - & 1, 2, 3 & \ref{tab:interfer}, \ref{tab:binary}\\   
$\zeta$ Her & 43255143 & 81693 & 2.81 & Hercules & $250.3$ & $31.6$ &  $718.5 \pm 24.4$ & $40.0 \pm 1.0$ & - & 4 & \ref{tab:plato}, \ref{tab:interfer}, \ref{tab:binary}\\   
$\beta$ Hyi & 267211065 & 2021 & 2.82 & Hydrus & $6.4$ & $-77.3$ &  $1038.1 \pm 14.9$ & $56.9 \pm 0.4$ & C & 5, 6, 7, 2, 8 & \ref{tab:interfer}\\   
$\theta$ UMa & 150226696 & 46853 & 3.17 & Ursa Major & $143.2$ & $51.7$ &  $779.3 \pm 17.8$ & $44.6 \pm 0.5$ & - & - & \ref{tab:interfer}\\   
$\xi$ Gem & 372480991 & 32362 & 3.35 & Gemini & $101.3$ & $12.9$ &  $2871.0 \pm 111.3$ & $126.2 \pm 2.5$ & - & - & \ref{tab:interfer}\\   
$\mu$$^1$ Her & 460067868 & 86974 & 3.42 & Hercules & $266.6$ & $27.7$ &  $1192.5 \pm 17.5$ & $63.8 \pm 0.2$ & - & 9, 10, 2 & \ref{tab:plato}, \ref{tab:interfer}, \ref{tab:binary}\\   
$\eta$ Cas & 445258206 & 3821 & 3.46 & Cassiopeia & $12.3$ & $57.8$ &  $2840.0 \pm 80.7$ & $128.0 \pm 0.4$ & A & 4 & \ref{tab:interfer}, \ref{tab:binary}\\   
$\delta$ Eri & 38511251 & 17378 & 3.52 & Eridanus & $55.8$ & $-9.8$ &  $677.6 \pm 8.3$ & $41.5 \pm 0.8$ & A & 11, 2, 8, 3 & \ref{tab:interfer}\\   
$\delta$ Pav & 409891396 & 99240 & 3.55 & Pavo & $302.2$ & $-66.2$ &  $2269.8 \pm 64.4$ & $107.9 \pm 0.2$ & A & 12 & \ref{tab:interfer}\\   
$\chi$ Dra & 341873045 & 89937 & 3.55 & Draco & $275.3$ & $72.7$ &  $2314.7 \pm 24.4$ & $108.4 \pm 0.1$ & - & - & \ref{tab:plato}, \ref{tab:binary}\\   
$\beta$ Vir & 366661076 & 57757 & 3.59 & Virgo & $177.7$ & $1.8$ &  $1446.3 \pm 75.9$ & $71.9 \pm 0.5$ & C & 13 & \ref{tab:interfer}\\   
$\gamma$ Lep & 93280676 & 27072 & 3.59 & Lepus & $86.1$ & $-22.4$ &  $2257.5 \pm 50.9$ & $102.0 \pm 0.9$ & A & - & \ref{tab:plato}, \ref{tab:interfer}\\   
$\beta$ Aql & 375621179 & 98036 & 3.71 & Aquila & $298.8$ & $6.4$ &  $414.7 \pm 6.8$ & $26.8 \pm 0.6$ & - & 14, 15, 8, 3 & \ref{tab:interfer}\\   
$\iota$ Peg & 357336603 & 109176 & 3.77 & Pegasus & $331.8$ & $25.3$ &  $2101.8 \pm 126.6$ & $100.7 \pm 1.4$ & - & - & \ref{tab:interfer}, \ref{tab:binary}\\   
$\alpha$ For & 88523071 & 14879 & 3.80 & Fornax & $48.0$ & $-29.0$ &  $1128.3 \pm 47.3$ & $61.0 \pm 0.9$ & C & 14 & \ref{tab:binary}\\   
$\gamma$ Ser & 377415363 & 78072 & 3.85 & Serpens & $239.1$ & $15.7$ &  $1744.6 \pm 38.2$ & $84.9 \pm 0.9$ & A & 14 & \ref{tab:interfer}\\   
$\theta$ Dra & 161825882 & 78527 & 4.01 & Draco & $240.5$ & $58.6$ &  $723.0 \pm 15.8$ & $40.2 \pm 0.3$ & - & - & \ref{tab:plato}, \ref{tab:binary}\\   
$\theta$ Boo & 441709021 & 70497 & 4.04 & Bo{\"o}tes & $216.3$ & $51.9$ &  $1354.0 \pm 108.4$ & $68.9 \pm 1.5$ & C & - & \ref{tab:interfer}\\   
$\iota$ Per & 116988032 & 14632 & 4.05 & Perseus & $47.3$ & $49.6$ &  $1855.3 \pm 33.9$ & $89.7 \pm 0.4$ & A & - & \ref{tab:interfer}\\   
$\alpha$ Cha & 287532010 & 40702 & 4.05 & Chamaleon & $124.6$ & $-76.9$ &  $1005.8 \pm 65.4$ & $54.8 \pm 0.5$ & - & - & -\\   
$\iota$ Vir & 6029884 & 69701 & 4.07 & Virgo & $214.0$ & $-6.0$ &  $644.8 \pm 24.0$ & $36.2 \pm 0.8$ & - & - & \ref{tab:interfer}, \ref{tab:binary}\\   
$\upsilon$ And & 189576919 & 7513 & 4.10 & Andromeda & $24.2$ & $41.4$ &  $1528.0 \pm 50.6$ & $76.2 \pm 0.6$ & C & - & \ref{tab:exo}, \ref{tab:interfer}\\   
$\theta$ Per & 302158903 & 12777 & 4.10 & Perseus & $41.0$ & $49.2$ &  $2314.2 \pm 166.2$ & $104.1 \pm 1.3$ & A & - & \ref{tab:interfer}, \ref{tab:binary}\\   
$\iota$ Psc & 419919445 & 116771 & 4.13 & Pisces & $355.0$ & $5.6$ &  $1416.4 \pm 53.6$ & $75.3 \pm 0.6$ & C & - & \ref{tab:interfer}\\   
$\psi$ Cap & 269995013 & 102485 & 4.13 & Capricornus & $311.5$ & $-25.3$ &  $1849.1 \pm 34.5$ & $89.1 \pm 2.3$ & C & - & -\\   
$110$ Her & 282038438 & 92043 & 4.19 & Hercules & $281.4$ & $20.5$ &  $1061.9 \pm 28.0$ & $56.2 \pm 1.6$ & - & - & \ref{tab:interfer}\\   
$\xi$ Peg & 60716322 & 112447 & 4.20 & Pegasus & $341.7$ & $12.2$ &  $986.8 \pm 16.7$ & $57.3 \pm 0.5$ & C & 8 & \ref{tab:interfer}\\   
$\gamma$ Pav & 265488188 & 105858 & 4.21 & Pavo & $321.6$ & $-65.4$ &  $2635.8 \pm 60.9$ & $119.6 \pm 0.6$ & A & 16, 17, 2 & -\\   
$\zeta$ Tuc & 425935521 & 1599 & 4.23 & Tucana & $5.0$ & $-64.9$ &  $2744.6 \pm 148.6$ & $125.6 \pm 0.5$ & B & 17, 2 & -\\   
$\beta$ CVn & 458445966 & 61317 & 4.24 & Canes Venatici & $188.4$ & $41.4$ &  $2385.4 \pm 136.0$ & $117.4 \pm 1.1$ & A & - & \ref{tab:interfer}\\   
$82$ Eri & 301051051 & 15510 & 4.26 & Eridanus & $50.0$ & $-43.1$ &  $3180.0 \pm 104.9$ & $147.5 \pm 1.5$ & B & - & \ref{tab:exo}\\   
$10$ Tau & 311092847 & 16852 & 4.29 & Taurus & $54.2$ & $0.4$ &  $1284.1 \pm 63.1$ & $69.7 \pm 1.2$ & B & - & \ref{tab:interfer}\\   
$\lambda$ Ser & 296740796 & 77257 & 4.42 & Serpens & $236.6$ & $7.4$ &  $1856.6 \pm 46.4$ & $89.3 \pm 0.6$ & A & 18 & \ref{tab:exo}, \ref{tab:interfer}\\   
o$^2$ Eri & 67772871 & 19849 & 4.43 & Eridanus & $63.8$ & $-7.7$ &  $3433.1 \pm 368.6$ & $131.9 \pm 3.1$ & A & - & \ref{tab:interfer}, \ref{tab:binary}\\   
HD $60532$ & 456871289 & 36795 & 4.44 & Puppis & $113.5$ & $-22.3$ &  $618.2 \pm 28.0$ & $33.5 \pm 1.2$ & - & 2, 8 & \ref{tab:plato}, \ref{tab:exo}\\   
$\theta$ Cyg & 27014182 & 96441 & 4.49 & Cygnus & $294.1$ & $50.2$ &  $1759.1 \pm 67.1$ & $82.8 \pm 1.2$ & - & 19 & \ref{tab:plato}, \ref{tab:interfer}\\   
$\mu$ Cyg & 452700589 & 107310 & 4.49 & Cygnus & $326.0$ & $28.7$ &  $1213.9 \pm 67.7$ & $62.3 \pm 1.8$ & - & - & \ref{tab:binary}\\   
$\upsilon$ Cep & 421444084 & 102431 & 4.52 & Cepheus & $311.3$ & $57.6$ &  $958.7 \pm 22.3$ & $53.6 \pm 0.4$ & - & 2 & \ref{tab:plato}, \ref{tab:binary}\\   
$\psi$$^1$ Dra A & 441804568 & 86614 & 4.57 & Draco & $265.5$ & $72.1$ &  $1232.4 \pm 19.8$ & $61.8 \pm 0.3$ & - & - & \ref{tab:plato}, \ref{tab:interfer}, \ref{tab:binary}\\   
$\tau$$^1$ Hya & 77549389 & 46509 & 4.59 & Hydra & $142.3$ & $-2.8$ &  $1572.0 \pm 44.5$ & $75.5 \pm 1.4$ & - & - & \ref{tab:binary}\\   
$\chi$ Her & 157364190 & 77760 & 4.60 & Hercules & $238.2$ & $42.5$ &  $1045.6 \pm 17.7$ & $58.8 \pm 0.7$ & A & 2, 8, 3 & \ref{tab:plato}, \ref{tab:interfer}\\   
$\sigma$ Dra & 259237827 & 96100 & 4.67 & Draco & $293.1$ & $69.7$ &  $4217.9 \pm 122.6$ & $182.2 \pm 0.5$ & A & 20 & \ref{tab:plato}, \ref{tab:interfer}\\   
$\lambda$ Aur & 409104974 & 24813 & 4.69 & Auriga & $79.8$ & $40.1$ &  $2152.0 \pm 54.2$ & $100.6 \pm 0.8$ & A & - & \ref{tab:interfer}\\   
$\kappa$ Ret & 262843771 & 16245 & 4.71 & Reticulum & $52.3$ & $-62.9$ &  $1552.1 \pm 48.6$ & $73.1 \pm 1.3$ & C & - & \ref{tab:plato}\\   
$61$ Vir & 422478973 & 64924 & 4.74 & Virgo & $199.6$ & $-18.3$ &  $3099.8 \pm 101.0$ & $138.6 \pm 1.3$ & B & - & \ref{tab:exo}, \ref{tab:interfer}\\   
HR $3220$ & 308844962 & 39903 & 4.74 & Carina & $122.3$ & $-61.3$ &  $1386.8 \pm 27.0$ & $71.6 \pm 0.6$ & - & - & \ref{tab:plato}, \ref{tab:binary}\\   
$\sigma$ Cet & 66604391 & 11783 & 4.74 & Cetus & $38.0$ & $-15.2$ &  $913.8 \pm 40.1$ & $50.9 \pm 0.9$ & - & 2 & \ref{tab:binary}\\   
$\lambda$ Ara & 96745915 & 86486 & 4.76 & Ara & $265.1$ & $-49.4$ &  $1476.1 \pm 43.0$ & $71.1 \pm 1.7$ & C & - & -\\   
$\omega$ Dra & 233195546 & 86201 & 4.77 & Draco & $264.2$ & $68.8$ &  $1999.5 \pm 55.0$ & $88.3 \pm 2.0$ & - & - & \ref{tab:plato}, \ref{tab:binary}\\   
$40$ Leo & 95431211 & 50564 & 4.78 & Leo & $154.9$ & $19.5$ &  $1405.8 \pm 57.9$ & $75.1 \pm 2.3$ & C & - & \ref{tab:interfer}\\   
HD $5015$ & 285544488 & 4151 & 4.80 & Cassiopeia & $13.3$ & $61.1$ &  $1399.3 \pm 52.9$ & $70.8 \pm 0.6$ & C & - & \ref{tab:interfer}\\   
$\sigma$$^2$ UMa & 219709102 & 45038 & 4.80 & Ursa Major & $137.6$ & $67.1$ &  $1354.9 \pm 64.6$ & $72.1 \pm 0.6$ & C & - & \ref{tab:binary}\\   
$12$ Boo & 418010485 & 69226 & 4.82 & Bo{\"o}tes & $212.6$ & $25.1$ &  $694.7 \pm 21.3$ & $41.6 \pm 1.2$ & - & 21 & \ref{tab:binary}\\   
$36$ UMa & 416519065 & 51459 & 4.82 & Ursa Major & $157.7$ & $56.0$ &  $2319.5 \pm 96.8$ & $105.0 \pm 2.1$ & A & - & \ref{tab:interfer}\\   
$\omega$ And & 191764184 & 6813 & 4.83 & Andromeda & $21.9$ & $45.4$ &  $1899.2 \pm 336.0$ & $96.1 \pm 1.8$ & - & - & \ref{tab:binary}\\   
HD $114613$ & 30293512 & 64408 & 4.85 & Centaurus & $198.0$ & $-37.8$ &  $875.7 \pm 18.6$ & $48.1 \pm 0.9$ & C & 2, 8 & -\\   
$19$ Dra & 289622310 & 82860 & 4.88 & Draco & $254.0$ & $65.1$ &  $2313.1 \pm 493.9$ & $104.8 \pm 3.5$ & - & - & \ref{tab:plato}, \ref{tab:binary}\\   
$6$ Cet & 289673491 & 910 & 4.89 & Cetus & $2.8$ & $-15.5$ &  $1369.8 \pm 37.8$ & $73.4 \pm 0.6$ & A & - & -\\   
HD $91324$ & 447823435 & 51523 & 4.89 & Vela & $157.8$ & $-53.7$ &  $1080.5 \pm 42.8$ & $58.9 \pm 0.4$ & C & 2, 8 & -\\   
HD $102365$ & 454082369 & 57443 & 4.89 & Centaurus & $176.6$ & $-40.5$ &  $3148.2 \pm 188.4$ & $137.3 \pm 1.5$ & A & 2 & \ref{tab:exo}\\   
$\zeta$ TrA & 362747897 & 80686 & 4.90 & Triangulum Australe & $247.1$ & $-70.1$ &  $3215.4 \pm 380.4$ & $148.8 \pm 1.4$ & - & - & \ref{tab:binary}\\   
HD $114837$ & 255854921 & 64583 & 4.90 & Centaurus & $198.6$ & $-59.1$ &  $1723.1 \pm 49.8$ & $84.8 \pm 0.9$ & C & - & -\\   
$104$ Tau & 27136704 & 23835 & 4.91 & Taurus & $76.9$ & $18.6$ &  $1261.4 \pm 46.5$ & $69.2 \pm 0.3$ & C & 8 & -\\   
$\epsilon$ Lib & 37018039 & 75379 & 4.92 & Libra & $231.1$ & $-10.3$ &  $775.9 \pm 51.4$ & $43.4 \pm 1.6$ & - & - & \ref{tab:binary}\\   
HD $84117$ & 11310083 & 47592 & 4.93 & Hydra & $145.6$ & $-23.9$ &  $2142.6 \pm 90.3$ & $107.4 \pm 1.8$ & A & - & -\\   
$17$ Crt & 429149606 & 56280 & 4.93 & Hydra & $173.1$ & $-29.3$ &  $1413.4 \pm 112.2$ & $70.8 \pm 1.4$ & - & - & -\\   
$\tau$ PsA & 97402436 & 109422 & 4.94 & Pisces Austrinus & $332.5$ & $-32.5$ &  $2112.0 \pm 145.3$ & $92.8 \pm 1.8$ & A & - & -\\   
HD $10307$ & 327507922 & 7918 & 4.96 & Andromeda & $25.4$ & $42.6$ &  $2490.1 \pm 96.0$ & $115.3 \pm 1.2$ & - & - & \ref{tab:binary}\\   
$70$ Vir & 95473936 & 65721 & 4.97 & Virgo & $202.1$ & $13.8$ &  $940.6 \pm 12.9$ & $51.5 \pm 1.0$ & - & 2, 8, 3 & \ref{tab:exo}, \ref{tab:interfer}\\   
$\nu$ Phe & 229092427 & 5862 & 4.97 & Phoenix & $18.8$ & $-45.5$ &  $2309.1 \pm 107.3$ & $106.9 \pm 1.2$ & B & 2 & -\\   
$36$ Dra & 233121747 & 89348 & 4.99 & Draco & $273.5$ & $64.4$ &  $1312.0 \pm 16.9$ & $69.6 \pm 0.2$ & C & - & \ref{tab:plato}, \ref{tab:interfer}\\   
$\nu$$^1$ Lup & 136915882 & 75206 & 4.99 & Lupus & $230.5$ & $-47.9$ &  $749.2 \pm 25.2$ & $41.0 \pm 1.2$ & - & 2 & -\\   
$17$ Cyg & 58445695 & 97295 & 5.00 & Cygnus & $296.6$ & $33.7$ &  $1484.4 \pm 36.1$ & $78.6 \pm 1.5$ & C & - & \ref{tab:plato}, \ref{tab:binary}\\   
$35$ Dra & 441813918 & 87234 & 5.02 & Draco & $267.4$ & $77.0$ &  $705.3 \pm 7.0$ & $42.1 \pm 0.1$ & - & 2, 8 & \ref{tab:plato}\\   
$47$ UMa & 21535479 & 53721 & 5.03 & Ursa Major & $164.9$ & $40.4$ &  $2327.9 \pm 55.1$ & $106.1 \pm 0.8$ & A & - & \ref{tab:exo}, \ref{tab:interfer}\\   
$53$ Vir & 308210818 & 64407 & 5.04 & Virgo & $198.0$ & $-16.2$ &  $796.5 \pm 13.3$ & $43.7 \pm 0.6$ & - & - & -\\   
HD $62644$ & 123699670 & 37606 & 5.04 & Puppis & $115.7$ & $-45.2$ &  $708.4 \pm 6.7$ & $41.1 \pm 0.3$ & - & 2, 8 & \ref{tab:plato}, \ref{tab:binary}\\   
$74$ Ori & 437886584 & 29800 & 5.04 & Orion & $94.1$ & $12.3$ &  $2115.3 \pm 156.2$ & $99.6 \pm 1.6$ & B & - & -\\   
$16$ Cep & 366412503 & 108535 & 5.04 & Cepheus & $329.8$ & $73.2$ &  $643.4 \pm 11.2$ & $39.4 \pm 0.6$ & - & - & \ref{tab:plato}, \ref{tab:interfer}\\   
$99$ Her & 22516402 & 88745 & 5.05 & Hercules & $271.8$ & $30.6$ &  $1950.7 \pm 41.0$ & $96.2 \pm 0.6$ & - & - & \ref{tab:plato}, \ref{tab:binary}\\   
$\kappa$ Del & 282254078 & 101916 & 5.07 & Delphinus & $309.8$ & $10.1$ &  $622.5 \pm 9.8$ & $36.2 \pm 0.7$ & - & - & \ref{tab:binary}\\   
$94$ Cet & 49845357 & 14954 & 5.07 & Cetus & $48.2$ & $-1.2$ &  $1267.2 \pm 99.6$ & $69.5 \pm 1.6$ & - & 22 & \ref{tab:exo}, \ref{tab:interfer}, \ref{tab:binary}\\   
$\alpha$ Men & 141810080 & 29271 & 5.08 & Mensa & $92.6$ & $-74.8$ &  $3237.0 \pm 77.1$ & $140.4 \pm 0.1$ & C & 23, 2 & \ref{tab:plato}\\   
HD $33564$ & 142103211 & 25110 & 5.08 & Camelopardalis & $80.6$ & $79.2$ &  $1736.0 \pm 61.1$ & $87.4 \pm 1.4$ & C & - & \ref{tab:exo}, \ref{tab:interfer}\\   
$15$ LMi & 23969522 & 48113 & 5.08 & Ursa Major & $147.1$ & $46.0$ &  $1402.7 \pm 90.1$ & $74.9 \pm 0.5$ & B & - & -\\   
$\phi$$^2$ Pav & 351604689 & 101983 & 5.11 & Pavo & $310.0$ & $-60.5$ &  $919.8 \pm 17.0$ & $51.3 \pm 0.7$ & - & 2, 8 & -\\   
$68$ Eri & 248411315 & 23941 & 5.11 & Eridanus & $77.2$ & $-4.5$ &  $1222.5 \pm 28.5$ & $65.9 \pm 1.7$ & - & - & -\\   
o Aql & 408842743 & 97675 & 5.12 & Aquila & $297.8$ & $10.4$ &  $1831.8 \pm 304.5$ & $89.2 \pm 1.7$ & B & 3 & -\\   
$\mu$ Ara & 362661163 & 86796 & 5.12 & Ara & $266.0$ & $-51.8$ &  $1936.8 \pm 75.6$ & $89.9 \pm 0.8$ & B & 24, 2, 8 & \ref{tab:exo}\\   
$\chi$ Cnc & 302188141 & 40843 & 5.13 & Cancer & $125.0$ & $27.2$ &  $1991.8 \pm 78.8$ & $96.4 \pm 0.7$ & B & - & \ref{tab:interfer}\\   
HD $50223$ & 170225363 & 32765 & 5.14 & Puppis & $102.5$ & $-46.6$ &  $1279.7 \pm 66.0$ & $69.0 \pm 0.9$ & - & 2 & \ref{tab:plato}\\   
HD $136064$ & 232563914 & 74605 & 5.15 & Ursa Minor & $228.7$ & $67.3$ &  $1081.9 \pm 23.8$ & $59.1 \pm 0.2$ & - & - & \ref{tab:plato}\\   
$\rho$ And & 288294358 & 1686 & 5.16 & Andromeda & $5.3$ & $38.0$ &  $390.3 \pm 11.4$ & $27.2 \pm 0.8$ & - & - & \ref{tab:interfer}\\   
$\sigma$ Peg & 318705095 & 112935 & 5.16 & Pegasus & $343.1$ & $9.8$ &  $1039.8 \pm 71.5$ & $58.1 \pm 0.9$ & - & - & -\\   
$31$ Aql & 359981217 & 95447 & 5.17 & Aquila & $291.2$ & $11.9$ &  $1791.8 \pm 225.2$ & $88.7 \pm 0.8$ & C & - & \ref{tab:interfer}\\   
$16$ UMa & 86177169 & 45333 & 5.18 & Ursa Major & $138.6$ & $61.4$ &  $1514.7 \pm 29.8$ & $77.6 \pm 0.6$ & - & - & \ref{tab:binary}\\   
$\kappa$ For & 279342892 & 11072 & 5.19 & Fornax & $35.6$ & $-23.8$ &  $1160.4 \pm 38.1$ & $63.2 \pm 0.4$ & - & 2 & \ref{tab:binary}\\   
HD $176051$ & 20601206 & 93017 & 5.20 & Lyra & $284.3$ & $32.9$ &  $2902.2 \pm 132.1$ & $127.4 \pm 1.7$ & - & - & \ref{tab:plato}, \ref{tab:binary}\\   
$94$ Aqr & 214664574 & 115126 & 5.20 & Aquarius & $349.8$ & $-13.5$ &  $886.9 \pm 12.0$ & $48.4 \pm 0.6$ & - & 25, 2, 8, 3 & \ref{tab:binary}\\   
HD $94388$ & 52160996 & 53252 & 5.23 & Hydra & $163.4$ & $-20.1$ &  $973.0 \pm 29.0$ & $57.4 \pm 1.6$ & - & - & -\\   
$\eta$ Ari & 306484795 & 10306 & 5.23 & Aries & $33.2$ & $21.2$ &  $1098.7 \pm 37.8$ & $56.6 \pm 0.9$ & - & - & -\\   
$26$ Dra & 219777482 & 86036 & 5.23 & Draco & $263.7$ & $61.9$ &  $3059.0 \pm 176.8$ & $132.9 \pm 0.9$ & - & - & \ref{tab:plato}, \ref{tab:binary}\\   
$\theta$ Scl & 70847587 & 950 & 5.24 & Sculptor & $2.9$ & $-35.1$ &  $1871.3 \pm 138.5$ & $91.2 \pm 2.0$ & C & - & -\\   
HD $210855$ & 329759640 & 109572 & 5.24 & Cepheus & $333.0$ & $56.8$ &  $711.4 \pm 13.2$ & $41.2 \pm 0.5$ & - & - & \ref{tab:plato}\\   
$\nu$$^2$ Col & 32500750 & 26460 & 5.28 & Columba & $84.4$ & $-28.7$ &  $691.3 \pm 43.5$ & $38.9 \pm 1.3$ & - & - & \ref{tab:plato}\\   
$\nu$ Ind & 317019578 & 110618 & 5.28 & Indus & $336.2$ & $-72.3$ &  $343.7 \pm 4.2$ & $25.6 \pm 0.4$ & - & 26, 27, 2, 8 & -\\   
$\mu$$^2$ Cnc & 446006789 & 39780 & 5.30 & Cancer & $121.9$ & $21.6$ &  $1075.5 \pm 26.7$ & $59.7 \pm 0.6$ & - & - & -\\   
HD $212330$ & 259291108 & 110649 & 5.31 & Tucana & $336.2$ & $-57.8$ &  $1181.2 \pm 20.6$ & $65.3 \pm 0.3$ & B & 2 & -\\   
$22$ Lyn & 328324648 & 36439 & 5.35 & Lynx & $112.5$ & $49.7$ &  $2167.1 \pm 147.6$ & $102.8 \pm 1.6$ & B & - & -\\   
$171$ Pup & 149672905 & 37853 & 5.36 & Puppis & $116.4$ & $-34.2$ &  $2107.7 \pm 59.8$ & $104.4 \pm 0.9$ & - & - & \ref{tab:plato}\\   
HD $81809$ & 46802551 & 46404 & 5.38 & Hydra & $141.9$ & $-6.1$ &  $691.0 \pm 19.8$ & $38.3 \pm 1.5$ & - & 2, 3 & \ref{tab:binary}\\   
$27$ Cyg & 41195655 & 99031 & 5.38 & Cygnus & $301.6$ & $36.0$ &  $702.8 \pm 44.4$ & $40.6 \pm 1.2$ & - & - & \ref{tab:plato}\\   
$\rho$ Tuc & 234498449 & 3330 & 5.38 & Tucana & $10.6$ & $-65.5$ &  $724.1 \pm 28.7$ & $39.2 \pm 0.9$ & - & - & \ref{tab:binary}\\   
$72$ Her & 9728611 & 84862 & 5.38 & Hercules & $260.2$ & $32.5$ &  $2241.4 \pm 85.1$ & $106.2 \pm 1.0$ & C & - & \ref{tab:plato}, \ref{tab:interfer}\\   
$\rho$ CrB & 458494003 & 78459 & 5.39 & Corona Borealis & $240.3$ & $33.3$ &  $1664.7 \pm 99.9$ & $87.4 \pm 1.7$ & B & 28 & \ref{tab:exo}, \ref{tab:interfer}\\   
CD29 $12513$ & 392326236 & 80399 & 5.40 & Scorpius & $246.2$ & $-29.7$ &  $1051.4 \pm 79.2$ & $58.1 \pm 1.6$ & - & - & -\\   
HD $46588$ & 141523112 & 32439 & 5.44 & Camelopardalis & $101.6$ & $79.6$ &  $2416.6 \pm 128.7$ & $113.9 \pm 1.5$ & B & - & \ref{tab:exo}\\   
$\zeta$ Pic & 219420836 & 24829 & 5.44 & Pictor & $79.8$ & $-50.6$ &  $853.0 \pm 42.4$ & $49.5 \pm 0.4$ & - & 2 & \ref{tab:plato}\\   
$51$ Peg & 139298196 & 113357 & 5.45 & Pegasus & $344.4$ & $20.8$ &  $2485.0 \pm 97.8$ & $114.9 \pm 1.1$ & - & 7 & \ref{tab:exo}, \ref{tab:interfer}\\   
HD $36553$ & 354552931 & 25768 & 5.46 & Pictor & $82.5$ & $-47.1$ &  $544.7 \pm 16.1$ & $33.7 \pm 0.2$ & - & 2, 8 & \ref{tab:plato}\\   
HD $100203$ & 100011351 & 56290 & 5.46 & Ursa Major & $173.1$ & $61.1$ &  $1532.7 \pm 38.3$ & $74.0 \pm 2.1$ & - & - & \ref{tab:binary}\\   
$\iota$ Pav & 303704858 & 89042 & 5.47 & Pavo & $272.6$ & $-62.0$ &  $2056.7 \pm 153.0$ & $101.3 \pm 1.5$ & B & - & -\\   
$\iota$ Crt & 40535161 & 56802 & 5.48 & Crater & $174.7$ & $-13.2$ &  $1282.8 \pm 51.3$ & $67.4 \pm 1.8$ & - & - & -\\   
HD $175225$ & 48194330 & 92549 & 5.51 & Draco & $282.9$ & $53.0$ &  $752.1 \pm 11.3$ & $43.5 \pm 0.3$ & - & - & \ref{tab:plato}\\   
$\iota$ Hyi & 431380163 & 15201 & 5.51 & Hydrus & $49.0$ & $-77.4$ &  $1685.4 \pm 75.6$ & $79.7 \pm 1.7$ & - & - & -\\   
$15$ Peg & 326202925 & 107975 & 5.52 & Pegasus & $328.1$ & $28.8$ &  $1389.9 \pm 70.7$ & $73.0 \pm 1.1$ & - & 2 & \ref{tab:interfer}\\   
HD $156098$ & 152449405 & 84551 & 5.53 & Scorpius & $259.3$ & $-32.7$ &  $631.4 \pm 42.9$ & $34.7 \pm 1.3$ & - & - & -\\   
$14$ Boo & 459250741 & 69536 & 5.53 & Bo{\"o}tes & $213.5$ & $13.0$ &  $929.6 \pm 40.5$ & $51.3 \pm 1.4$ & - & - & -\\   
$\phi$ Gru & 175400332 & 115054 & 5.54 & Grus & $349.5$ & $-40.8$ &  $1330.9 \pm 60.4$ & $65.0 \pm 1.6$ & - & - & -\\   
HD $55575$ & 156890613 & 35136 & 5.54 & Lynx & $109.0$ & $47.2$ &  $1934.9 \pm 111.0$ & $94.6 \pm 1.7$ & B & - & -\\   
HD $104304$ & 152738837 & 58576 & 5.54 & Virgo & $180.2$ & $-10.4$ &  $3224.3 \pm 52.3$ & $136.2 \pm 0.7$ & - & - & \ref{tab:binary}\\   
HD $53705$ & 130645536 & 34065 & 5.56 & Puppis & $106.0$ & $-43.6$ &  $1989.9 \pm 98.4$ & $101.8 \pm 0.7$ & B & - & \ref{tab:plato}, \ref{tab:binary}\\   
HD $46569$ & 255630992 & 31079 & 5.58 & Carina & $97.8$ & $-51.8$ &  $932.2 \pm 20.5$ & $49.7 \pm 0.3$ & - & - & \ref{tab:plato}, \ref{tab:binary}\\   
HD $17948$ & 390733496 & 13665 & 5.59 & Cassiopeia & $44.0$ & $61.5$ &  $1785.4 \pm 51.4$ & $86.0 \pm 1.4$ & - & - & -\\   
HD $65907$ & 372914091 & 38908 & 5.59 & Carina & $119.4$ & $-60.3$ &  $3006.9 \pm 165.7$ & $128.5 \pm 1.2$ & B & - & \ref{tab:plato}\\   
HD $14214$ & 419994887 & 10723 & 5.60 & Cetus & $34.5$ & $1.8$ &  $1667.2 \pm 51.5$ & $83.4 \pm 0.9$ & - & - & \ref{tab:binary}\\   
HD $43318$ & 242250358 & 29716 & 5.62 & Orion & $93.9$ & $-0.5$ &  $863.0 \pm 51.8$ & $50.1 \pm 1.8$ & - & - & -\\   
HD $132254$ & 309765388 & 73100 & 5.63 & Bo{\"o}tes & $224.1$ & $49.6$ &  $1764.3 \pm 80.8$ & $87.8 \pm 2.2$ & - & - & -\\   
HD $133002$ & 288183829 & 72573 & 5.63 & Ursa Minor & $222.6$ & $82.5$ &  $424.7 \pm 15.9$ & $26.0 \pm 0.5$ & - & 8 & -\\   
$10$ Ari & 118247219 & 9621 & 5.64 & Aries & $30.9$ & $25.9$ &  $574.7 \pm 19.2$ & $34.6 \pm 1.2$ & - & 2 & \ref{tab:binary}\\   
$72$ Psc & 384881668 & 5081 & 5.64 & Pisces & $16.3$ & $14.9$ &  $946.2 \pm 77.0$ & $52.0 \pm 1.6$ & - & - & \ref{tab:binary}\\   
$64$ Cet & 337046898 & 10212 & 5.64 & Cetus & $32.8$ & $8.6$ &  $760.9 \pm 24.6$ & $43.0 \pm 0.6$ & - & - & -\\   
HD $159332$ & 462666639 & 85912 & 5.65 & Hercules & $263.3$ & $19.3$ &  $869.8 \pm 28.2$ & $49.2 \pm 0.9$ & - & 2 & -\\   
$\pi$ Men & 261136679 & 26394 & 5.65 & Mensa & $84.3$ & $-80.5$ &  $2621.4 \pm 59.7$ & $116.6 \pm 0.5$ & B & 17, 2 & \ref{tab:exo}\\   
$\nu$$^2$ Lup & 136916387 & 75181 & 5.65 & Lupus & $230.5$ & $-48.3$ &  $2689.9 \pm 138.1$ & $123.0 \pm 0.9$ & B & 29 & \ref{tab:exo}\\   
HD $136359$ & 456061659 & 75308 & 5.65 & Circinus & $230.8$ & $-60.7$ &  $843.2 \pm 37.1$ & $47.1 \pm 1.9$ & - & - & -\\   
HD $195564$ & 205591703 & 101345 & 5.66 & Capricornus & $308.1$ & $-9.9$ &  $1167.3 \pm 46.9$ & $59.3 \pm 1.8$ & - & - & \ref{tab:interfer}\\   
$44$ And & 196692178 & 5493 & 5.67 & Andromeda & $17.6$ & $42.1$ &  $424.3 \pm 24.7$ & $27.4 \pm 0.8$ & - & - & -\\   
$5$ And & 252676979 & 114210 & 5.68 & Andromeda & $346.9$ & $49.3$ &  $1291.0 \pm 35.5$ & $65.9 \pm 1.5$ & - & - & -\\   
$38$ Cet & 248383125 & 5833 & 5.70 & Cetus & $18.7$ & $-1.0$ &  $831.3 \pm 34.9$ & $45.6 \pm 1.4$ & - & - & -\\   
HD $142$ & 389757979 & 522 & 5.70 & Phoenix & $1.6$ & $-49.1$ &  $1890.9 \pm 163.6$ & $91.1 \pm 2.0$ & - & - & \ref{tab:exo}\\   
$68$ Dra & 236871353 & 99500 & 5.70 & Draco & $302.9$ & $62.1$ &  $716.2 \pm 14.7$ & $40.2 \pm 0.6$ & - & - & \ref{tab:plato}\\   
HD $184960$ & 26884478 & 96258 & 5.71 & Cygnus & $293.6$ & $51.2$ &  $1870.6 \pm 73.9$ & $91.9 \pm 0.9$ & - & - & \ref{tab:plato}, \ref{tab:exo}\\   
HD $214850$ & 379660147 & 111974 & 5.72 & Pegasus & $340.2$ & $14.5$ &  $637.9 \pm 14.3$ & $39.4 \pm 1.5$ & - & - & \ref{tab:binary}\\   
HD $190360$ & 105999792 & 98767 & 5.73 & Cygnus & $300.9$ & $29.9$ &  $2358.2 \pm 74.1$ & $112.2 \pm 0.7$ & B & - & \ref{tab:plato}, \ref{tab:exo}, \ref{tab:interfer}\\   
HD $89744$ & 8154501 & 50786 & 5.73 & Ursa Major & $155.5$ & $41.2$ &  $1028.4 \pm 108.5$ & $55.2 \pm 1.6$ & - & - & \ref{tab:exo}, \ref{tab:interfer}\\   
$84$ Her & 460022840 & 86731 & 5.73 & Hercules & $265.8$ & $24.3$ &  $631.4 \pm 16.0$ & $37.6 \pm 1.5$ & - & - & -\\   
$37$ Gem & 80226651 & 33277 & 5.74 & Gemini & $103.8$ & $25.4$ &  $2436.1 \pm 418.7$ & $109.8 \pm 1.3$ & B & - & -\\   
HD $9562$ & 29845542 & 7276 & 5.75 & Cetus & $23.4$ & $-7.0$ &  $1030.5 \pm 36.2$ & $58.8 \pm 0.9$ & - & - & -\\   
HD $199623$ & 79015105 & 103673 & 5.76 & Indus & $315.1$ & $-51.3$ &  $1631.1 \pm 59.9$ & $81.1 \pm 1.6$ & - & - & -\\   
$59$ Eri & 118034753 & 22325 & 5.76 & Eridanus & $72.1$ & $-16.3$ &  $781.6 \pm 24.0$ & $46.6 \pm 1.7$ & - & - & -\\   
HD $109409$ & 72814920 & 61379 & 5.76 & Centaurus & $188.7$ & $-44.7$ &  $1172.3 \pm 23.9$ & $61.5 \pm 0.3$ & - & 2, 8 & -\\   
$22$ UMa & 358013866 & 47013 & 5.77 & Ursa Major & $143.7$ & $72.2$ &  $744.5 \pm 26.3$ & $41.0 \pm 0.9$ & - & - & -\\   
HD $30562$ & 176379354 & 22336 & 5.77 & Eridanus & $72.2$ & $-5.7$ &  $1447.6 \pm 216.6$ & $74.1 \pm 1.0$ & - & 2 & \ref{tab:exo}\\   
HD $49933$ & 281812116 & 32851 & 5.78 & Monoceros & $102.7$ & $-0.5$ &  $1997.7 \pm 117.8$ & $92.5 \pm 1.5$ & - & 30, 31 & \ref{tab:interfer}\\   
$\lambda$$^2$ For & 122555698 & 12186 & 5.78 & Fornax & $39.2$ & $-34.6$ &  $1374.3 \pm 28.0$ & $69.7 \pm 0.7$ & - & 32, 2 & \ref{tab:exo}\\   
HD $11007$ & 20931913 & 8433 & 5.78 & Triangulum & $27.2$ & $32.7$ &  $1239.7 \pm 75.9$ & $67.0 \pm 0.7$ & - & - & -\\   
HD $89569$ & 463794734 & 50493 & 5.80 & Vela & $154.7$ & $-56.1$ &  $1143.8 \pm 32.2$ & $61.1 \pm 1.8$ & - & - & -\\   
HD $76932$ & 348837162 & 44075 & 5.80 & Hydra & $134.7$ & $-16.1$ &  $1548.6 \pm 56.4$ & $86.0 \pm 0.8$ & - & 2 & -\\   
HD $191195$ & 405902259 & 99026 & 5.81 & Cygnus & $301.6$ & $53.2$ &  $1353.0 \pm 146.7$ & $69.2 \pm 2.3$ & - & - & \ref{tab:plato}\\   
$39$ Leo & 95360514 & 50384 & 5.81 & Leo & $154.3$ & $23.1$ &  $2027.5 \pm 100.2$ & $104.4 \pm 1.4$ & - & - & -\\   
HD $221420$ & 277890728 & 116250 & 5.82 & Octans & $353.3$ & $-77.4$ &  $1043.0 \pm 17.5$ & $56.1 \pm 0.2$ & - & 2, 8 & \ref{tab:exo}\\   
$\iota$$^2$ For & 122556844 & 12288 & 5.84 & Fornax & $39.6$ & $-30.2$ &  $1227.8 \pm 37.5$ & $63.2 \pm 1.5$ & - & - & -\\   
HD $103026$ & 141969034 & 57841 & 5.85 & Hydra & $177.9$ & $-30.8$ &  $1189.9 \pm 86.2$ & $66.1 \pm 1.2$ & - & - & -\\   
$\epsilon$ For & 88405751 & 14086 & 5.88 & Fornax & $45.4$ & $-28.1$ &  $402.1 \pm 7.2$ & $27.8 \pm 0.4$ & - & 2, 8 & -\\   
HD $45067$ & 42334982 & 30545 & 5.88 & Monoceros & $96.3$ & $-0.9$ &  $1117.1 \pm 116.1$ & $59.9 \pm 0.9$ & - & 2 & -\\   
HD $3823$ & 281667171 & 3170 & 5.89 & Tucana & $10.1$ & $-59.5$ &  $1593.8 \pm 44.7$ & $82.6 \pm 0.6$ & - & 2 & -\\   
$13$ Tri & 286176467 & 11548 & 5.89 & Triangulum & $37.2$ & $29.9$ &  $1049.1 \pm 38.7$ & $57.8 \pm 0.9$ & - & - & -\\   
HD $112164$ & 248127641 & 63033 & 5.89 & Centaurus & $193.7$ & $-44.2$ &  $897.2 \pm 23.6$ & $50.6 \pm 0.9$ & - & 2 & -\\   
$112$ Psc & 422679659 & 9353 & 5.89 & Pisces & $30.0$ & $3.1$ &  $1248.9 \pm 86.8$ & $67.4 \pm 2.1$ & - & - & -\\   
HD $59984$ & 6677245 & 36640 & 5.90 & Monoceros & $113.0$ & $-8.9$ &  $1139.0 \pm 35.5$ & $59.3 \pm 1.0$ & - & 2 & -\\   
HD $193664$ & 403585118 & 100017 & 5.91 & Draco & $304.4$ & $66.9$ &  $3174.6 \pm 201.7$ & $138.5 \pm 2.8$ & C & - & \ref{tab:plato}\\   
HD $57006$ & 453144504 & 35509 & 5.91 & Canis Minor & $109.9$ & $7.1$ &  $591.6 \pm 29.5$ & $35.0 \pm 1.2$ & - & - & -\\   
HD $52711$ & 91686663 & 34017 & 5.93 & Gemini & $105.9$ & $29.3$ &  $2583.3 \pm 99.1$ & $122.3 \pm 0.9$ & - & - & -\\   
$58$ UMa & 56939464 & 56148 & 5.94 & Ursa Major & $172.6$ & $43.2$ &  $586.7 \pm 29.4$ & $34.8 \pm 1.3$ & - & - & -\\   
$35$ Leo & 95360265 & 50319 & 5.95 & Leo & $154.1$ & $23.5$ &  $985.2 \pm 55.0$ & $54.3 \pm 0.9$ & - & - & \ref{tab:binary}\\   
HD $38529$ & 200093173 & 27253 & 5.95 & Orion & $86.6$ & $1.2$ &  $622.5 \pm 27.9$ & $35.6 \pm 0.9$ & - & 33, 2 & \ref{tab:exo}, \ref{tab:interfer}\\   
HD $21722$ & 31799975 & 15968 & 5.96 & Hydrus & $51.4$ & $-69.3$ &  $1618.1 \pm 175.1$ & $78.6 \pm 2.0$ & - & - & \ref{tab:plato}\\   
HD $33093$ & 169396790 & 23831 & 5.97 & Lepus & $76.9$ & $-12.5$ &  $856.0 \pm 48.8$ & $49.1 \pm 1.5$ & - & 2 & -\\   
$21$ Eri & 301558151 & 17027 & 5.97 & Eridanus & $54.8$ & $-5.6$ &  $460.3 \pm 5.6$ & $27.3 \pm 0.9$ & - & 2, 8 & -\\   
HD $29645$ & 156157340 & 21847 & 5.97 & Perseus & $70.5$ & $38.3$ &  $1392.7 \pm 44.6$ & $69.2 \pm 1.0$ & - & - & -\\   
HD $18262$ & 387541497 & 13679 & 5.97 & Cetus & $44.1$ & $8.4$ &  $1184.6 \pm 42.3$ & $62.4 \pm 1.7$ & - & - & -\\   
HD $98560$ & 467957174 & 55280 & 5.99 & Carina & $169.8$ & $-64.6$ &  $1260.6 \pm 167.7$ & $66.1 \pm 1.6$ & - & - & -\\   
$16$ Cyg A & 27533341 & 96895 & 5.99 & Cygnus & $295.5$ & $50.5$ &  $2236.5 \pm 97.9$ & $103.5 \pm 1.1$ & - & 34 & \ref{tab:plato}, \ref{tab:interfer}, \ref{tab:binary}\\   
HD $152303$ & 233503400 & 81854 & 5.99 & Ursa Minor & $250.8$ & $77.5$ &  $1689.5 \pm 49.6$ & $82.2 \pm 1.5$ & - & - & \ref{tab:plato}\\   
HD $121384$ & 208789160 & 68101 & 6.00 & Centaurus & $209.1$ & $-54.7$ &  $425.6 \pm 8.2$ & $28.6 \pm 0.6$ & - & 2, 8 & \ref{tab:binary}   
\end{longtable}
\tablefoot{\tiny Table~\ref{tab:all_seis} provides the measured global asteroseismic parameters for the sample, with stars sorted according to their visual magnitude (``$V$"). The first three columns provide identifiers for the stars in the form of their Bayer/Flamsteed designation (or primary name according to \texttt{SIMBAD}) in addition to their TESS (``TIC") and Hipparcos (``HIP") IDs. The ``HWO" column indicates if the star is found in the HWO target list and provides    the tier (A, B, or C). The last two columns provide references to the primary sources of previous seismic detections if they exits (in all cases we have indicated if a given star is contained in the \citet{Hatt2023}, \citet{Zhou2024}, and/or \citet{Corsaro2024} samples), and if a given star is listed in one of the other tables in the paper on ``PLATO" (\tref{tab:plato}), ``Exoplanets" (\tref{tab:exo}),    ``Interferometry" (\tref{tab:interfer}), or ``Binarity" (\tref{tab:binary}).}
\tablebib{\tiny (1) \citet{Kjeldsen1995_etaboo}; (2) \citet{Hatt2023}; (3) \citet{Corsaro2024}; (4) \citet{Martic2001}; (5) \citet{Bedding2001}; (6) \citet{Carrier2001}; (7) \citet{Metcalfe2024}; (8) \citet{Zhou2024}; (9) \citet{Bonanno2008}; (10) \citet{Grundahl2017}; (11) \citet{Carrier2006}; (12) \citet{Kjeldsen2005}; (13) \citet{Carrier2005}; (14) \citet{Kjeldsen2008}; (15) \citet{Kjeldsen2025}; (16) \citet{Mosser2008}; (17) \citet{Huber2022}; (18) \citet{Metcalfe2023}; (19) \citet{Guzik2016}; (20) \citet{Hon2024}; (21) \citet{Ball2022}; (22) \citet{Deal2017}; (23) \citet{Chontos2021}; (24) \citet{Bouchy2005}; (25) \citet{Metcalfe2020}; (26) \citet{Bedding2006}; (27) \citet{Carrier2007}; (28) \citet{Metcalfe2021}; (29) \citet{Delrez2021}; (30) \citet{Mosser2005}; (31) \citet{Appourchaux2008}; (32) \citet{Nielsen2020}; (33) \citet{Ball2020}; (34) \citet{Metcalfe2012}.} 
\endgroup}

\setlength\tabcolsep{0pt} 
\begin{table*} 
\centering 
\caption{TLS stars in the PLATO fields} 
\label{tab:plato}
\scriptsize 
\begin{tabular*}{\linewidth}{@{\extracolsep{\fill}}lccccccc@{}} 
\toprule 
Name & TIC & HIP & RA & DEC & $V$ & Sep & SpT \\ 
  &    &   & (deg)  & (deg) & (mag) & (deg) & \\ 
\midrule [0.3ex] \\ 
 \multicolumn{8}{c}{Inside PLATO LOPS 2} \\ 
\midrule 
HR $3220$ & 308844962 & 39903 & 122.253 & -61.302 & 4.74 & 20.3 & F6VFe-0.8CH-0.4 \\  [0.1ex] 
HD $62644$ & 123699670 & 37606 & 115.738 & -45.173 & 5.04 & 14.3 & G8IV-V \\  [0.1ex] 
HD $50223$ & 170225363 & 32765 & 102.478 & -46.615 & 5.14 & 5.0 & F5.5V \\  [0.1ex] 
$\nu$$^2$ Col & 32500750 & 26460 & 84.436 & -28.690 & 5.28 & 21.0 & F5V \\  [0.1ex] 
$171$ Pup & 149672905 & 37853 & 116.396 & -34.173 & 5.36 & 20.9 & F9V \\  [0.1ex] 
$\zeta$ Pic & 219420836 & 24829 & 79.842 & -50.606 & 5.44 & 10.4 & F6IV \\  [0.1ex] 
HD $36553$ & 354552931 & 25768 & 82.539 & -47.078 & 5.46 & 8.7 & F8/G2 \\  [0.1ex] 
HD $53705$ & 130645536 & 34065 & 105.989 & -43.608 & 5.56 & 8.6 & G1.5V \\  [0.1ex] 
HD $46569$ & 255630992 & 31079 & 97.826 & -51.826 & 5.58 & 4.3 & F8VFe-0.4 \\  [0.1ex] 
HD $65907$ & 372914091 & 38908 & 119.445 & -60.303 & 5.59 & 18.6 & F9.5V \\  [0.1ex] 
\midrule 
\multicolumn{8}{c}{Near PLATO LOPS 2} \\ 
\midrule 
$\gamma$ Lep & 93280676 & 27072 & 86.116 & -22.448 & 3.59 & 26.5 & F6V \\  [0.1ex] 
HD $60532$ & 456871289 & 36795 & 113.513 & -22.296 & 4.44 & 29.4 & F6IV-V \\  [0.1ex] 
$\kappa$ Ret & 262843771 & 16245 & 52.344 & -62.938 & 4.71 & 27.9 & F3IV/V \\  [0.1ex] 
$\alpha$ Men & 141810080 & 29271 & 92.560 & -74.753 & 5.08 & 26.9 & G7V \\  [0.1ex] 
HD $21722$ & 31799975 & 15968 & 51.401 & -69.336 & 5.96 & 30.2 & F5V \\  [0.1ex] 
\midrule \\ [0.3ex]
\multicolumn{8}{c}{Inside PLATO LOPN 1} \\ 
\midrule 
$\chi$ Dra & 341873045 & 89937 & 275.264 & 72.733 & 3.55 & 19.9 & F7V \\  [0.1ex] 
$\theta$ Dra & 161825882 & 78527 & 240.472 & 58.565 & 4.01 & 21.2 & F9V \\  [0.1ex] 
$\theta$ Cyg & 27014182 & 96441 & 294.111 & 50.221 & 4.49 & 10.8 & F3+V \\  [0.1ex] 
$\upsilon$ Cep & 421444084 & 102431 & 311.338 & 57.580 & 4.52 & 19.8 & F8IV-V+F9IV-V \\  [0.1ex] 
$\psi$$^1$ Dra A & 441804568 & 86614 & 265.484 & 72.149 & 4.57 & 19.9 & F5IV-V \\  [0.1ex] 
$\sigma$ Dra & 259237827 & 96100 & 293.090 & 69.661 & 4.67 & 18.3 & K0V \\  [0.1ex] 
$\omega$ Dra & 233195546 & 86201 & 264.238 & 68.758 & 4.77 & 17.0 & F4V \\  [0.1ex] 
$19$ Dra & 289622310 & 82860 & 254.007 & 65.135 & 4.88 & 16.9 & F8V \\  [0.1ex] 
$36$ Dra & 233121747 & 89348 & 273.474 & 64.397 & 4.99 & 11.7 & F5V \\  [0.1ex] 
$17$ Cyg & 58445695 & 97295 & 296.607 & 33.728 & 5.00 & 23.6 & F5.5IV-V \\  [0.1ex] 
$35$ Dra & 441813918 & 87234 & 267.363 & 76.963 & 5.02 & 24.4 & F6IV-Vs \\  [0.1ex] 
$99$ Her & 22516402 & 88745 & 271.757 & 30.562 & 5.05 & 22.6 & F6V \\  [0.1ex] 
HD $136064$ & 232563914 & 74605 & 228.660 & 67.347 & 5.15 & 27.2 & F8V \\  [0.1ex] 
HD $176051$ & 20601206 & 93017 & 284.257 & 32.901 & 5.20 & 20.6 & F9V+K1V \\  [0.1ex] 
$26$ Dra & 219777482 & 86036 & 263.748 & 61.875 & 5.23 & 11.5 & G0IV-V \\  [0.1ex] 
$27$ Cyg & 41195655 & 99031 & 301.591 & 35.972 & 5.38 & 24.0 & G8.5IVa \\  [0.1ex] 
$72$ Her & 9728611 & 84862 & 260.165 & 32.468 & 5.38 & 23.8 & G0V \\  [0.1ex] 
HD $175225$ & 48194330 & 92549 & 282.896 & 52.975 & 5.51 & 3.4 & G9IVa \\  [0.1ex] 
$68$ Dra & 236871353 & 99500 & 302.895 & 62.079 & 5.70 & 16.4 & F5V \\  [0.1ex] 
HD $184960$ & 26884478 & 96258 & 293.582 & 51.237 & 5.71 & 10.2 & F7V \\  [0.1ex] 
HD $191195$ & 405902259 & 99026 & 301.558 & 53.166 & 5.81 & 14.6 & F5V \\  [0.1ex] 
HD $193664$ & 403585118 & 100017 & 304.381 & 66.854 & 5.91 & 19.3 & G3V \\  [0.1ex] 
$16$ Cyg A & 27533341 & 96895 & 295.454 & 50.525 & 5.99 & 11.5 & G1.5Vb \\  [0.1ex] 
HD $152303$ & 233503400 & 81854 & 250.776 & 77.514 & 5.99 & 26.5 & F4V \\  [0.1ex] 
\midrule 
\multicolumn{8}{c}{Near PLATO LOPN 1} \\ 
\midrule 
$\zeta$ Her & 43255143 & 81693 & 250.322 & 31.603 & 2.81 & 28.8 & G0IV \\  [0.1ex] 
$\mu$$^1$ Her & 460067868 & 86974 & 266.615 & 27.721 & 3.42 & 26.3 & G5IV \\  [0.1ex] 
$\chi$ Her & 157364190 & 77760 & 238.169 & 42.452 & 4.60 & 27.8 & F8VFe-2Hdel-1 \\  [0.1ex] 
$16$ Cep & 366412503 & 108535 & 329.812 & 73.180 & 5.04 & 29.6 & F5V \\  [0.1ex] 
HD $210855$ & 329759640 & 109572 & 332.953 & 56.839 & 5.24 & 31.5 & F8V \\  [0.1ex] 
HD $190360$ & 105999792 & 98767 & 300.906 & 29.897 & 5.73 & 28.8 & G7IV-V \\  [0.1ex] 
\bottomrule
\end{tabular*} 
\tablefoot{\small The table provides the stars identified within or near (within 5 deg of the field edge) the PLATO northern         (LOPN1) and southern (LOPS2) long stare fields, with stars within each grouping sorted according to their visual magnitude (``$V$").     The first three columns provide identifiers for the stars in the form of their Bayer/Flamsteed designation (or primary     name according to \texttt{SIMBAD}) in addition to their TESS (``TIC") and Hipparcos (``HIP") IDs. ``Sep" gives the separation in degrees between the         stars and the fields centres, while ``SpT gives the spectral type of the stars according to \texttt{Simbad}.} 
\end{table*}

\setlength\tabcolsep{0pt} 
\begin{table*} 
\centering 
\caption{TLS exoplanets and sub-stellar objects} 
\label{tab:exo}
\scriptsize 
\renewcommand{\arraystretch}{1}  
\begin{tabular*}{\linewidth}{@{\extracolsep{\fill}}llccccccc@{}} 
\toprule 
Name  & HIP & $V$ & $\nu_{\rm max}$ & Known Seismic & Binary & HWO & Planets &  Reference\\  &      &   (mag) & ($\mu$Hz) & (X/O) & (X/O) & (tier) &(No.) & (discovery)\\ 
\midrule 
$\upsilon$ And &  7513 &  4.10 & 1528 & O & X\tablefootmark{\textdagger} & C & 3  &  {1, 2}\\   
$82$ Eri &  15510 &  4.26 & 3180 & O & O & B & 4  &  {3, 4, 5}\\   
$\lambda$ Ser &  77257 &  4.42 & 1856 & X & O & A & 1  &  {6}\\   
HD $60532$$^{(PS)}$ &  36795 &  4.44 & 618 & O & O & - & 2  &  {7}\\   
$61$ Vir &  64924 &  4.74 & 3099 & O & O & B & 3  &  {8}\\   
HD $102365$ &  57443 &  4.89 & 3148 & X & X\tablefootmark{(a)} & A & 1  &  {9}\\   
$70$ Vir &  65721 &  4.97 & 940 & X & O & - & 1  &  {10}\\   
$47$ UMa &  53721 &  5.03 & 2327 & O & O & A & 3  &  {11, 12, 13}\\   
$94$ Cet &  14954 &  5.07 & 1267 & X & X$^*$\tablefootmark{\textdagger} & - & 1  &  {14}\\   
HD $33564$ &  25110 &  5.08 & 1736 & O & O & C & 1  &  {15}\\   
$\mu$ Ara &  86796 &  5.12 & 1936 & X & O & B & 4  &  {16, 17, 18, 19}\\   
$\rho$ CrB &  78459 &  5.39 & 1664 & X & O\tablefootmark{\textdagger} & B & 4  &  {20, 21, 22}\\   
HD $46588$ &  32439 &  5.44 & 2416 & O & X\tablefootmark{(b)} & B & 1  &  {23}\\   
$51$ Peg &  113357 &  5.45 & 2485 & X & O & - & 1  &  {24}\\   
$\nu$$^2$ Lup &  75181 &  5.65 & 2689 & X & O & B & 3  &  {25}\\   
$\pi$ Men &  26394 &  5.65 & 2621 & X & O & B & 3  &  {26, 27, 28}\\   
HD $142$ &  522 &  5.70 & 1890 & O & X\tablefootmark{(c)} & - & 3  &  {29, 30, 31}\\   
HD $184960$$^{PN}$ &  96258 &  5.71 & 1870 & O & O & - & 1  &  {32}\\   
HD $190360$$^{(PN)}$ &  98767 &  5.73 & 2358 & O & X\tablefootmark{(d)} & B & 2\tablefootmark{(e)}  &  {33, 34}\\   
HD $89744$ &  50786 &  5.73 & 1028 & O & X\tablefootmark{(f)} & - & 1\tablefootmark{(g)}  &  {35}\\   
HD $30562$ &  22336 &  5.77 & 1447 & X & O & - & 1  &  {36}\\   
$\lambda$$^2$ For &  12186 &  5.78 & 1374 & X & X\tablefootmark{(h)} & - & 1  &  {37}\\   
HD $221420$ &  116250 &  5.82 & 1042 & O & X\tablefootmark{(i)} & - & 1  &  {38}\\   
HD $38529$ &  27253 &  5.95 & 622 & X & X\tablefootmark{(j)} & - & 2\tablefootmark{(k)}  &  {39, 40}\\   
\bottomrule
\end{tabular*} 
\tablefoot{\tiny The table provides an identification of the exoplanets and their hosts for the seismic stars in the sample. The stars are                 ordered according to their $V$-band brightness. The first two columns provide identifiers for the stars in the form of their Bayer/Flamsteed designation (or primary name according to \texttt{SIMBAD}) and their                 ```HIP'' IDs. Super-scripts of $PN$ or $PS$ on the star's name refer to their                     position within (or near when in parenthesis) the northern ($N$) or southern ($S$) PLATO fields (see Table~\ref{tab:plato}). ``$V$'' gives the $V$-band magnitude, ``$\nu_{\rm max}$'' (central value only) is adopted from Table~\ref{tab:all_seis}.                ``Known Seismic'' denotes with an X (otherwise an O) if the star is a know asteroseismic target (see \tref{tab:all_seis}); ``Binary'' denotes with an X (otherwise an O) if the star is                    identified as being part of a binary or multiple star system, and here an asterics (*) indicates that orbital information is available in                   \tref{tab:binary}; ``HWO'' indicates if the star is part of the \citet{Mamajek2024} HWO target list, and if so provides its tier; ``Planets'' gives the number of identified exoplanets; ``Reference'' provides the reference(s) for the identified exoplanet(s).\\ 
\tablefoottext{$\dagger$}{see \aref{app:bin_indv};}\tablefoottext{a}{A faint M4 dwarf companion star (GJ 442 B) at a distance of ${\sim}212$ AU \citep{Raghavan2010};}\tablefoottext{b}{\citet{Loutrel2011} identified a brown dwarf companion (HD 46588 B; spT L9), one of the few known brown dwarfs at the L/T transition for which both age and distance estimates are available;}\tablefoottext{c}{A low-mass K-M dwarf companion star (LHS 1021 / HD 142 B) \citep{Eggenberger2007,Mugrauer2019};}\tablefoottext{d}{A faint M4.5V dwarf companion star (G 125-55 / GJ 777 B) at a distance of ${\sim}2849$ AU, which in turn could be an unresolved binary with a similar companion \citep{2014WDS};}\tablefoottext{e}{A potential third planet was identified by \citet{Hirsch2021};}\tablefoottext{f}{A wide-separation ($a{\sim}2460$ AU) L-type companion \citep{Wilson2001,Mugrauer2004}, and another ($a{\sim}219$ AU) proposed        candidate companion \citep{Roberts2011_94Cet,Fontanive2019};}\tablefoottext{g}{\citet{Wittenmyer2019} finds indications of another Jupiter-mass exoplanet (HD 89744 c) at a separation of $8.3\pm1.8$ AU and with an orbital period of $6974\pm2161$ days;}\tablefoottext{h}{A co-moving mid-M-dwarf companion (HD 16417 B) detected by \citet{Mugrauer2004};}\tablefoottext{i}{A wide-separation ($a{\sim}21,756$ AU) companion candidate (likely a mid-M-dwarf) was identified by \citet{Venner2021} \citep[see also][]{ElBadry2021};}\tablefoottext{j}{A wide-separation ($a{\sim}11,000$ AU) M2.5V companion \citep{Raghavan2006,Montes2018};}\tablefoottext{k}{\citet{Benedict2010} finds indications in the RV residuals of a possible third planetary companion at a period of ${\sim}194$ days and an inferred $a{\sim}0.75$ AU.}} 
\tablebib{\tiny (1) \citet{Butler1997}; (2) \citet{Butler1999}; (3) \citet{Pepe2011}; (4) \citet{Feng2017}; (5) \citet{Nari2025}; (6) \citet{Rosenthal2021}; (7) \citet{Desort2008}; (8) \citet{Vogt2010}; (9) \citet{Tinney2011}; (10) \citet{Marcy1996}; (11) \citet{Butler1996}; (12) \citet{Fischer2002}; (13) \citet{Gregory2010}; (14) \citet{Mayor2004}; (15) \citet{Galland2005}; (16) \citet{Butler2001}; (17) \citet{McCarthy2004}; (18) \citet{Santos2004}; (19) \citet{Pepe2007}; (20) \citet{Noyes1997}; (21) \citet{Fulton2016}; (22) \citet{Brewer2023}; (23) \citet{Subjak2023}; (24) \citet{Mayor1995}; (25) \citet{Udry2019}; (26) \citet{Gandolfi2018}; (27) \citet{Jones2002}; (28) \citet{Hatzes2022}; (29) \citet{Feng2022}; (30) \citet{Tinney2002}; (31) \citet{Wittenmyer2012}; (32) \citet{Barnes2023}; (33) \citet{Naef2003}; (34) \citet{Vogt2005}; (35) \citet{Korzennik2000}; (36) \citet{Fischer2009}; (37) \citet{OToole2009}; (38) \citet{Kane2019}; (39) \citet{Fischer2001}; (40) \citet{Fischer2003}. } 
\end{table*}

\setlength\tabcolsep{0pt} 
\begin{table*} 
\centering 
\caption{TLS long-baseline interferometry overview} 
\label{tab:interfer}
\scriptsize 
\renewcommand{\arraystretch}{0.8}  
\begin{tabular*}{\linewidth}{@{\extracolsep{\fill}}lllccccc@{}} 
\toprule 
Name & TIC & HIP & $V$ &  RA & DEC & $\nu_{\rm max}$ & References \\ 
  &      &  & (mag) & (deg) & (deg) & ($\mu$Hz) &  \\  
\midrule 
$\eta$ Boo & 367758676 & 67927 & 2.68 & $208.7$ & $18.4$ &  $697.9 \pm 17.8$ & 1, 2, 3, 4 \\   
$\zeta$ Her & 43255143 & 81693 & 2.81 & $250.3$ & $31.6$ &  $718.5 \pm 24.4$ & 1, 2, 4 \\   
$\beta$ Hyi & 267211065 & 2021 & 2.82 & $6.4$ & $-77.3$ &  $1038.1 \pm 14.9$ & 5 \\   
$\theta$ UMa & 150226696 & 46853 & 3.17 & $143.2$ & $51.7$ &  $779.3 \pm 17.8$ & 6, 7 \\   
$\xi$ Gem & 372480991 & 32362 & 3.35 & $101.3$ & $12.9$ &  $2871.0 \pm 111.3$ & 6 \\   
$\mu$$^1$ Her & 460067868 & 86974 & 3.42 & $266.6$ & $27.7$ &  $1192.5 \pm 17.5$ & 2, 4, 7 \\   
$\eta$ Cas & 445258206 & 3821 & 3.46 & $12.3$ & $57.8$ &  $2840.0 \pm 80.7$ & 6, 8 \\   
$\delta$ Eri & 38511251 & 17378 & 3.52 & $55.8$ & $-9.8$ &  $677.6 \pm 8.3$ & 3, 9 \\   
$\delta$ Pav & 409891396 & 99240 & 3.55 & $302.2$ & $-66.2$ &  $2269.8 \pm 64.4$ & 9 \\   
$\beta$ Vir & 366661076 & 57757 & 3.59 & $177.7$ & $1.8$ &  $1446.3 \pm 75.9$ & 6, 8, 10 \\   
$\gamma$ Lep & 93280676 & 27072 & 3.59 & $86.1$ & $-22.4$ &  $2257.5 \pm 50.9$ & 8 \\   
$\beta$ Aql & 375621179 & 98036 & 3.71 & $298.8$ & $6.4$ &  $414.7 \pm 6.8$ & 11, 4, 9 \\   
$\iota$ Peg & 357336603 & 109176 & 3.77 & $331.8$ & $25.3$ &  $2101.8 \pm 126.6$ & 8 \\   
$\gamma$ Ser & 377415363 & 78072 & 3.85 & $239.1$ & $15.7$ &  $1744.6 \pm 38.2$ & 6, 8 \\   
$\theta$ Boo & 441709021 & 70497 & 4.04 & $216.3$ & $51.9$ &  $1354.0 \pm 108.4$ & 6, 10 \\   
$\iota$ Per & 116988032 & 14632 & 4.05 & $47.3$ & $49.6$ &  $1855.3 \pm 33.9$ & 6, 8, 12 \\   
$\iota$ Vir & 6029884 & 69701 & 4.07 & $214.0$ & $-6.0$ &  $644.8 \pm 24.0$ & 10 \\   
$\upsilon$ And & 189576919 & 7513 & 4.10 & $24.2$ & $41.4$ &  $1528.0 \pm 50.6$ & 13, 14, 15, 8, 16, 12 \\   
$\theta$ Per & 302158903 & 12777 & 4.10 & $41.0$ & $49.2$ &  $2314.2 \pm 166.2$ & 6, 8 \\   
$\iota$ Psc & 419919445 & 116771 & 4.13 & $355.0$ & $5.6$ &  $1416.4 \pm 53.6$ & 6, 8 \\   
$110$ Her & 282038438 & 92043 & 4.19 & $281.4$ & $20.5$ &  $1061.9 \pm 28.0$ & 6, 10 \\   
$\xi$ Peg & 60716322 & 112447 & 4.20 & $341.7$ & $12.2$ &  $986.8 \pm 16.7$ & 6 \\   
$\beta$ CVn & 458445966 & 61317 & 4.24 & $188.4$ & $41.4$ &  $2385.4 \pm 136.0$ & 6, 7 \\   
$10$ Tau & 311092847 & 16852 & 4.29 & $54.2$ & $0.4$ &  $1284.1 \pm 63.1$ & 6, 8 \\   
$\lambda$ Ser & 296740796 & 77257 & 4.42 & $236.6$ & $7.4$ &  $1856.6 \pm 46.4$ & 10 \\   
o$^2$ Eri & 67772871 & 19849 & 4.43 & $63.8$ & $-7.7$ &  $3433.1 \pm 368.6$ & 17, 18, 9 \\   
$\theta$ Cyg & 27014182 & 96441 & 4.49 & $294.1$ & $50.2$ &  $1759.1 \pm 67.1$ & 6, 15, 19, 16, 20 \\   
$\psi$$^1$ Dra A & 441804568 & 86614 & 4.57 & $265.5$ & $72.1$ &  $1232.4 \pm 19.8$ & 6 \\   
$\chi$ Her & 157364190 & 77760 & 4.60 & $238.2$ & $42.5$ &  $1045.6 \pm 17.7$ & 10 \\   
$\sigma$ Dra & 259237827 & 96100 & 4.67 & $293.1$ & $69.7$ &  $4217.9 \pm 122.6$ & 21, 6 \\   
$\lambda$ Aur & 409104974 & 24813 & 4.69 & $79.8$ & $40.1$ &  $2152.0 \pm 54.2$ & 6, 8 \\   
$61$ Vir & 422478973 & 64924 & 4.74 & $199.6$ & $-18.3$ &  $3099.8 \pm 101.0$ & 22, 8 \\   
$40$ Leo & 95431211 & 50564 & 4.78 & $154.9$ & $19.5$ &  $1405.8 \pm 57.9$ & 23 \\   
HD $5015$ & 285544488 & 4151 & 4.80 & $13.3$ & $61.1$ &  $1399.3 \pm 52.9$ & 6 \\   
$36$ UMa & 416519065 & 51459 & 4.82 & $157.7$ & $56.0$ &  $2319.5 \pm 96.8$ & 6 \\   
$70$ Vir & 95473936 & 65721 & 4.97 & $202.1$ & $13.8$ &  $940.6 \pm 12.9$ & 13, 8, 24 \\   
$36$ Dra & 233121747 & 89348 & 4.99 & $273.5$ & $64.4$ &  $1312.0 \pm 16.9$ & 25, 16 \\   
$47$ UMa & 21535479 & 53721 & 5.03 & $164.9$ & $40.4$ &  $2327.9 \pm 55.1$ & 8 \\   
$16$ Cep & 366412503 & 108535 & 5.04 & $329.8$ & $73.2$ &  $643.4 \pm 11.2$ & 16 \\   
$94$ Cet & 49845357 & 14954 & 5.07 & $48.2$ & $-1.2$ &  $1267.2 \pm 99.6$ & 13, 16 \\   
HD $33564$ & 142103211 & 25110 & 5.08 & $80.6$ & $79.2$ &  $1736.0 \pm 61.1$ & 22 \\   
$\chi$ Cnc & 302188141 & 40843 & 5.13 & $125.0$ & $27.2$ &  $1991.8 \pm 78.8$ & 25 \\   
$\rho$ And & 288294358 & 1686 & 5.16 & $5.3$ & $38.0$ &  $390.3 \pm 11.4$ & 16 \\   
$31$ Aql & 359981217 & 95447 & 5.17 & $291.2$ & $11.9$ &  $1791.8 \pm 225.2$ & 6 \\   
$72$ Her & 9728611 & 84862 & 5.38 & $260.2$ & $32.5$ &  $2241.4 \pm 85.1$ & 25 \\   
$\rho$ CrB & 458494003 & 78459 & 5.39 & $240.3$ & $33.3$ &  $1664.7 \pm 99.9$ & 13, 22 \\   
$51$ Peg & 139298196 & 113357 & 5.45 & $344.4$ & $20.8$ &  $2485.0 \pm 97.8$ & 13, 25, 16 \\   
$15$ Peg & 326202925 & 107975 & 5.52 & $328.1$ & $28.8$ &  $1389.9 \pm 70.7$ & 20 \\   
HD $195564$ & 205591703 & 101345 & 5.66 & $308.1$ & $-9.9$ &  $1167.3 \pm 46.9$ & 25 \\   
HD $190360$ & 105999792 & 98767 & 5.73 & $300.9$ & $29.9$ &  $2358.2 \pm 74.1$ & 13, 16, 20 \\   
HD $89744$ & 8154501 & 50786 & 5.73 & $155.5$ & $41.2$ &  $1028.4 \pm 108.5$ & 26 \\   
HD $49933$ & 281812116 & 32851 & 5.78 & $102.7$ & $-0.5$ &  $1997.7 \pm 117.8$ & 27 \\   
HD $38529$ & 200093173 & 27253 & 5.95 & $86.6$ & $1.2$ &  $622.5 \pm 27.9$ & 13 \\   
$16$ Cyg A & 27533341 & 96895 & 5.99 & $295.5$ & $50.5$ &  $2236.5 \pm 97.9$ & 25, 19, 20 \\   
\bottomrule
\end{tabular*} 
\tablefoot{\tiny The table provides an overview of the stars with published measurements of the stellar diameter from long-baseline interferometry, with stars sorted according to their visual magnitude (``$V$"). The first three columns provide identifiers for the stars in the form of their Bayer/Flamsteed designation (or primary name according to \texttt{SIMBAD}) in addition to their TESS (``TIC") and Hipparcos (``HIP") IDs. The last column provides references to the interferometric sources.}
\tablebib{\tiny (1) \citet{Nordgren2001}; (2) \citet{Mozurkewich2003}; (3) \citet{Thevenin2005}; (4) \citet{Baines2014}; (5) \citet{North2007}; (6) \citet{Boyajian2012}; (7) \citet{Baines2018}; (8) \citet{Mennesson2014}; (9) \citet{Rains2020}; (10) \citet{Baines2023}; (11) \citet{Nordgren1999}; (12) \citet{Baines2021}; (13) \citet{Baines2008}; (14) \citet{Zhao2011}; (15) \citet{Ligi2012}; (16) \citet{Ligi2016}; (17) \citet{Kervella2004}; (18) \citet{Boyajian2012b}; (19) \citet{White2013}; (20) \citet{Karovicova2022}; (21) \citet{Boyajian2008}; (22) \citet{vonBraun2014}; (23) \citet{Maestro2013}; (24) \citet{Kane2015}; (25) \citet{Boyajian2013}; (26) \citet{Schaefer2018}; (27) \citet{Bigot2011}.} 
\end{table*}

\setlength\tabcolsep{0pt} 
\begin{table*} 
\centering 
\caption{TLS and binarity} 
\label{tab:binary}
\scriptsize 
\renewcommand{\arraystretch}{0.8}  
\begin{tabular*}{\linewidth}{@{\extracolsep{\fill}}llllccccccc@{}} 
\toprule 
Name  & HIP & WDS & SB9 & $V$ & $\nu_{\rm max}$ & $P$ (Spec) & $P$ (Orb) & $K_1(/K_2)$ & $v\sin i$ &SONG \\  &      &  &  & (mag) & ($\mu$Hz) & (year/day) & (year/day) & km/s & km/s  & (\# spectra) \\ 
\midrule 
$\eta$ Boo &  67927 & 13547+1824A & 794 & 2.68 & 697 & 494.2 d$^{(1)}$ & 494.2 d & 8.4 & 11.3$^{(2)}$ & 2075\\   
$\zeta$ Her$^{(PN)}$ &  81693 & 16413+3136AB & 915 & 2.81 & 718 & 34.4 yr$^{(3)}$ & 34.5 yr & 4.0 & 6.2$^{(4)}$ & 377\\   
$\mu$$^1$ Her$^{(PN)}$ &  86974 & 17465+2743Aa,Ab & - & 3.42 & 1192 & $(m)$ & 98.9$\pm$22.7 yr\tablefootmark{(m)} & 1.12\tablefootmark{(m)} & 5.2$^{(4)}$ & >106000\\   
$\chi$ Dra$^{PN}$ &  89937 & 18211+7244Aa,Ab & 1058 & 3.55 & 2314 & 280.5 d$^{(5)}$ & 280.5 d & 17.3/24.3 & 5.5$^{(4)}$ & 1079\\   
$\iota$ Peg &  109176 & 22070+2521A & 1354 & 3.77 & 2101 & 10.2 d$^{(6)}$ & 10.2 d & 48.4/77.6 & 9.3$^{(4)}$ & 95\\   
$\upsilon$ Cep$^{PN}$ &  102431 & 20454+5735Aa,Ab & 1489 & 4.52 & 958 & 523.4 d$^{(7)}$ & 523.4 d & 8.3/8.5 & 8.7$^{(4)}$ & 3\\   
HR $3220$$^{PS}$ &  39903 & 08090-6118A & 3654 & 4.74 & 1386 & 899.3 d$^{(8)}$ & 925.0($\pm$10.6) d & 3.2 & 8.8$^{(9)}$ & -\\   
$\omega$ Dra$^{PN}$ &  86201 & 17370+6845A & 981 & 4.77 & 1999 & 5.3 d$^{(6)}$ & 5.3 d & 36.3/44.7 & 5.9$^{(10)}$ & 248\\   
$12$ Boo &  69226 & 14104+2506AB & 804 & 4.82 & 694 & 9.6 d$^{(6)}$ & 9.6 d & 67.1/69.1 & 17.0$^{(10)}$ & 391\\   
$\omega$ And &  6813 & 01277+4524A & 3863 & 4.83 & 1899 & 254.8 d$^{(11)}$ & 254.9 d & 17.8/18.9 & 57.1$^{(10)}$ & 3\\   
$19$ Dra$^{PN}$ &  82860 & 16564+6502C & 938 & 4.88 & 2313 & 52.1 d$^{(12)}$ & $(d)$ & 17.2/$(e)$ & 13.0$^{(13)}$ & 4\\   
$\zeta$ TrA &  80686 & 16285-7005A & 898 & 4.90 & 3215 & 13.0 d\tablefootmark{(a)} & 12.9 d & 7.4 & 3.23$^{(26)}$ & -\\   
$\epsilon$ Lib &  75379 & - & 841 & 4.92 & 775 & 226.9 d$^{(12)}$ & $(b)$ & 14.2 & 10.0$^{(13)}$ & 5\\   
HD $10307$ &  7918 & 01418+4237AB & 2546 & 4.96 & 2490 & 19.7 yr$^{(14)}$ & 19.5 yr & 3.0 & 4.9$^{(4)}$ & 6\\   
$99$ Her$^{PN}$ &  88745 & 18070+3034Aa,Ab & 3658 & 5.05 & 1950 & 56.4 yr$^{(3)}$ & 56.4($\pm$0.2) yr & 3.2 & 4.5$^{(4)}$ & 4\\   
$\kappa$ For &  11072 & 02225-2349A & - & 5.19 & 1160 & $26.54\pm0.05$ yr\tablefootmark{(f)} & 26.5 yr & $5.45\pm0.04$\tablefootmark{(f)} & 6.25$^{(15)}$ & -\\   
$94$ Aqr &  115126 & 23191-1328Aa,Ab & 1438 & 5.20 & 886 & 6.3 yr$^{(12)}$ & 6.3 yr & 6.0 & 4.9$^{(4)}$ & 4\\   
HD $176051$$^{PN}$ &  93017 & 18570+3254AB & 2559 & 5.20 & 2902 & 61.4 yr$^{(16)}$ & 61.4($\pm$0.1) yr & 3.5 & 5.7$^{(4)}$ & 5\\   
$26$ Dra$^{PN}$ &  86036 & 17350+6153AB & 2557 & 5.23 & 3058 & 74.2 yr$^{(16)}$ & 76.1 yr & 3.8 & 6.0$^{(4)}$ & 5\\   
HD $81809$ &  46404 & 09278-0604AB & 1474 & 5.38 & 691 & 34.5 yr$^{(5)}$ & 32.0 yr & 4.8/8.2 & 5.5$^{(4)}$ & 2\\   
HD $14214$ &  10723 & - & 118 & 5.60 & 1667 & 93.3 d$^{(12)}$ & $(g)$ & 19.3 & 5.6$^{(4)}$ & 4\\   
HD $214850$ &  111974 & 22409+1433AB & 1395 & 5.72 & 637 & 20.9 yr$^{(17)}$ & 20.8 yr & 13.6 & 5.3$^{(4)}$ & 4\\   
$35$ Leo &  50319 & 10167+2325B & - & 5.95 & 985 & $531$ d\tablefootmark{(h)} & $(h)$ & NA & 5.5$^{(4)}$ & -\\   
\midrule 
$\theta$ Dra$^{PN}$ &  78527 & - & 882 & 4.01 & 722 & 3.1 d$^{(18)}$ & - & 25.1/66.0 & 30.7$^{(4)}$ & 2\\   
$\tau$$^1$ Hya &  46509 & 09291-0246A & 2549 & 4.59 & 1572 & 7.7 yr$^{(19)}$ & - & 2.8 & 32.2$^{(4)}$ & 3\\   
$\sigma$ Cet &  11783 & 02321-1515Aa,Ab & - & 4.74 & 913 & $3.2$ yr\tablefootmark{(j)} & - & $27/19.5$\tablefootmark{(j)} & - & 3\\   
$16$ UMa &  45333 & - & 558 & 5.18 & 1514 & 16.2 d$^{(20)}$ & - & 35.3/65.0 & 6.5$^{(4)}$ & 3\\   
$\rho$ Tuc &  3330 & - & 40 & 5.38 & 724 & 4.8 d$^{(21)}$ & - & 26.1 & 23.2$^{(9)}$ & -\\   
HD $104304$ &  58576 & 12007-1028A & - & 5.54 & 3224 & $48.5$ yr\tablefootmark{(n)} & - & NA & 2.87$^{(22)}$ & -\\   
HD $46569$$^{PS}$ &  31079 & - & - & 5.58 & 932 & $25.6\pm1.5$ yr\tablefootmark{(i)} & - & $1.7\pm0.08$\tablefootmark{(i)} & 5.6$^{(9)}$ & -\\   
$72$ Psc &  5081 & 01051+1457A & 1654 & 5.64 & 946 & 50.4 d$^{(23)}$ & - & 36.0/39.9 & 5.0$^{(24)}$ & 6\\   
HD $121384$ &  68101 & 13565-5442A & - & 6.00 & 425 & $178.7\pm0.1$ d\tablefootmark{(o)} & - & $10.9\pm2.2$\tablefootmark{(o)} & NA & -\\   
\midrule 
$\eta$ Cas &  3821 & 00491+5749A & - & 3.46 & 2839 & - & 480.0 yr & NA & 5.4$^{(4)}$ & 6058\\   
$\alpha$ For &  14879 & 03121-2859A & - & 3.80 & 1128 & - & 269.0($\pm$22.5) yr & NA & 6.24$^{(15)}$ & 3\\   
$\iota$ Vir &  69701 & 14190-0636C & - & 4.07 & 644 & - & 55 yr\tablefootmark{(k)} & NA & - & 1\\   
$\theta$ Per &  12777 & 02442+4914A & - & 4.10 & 2314 & - & 2720 yr & NA & 10.2$^{(4)}$ & 4\\   
o$^2$ Eri &  19849 & 04153-0739A & - & 4.43 & 3433 & - & 8000 yr\tablefootmark{(p)} & NA & 1.23$^{(25)}$ & 3\\   
$\mu$ Cyg &  107310 & 21441+2845AB & - & 4.49 & 1213 & - & 789.0($\pm$37.9) yr & NA & 5.45$^{(26)}$ & 2\\   
$\psi$$^1$ Dra A$^{PN}$ &  86614 & 17419+7209A & - & 4.57 & 1232 & - & 10000 yr & NA & 12.5$^{(27)}$ & 6\\   
 &   &  &  &  &  & 18.5 yr\tablefootmark{(c)} & -  & 5.2/11.1\tablefootmark{(c)} & \\  
$\sigma$$^2$ UMa &  45038 & 09104+6708AB & - & 4.80 & 1354 & - & 921.0 yr & NA & 7.55$^{(26)}$ & 3\\   
$17$ Cyg$^{PN}$ &  97295 & 19464+3344A & - & 5.00 & 1484 & - & 232.0($\pm$3.4) yr & NA & 9.0$^{(13)}$ & 1\\   
HD $62644$$^{PS}$ &  37606 & 07430-4511AB & - & 5.04 & 708 & - & 380.0($\pm$1.5) d & NA & 2.49$^{(22)}$ & -\\   
$\kappa$ Del &  101916 & 20391+1005A & - & 5.07 & 622 & - & 45.0 yr & NA & 4.42$^{(22)}$ & 3\\   
$94$ Cet &  14954 & 03128-0112A & - & 5.07 & 1267 & - & 2029\tablefootmark{(l)}($\pm$304.5) yr & NA & 10.3$^{(4)}$ & 149\\   
HD $100203$ &  56290 & 11323+6105AB & - & 5.46 & 1532 & - & 72.7($\pm$0.2) yr & NA & 6.1$^{(10)}$ & -\\   
HD $53705$$^{PS}$ &  34065 & 07040-4337A & - & 5.56 & 1989 & - & 4.1 yr & NA & 4.3$^{(9)}$ & -\\   
$10$ Ari &  9621 & 02037+2556AB & - & 5.64 & 574 & - & 325.0($\pm$8.0) yr & NA & 7.482$^{(28)}$ & 1\\   
$16$ Cyg A$^{PN}$ &  96895 & 19418+5032Aa & - & 5.99 & 2236 & - & 13513 yr & NA & - & -\\   
\bottomrule
\end{tabular*} 
\tablefoot{\tiny The table provides an identification of the binarity for the seismic stars in the sample. The first four columns provide identifiers for the stars in the form of their Bayer/Flamsteed designation (or primary name according to \texttt{SIMBAD}) and their             ``HIP'', ``WDS'', and ``SB9'' IDs. Super-scripts of $PN$ or $PS$ on the star's name refer to their                 position within (or near when in parenthesis) the northern ($N$) or southern ($S$) PLATO fields (see Table~\ref{tab:plato}). ``$V$'' gives the $V$-band magnitude, ``$\nu_{\rm max}$'' (central value only) is adopted from Table~\ref{tab:all_seis}.            ``$P$'' refers to the binary period, where spectroscopic (``Spec'') periods are obtained from SB9 (unless otherwise stated with a letter reference), with reference to the adopted most recent period determination. Orbital (``Orb'') binary periods are obtained from                ORB6 (unless otherwise stated with a letter reference), with non-zero RMSDs between ORB6 and OARMAC entries in parenthesis. ``$K_1(/K_2)$'' gives the semi-amplitude(s) corresponding to the $P$ (Spec) reference.                        ``SONG'' gives the number of spectra available for the target in the SONG database SODA (\url{https://soda.phys.au.dk/}). \\ 
\tablefoottext{a}{Included in SB9 with reference to \citet{1928AnCap..10....8S}, but more recent spectroscopic analysis is provided by \citet{Skuljan2004};}\tablefoottext{b}{Adopting the spectroscopic orbital information, \citet{Castelaz1989} provides astrometric binary orbit solution;}\tablefoottext{c}{$\psi^1$ Dra is a hierarchical multiple system where the AB system is determined astrometrically (in ORB6), while the AC system is determined spectroscopically by \citet{Gullikson2015};}\tablefoottext{d}{Adopting the spectroscopic orbital information, \citet{Ren2013} provides astrometric binary orbit solution;}\tablefoottext{e}{Found to be SB2 by \citet{2013AJ....145...41K}, but $K_2$ is not provided due to weakness of lines;}\tablefoottext{f}{\citet{Fekel2018} \citep[see also][]{Tokovinin2013} provides a joint analysis of spectroscopic, astrometric, and visual orbit data;}\tablefoottext{g}{Adopting the spectroscopic orbital information, \citet{Fekel2007} provides astrometric binary orbit solution;}\tablefoottext{h}{\citet{Tokovinin2014a} provides an orbital period for this SB1 binary (but no semi-amplitude), in agreement with the Gaia DR3 Non-single-star (NNS) period of $524\pm6$ days \citep{Gaia_NNS2023};}\tablefoottext{i}{\citet{Zechmeister2013};}\tablefoottext{j}{\citet{McLaughlin1962};}\tablefoottext{k}{\citet{Gontcharov2010};}\tablefoottext{l}{Adopting the value in OARMAC from \citet{Roberts2011_94Cet};}\tablefoottext{m}{\citet{Roberts2016muHer} provides a combined astrometric ans spectroscopic analysis;}\tablefoottext{n}{Preliminary period by \citet{Schnupp2010} from 9 yr RV data (no semi-amplitude provided);}\tablefoottext{o}{\citet{Jenkins2015} provides spectroscopic solution and finds indications of a high eccentricity ($ecc{\sim}0.84$) low-mass companion;}\tablefoottext{p}{Estimate by \citet{Heintz1974}, ORB6 only lists period for BC component of this triple system.}} 
\tablebib{\tiny (1) \citet{1957ApJ...125..696B}; (2) \citet{2023A&A...676A.129H}; (3) \citet{2015A&A...574A...6A}; (4) \citet{2017AJ....153...21L}; (5) \citet{2000A&AS..145..215P}; (6) \citet{2011AJ....142....6B}; (7) \citet{1999Obs...119..272G}; (8) \citet{1993Obs...113..126M}; (9) \citet{2012A&A...542A.116A}; (10) \citet{2009A&A...493.1099S}; (11) \citet{2011Obs...131..225G}; (12) \citet{2013AJ....145...41K}; (13) \citet{2005PASJ...57...13T}; (14) \citet{2018A&A...619A..81H}; (15) \citet{2023ApJS..266...11B}; (16) \citet{1991A&A...248..485D}; (17) \citet{1985JRASC..79..167B}; (18) \citet{2002ApJ...564.1007M}; (19) \citet{2012MNRAS.422...14H}; (20) \citet{2015AJ....149...63F}; (21) \citet{1929PASP...41..371N}; (22) \citet{2019A&A...629A..80H}; (23) \citet{2001Obs...121..162G}; (24) \citet{1970CoAsi.239....1B}; (25) \citet{2018AJ....155..126D}; (26) \citet{2010A&A...520A..79M}; (27) \citet{2023A&A...671A...7S}; (28) \citet{2020AJ....160..120J} } 
\end{table*} 

\FloatBarrier


\section{Notes on binarity for individual stars}\label{app:bin_indv}

In addition to the stars listed in \tref{tab:binary} we identified 8 stars ($\theta$ UMa, $\eta$ Cas, $\lambda$ Ser, HD 5015, 14 Boo, $\upsilon$ And, HD 46588, and $\beta$ CVn) that are labeled as spectroscopic binaries on \texttt{SimBad} from appearing in the 
Seventh Catalogue of Spectroscopic Binary Orbits \citep[SBC7;][]{Batten1978_SBC7}. In all cases, the discovery reference in SBC7 is to \citet{Abt1976_thetaUma} where the stars are given as new (first orbit) binary detections. However, nearly all new detections in \citet{Abt1976_thetaUma} were effectively refuted by \citet{Morbey1987} (and acknowledged by \citet{Abt1987}), including all the above cases identified in our sample.

\subsubsection*{Below, we provide notes on the binarity of individual stars in our sample:} 

\paragraph{\object{104 Tau} (\object{HIP 23835}/\object{m Tau})} has been intensely studied for binarity and is included in ORB6 with reference to \citet{Eggen1956} who finds two possible period solutions of 1.19 yr and 2.38 yr for the system. However, as discussed in \citet{Tokovinin2012} \citep[see also][]{Heintz1984} the system has remained unresolved in many speckle interferometric studies and found, \eg, by \citet{Nidever2002} to be RV stable over a baseline ruling out the previously published period(s).

\paragraph{\object{$\upsilon$ And} (\object{HIP 7513}; \object{Titawin})} appears in WDS and is listed in \citet{Tokovinin2014a} \citep[see also][]{Lowrance2002,Raghavan2010} as a multiple hierarchical system with a wide common proper motion companion (WDS component D) in a $>16.000$ yr orbit. $\upsilon$ And is the host of at least three confirmed exoplanets \citep{Butler1997,Butler1999}.

\paragraph{\object{$\rho$ CrB} (\object{HIP 78459})} is listed in ORB6 with reference to \citet{Gatewood2001} who argued, based on an analysis of Hipparcos and their astrometric data, that the first claimed planetary companion by \citet{Noyes1997} at a period of $39.6$ days must instead be a stellar-mass object. However, the significance of this claim was questioned by \citet{Zucker2001} and later refuted by \citet{Bender2005}, who was unable to detect the alleged M dwarf companion from high-resolution infrared spectroscopy. Later RV follow-up studies (of which there are many) have currently identified four exoplanets \citep[\eg,][]{Brewer2023}.

\paragraph{\object{$\kappa$ For} (\object{HIP 11072})} is a triple star system consisting of a tight binary (radio emitting) M-dwarf pair in a $3.7$ day orbit, which orbits the main star with a period of $26.54\pm0.05$ years \citep[][and references therein]{Fekel2018, Tokovinin2013}. It is listed on \texttt{SimBad} as being part of the young moving group IC 2391 \citep{Nakajima2012}, but according to \citet{Tokovinin2013}, the calculation of the kinematic parameters leading to this conclusion were biased in that they overlooked the companion.

\paragraph{\object{$\sigma$ Cet} (\object{HIP 11783})} is given by \citet{McLaughlin1947} and \citet{McLaughlin1962} as a triple spectroscopic binary, consisting of an A-star pair in a $3.76$ day orbit ($K_1=K_2=110$ km/s), which orbits the main G-type star at a period of $3.3$ yr. We note that the above references are short notes with limited information, and no later or follow-up studies have been identified.

\paragraph{\object{$\iota$ Vir} (\object{HIP 69701})} appears to be part of a hierarchical quadruple system, consisting of two binary pairs in orbit around a common centre of mass. $\iota$ Vir (WDS C component) is an astrometric binary with a low-mass companion in an orbit with a preliminary period of ${\sim}55$ yr \citep{Gontcharov2010}. The binary system HIP 69962 (the WDS AB component listed in ORB6 with reference to \citet{Videla2022}, and also included in SB9 with a period of ${\sim}18.7$ yr \citep{Halbwachs2018}) is a likely wide ($\rho =$ 57.1 arcmin (0.37 pc)) binary component to $\iota$ Vir \citep{Shaya2011}. \citet{Fuhrmann2015} discuss $\iota$ Vir as a possible blue straggler given its higher X-ray luminosity compared to the wide companion. 

\paragraph{\object{$\mu$ Her} (\object{HIP 86974})} is a quadruple system consisting of an M-dwarf Ab component orbiting the primary G5IV star at a period of ${\sim}99$ yr \citep{Roberts2016muHer}, and a faint BC binary component consisting of an M-dwarfs pair with a period of ${\sim}43.5$ yr \citep{Prieur2014,Mann2019}. We note that currently, the SONG project has observed $\mu$ Her for more than ten years for a detailed asteroseismic characterisation \citep{Grundahl2017}, and these observations can also greatly help to improve the Aa-Ab orbital solution and yield an independent mass constraint (Marcussen et al., in prep.). 

\paragraph{\object{94 Cet} (\object{HIP 14954})} is a triple-star system \citep{Wiegert2016} consisting of an M-dwarf binary pair with a period of ${\sim}1$ yr \citep{Roell2012} in a wide $2029$-year-long orbit around 94 Cet A \citep{Roberts2011_94Cet}. 94 Cet is furthermore a known asteroseismic target \citep{Deal2017} and an exoplanet host \citep{Mayor2004} (see also \tref{tab:exo}).

\paragraph{\object{$\lambda$ Ara} (\object{HIP 86486})} is speculated by \citet{Fuhrmann2011} to be an equal mass binary from the discrepancy between a spectroscopic \logg with that of astrometry from Hipparcos. Only vague lower limits on period are provided, and $\lambda$ Ara has not been included in \tref{tab:binary}. We note, however, that if we use \eqref{eq:numax} with our measured \numax of $\rm 1476\pm 43\, \mu Hz$ and the \teff of $6532 \pm 80\rm\,  K$ from \citet{Fuhrmann2011} we obtain $\logg= 4.15 \pm 0.01$ dex, which is fully consistent with the Hipparcos \logg of $4.1\pm 0.1$ dex.

\end{appendix}

\end{document}